# Revealing the State of the Art of Large-Scale Agile Development Research: A Systematic Mapping Study

Ömer Uludağ[a,*], Pascal Philipp[a], Abheeshta Putta[b], Maria Paasivaara[b,c], Casper Lassenius[b,d], Florian Matthes[a]

[a] *Technical University of Munich, Department of Informatics, Munich, Germany*
[b] *Aalto University, Department of Computer Science, Espoo, Finland*
[c] *LUT University, Finland*
[d] *Simula Research Laboratory, Oslo, Norway*

## Abstract

*Context:* Success with agile methods in the small scale has led to an increasing adoption also in large development undertakings and organizations. Recent years have also seen an increasing amount of primary research on the topic, as well as a number of systematic literature reviews. However, there is no systematic overview of the whole research field. *Objective:* This work identifies, classifies, and evaluates the state of the art of research in large-scale agile development.
*Method:* We conducted a systematic mapping study and rigorously selected 136 studies. We designed a classification framework and extracted key information from the studies. We synthesized the obtained data and created an overview of the state of the art.
*Results:* This work contributes with (i) a description of large-scale agile endeavors reported in the industry, (ii) a systematic map of existing research in the field, (iii) an overview of influential studies, (iv) an overview of the central research themes, and (v) a research agenda for future research.
*Conclusion:* This study portrays the state of the art in large-scale agile development and offers researchers and practitioners a reflection of the past thirteen years of research and practice on the large-scale application of agile methods.

## 1. Introduction

Contemporary business environments are characterized by high unpredictability due to rapidly shifting customer demands and technological advancements, implying that flexibility, adaptability, and learning are crucial to business success [76]. Over the past decades, software has become an integral part of many products and services [81]. To react quickly to changing environments and fluctuating customer requirements, the agile movement emerged in the 1990s, leading to the creation of the Agile Manifesto[1] and many agile methods, e.g., Extreme Programming (XP) [14] and Scrum [94] [54, 81]. Agile methods were originally designed for small, co-located, and self-organizing teams that produce software in close collaboration with business customers, using regular feedback and rapid development iterations [32, 36]. The successful application of agile methods in small projects inspired companies to increasingly adopt agile methods also in large-scale projects and organizations [35]. During the last decade, agile methods have been extended to better fit large-scale settings. Several scaling frameworks have been created both by some custodians of existing agile methods and by others who have worked with companies in scaling agile methods to their settings [101]. As the frameworks claim to provide off-the-shelf solutions to the problems of scaling, their adoption has rapidly increased in practice, as confirmed by the latest State of Agile survey [31]. The survey shows that many large software-intensive companies have adopted scaling frameworks to address challenges accompanied by the scaling of agile methods [5, 24]. While there are more than 20 available frameworks [101], the most popular ones are [31]: Scaled Agile Framework (SAFe) [51], Large-Scale Scrum (LeSS) [23], and Disciplined Agile Delivery (DAD) [8].

As the popularity of applying agile methods at scale has increased in the industry, scientific research on the topic has emerged in recent years. Eleven years ago, at the International Conference on Agile Software Development, industrial practitioners were asked to create a backlog of topics they think should be studied. They voted "*Agile and large projects*" as a top burning research question [41]. After that, nine International Workshops on Large-Scale Agile Development at the yearly International Conferences on Agile Software Development have gathered researchers and practitioners to discuss recent large-scale agile development studies and create research agendas for future research on that area (cf. [33, 70]). In recent years, research publications on large-scale agile have been published at scientific conferences and journals, and the body of knowledge has grown immensely (cf. [32, 40]). Although scientific articles on large-scale agile development are mainly primary studies, we also identified several secondary studies that synthesize the scientific knowledge on large-scale agile devel-

*Corresponding author
*Email address:* `oemer.uludag@tum.de` (Ömer Uludağ)

[1] `https://agilemanifesto.org/`, last accessed on: 19-04-2022.

*Preprint submitted to Journal of Systems and Software* *July 29, 2022*

opment related to specific topics, e.g., the systematic literature review on "*challenges and success factors for large-scale agile transformations*" [32]. A steadily growing number of primary studies is a valuable indicator of the increasing maturity of a research field. It leads to a critical tipping point when secondary studies can be conducted, which then facilitates the aggregation of the results of the primary studies. As the research area of large-scale agile development matures and the number of studies increases, there is a need to systematically identify, analyze, and classify the state of the art of this research field. While the number of primary studies is sufficient, so far, no systematic mapping study has been published to provide a comprehensive overview of the state of research in this area. This mapping study aims to close this gap and provide an overview of the research activities pertaining to large-scale agile development. The main contributions of this study are as follows:

- A *description* of large-scale agile endeavors and their characteristics reported in the industry,
- A *systematic map* classifying, comparing, and evaluating existing research on large-scale agile development,
- An *overview* of the most influential studies on large-scale agile development,
- An *overview* of the central research themes and main findings on large-scale agile development, and
- A *research agenda* consisting of a detailed description of research gaps and a list of open research questions in large-scale agile development.

The remainder of this paper is structured as follows. Section 2 presents background and related work. Section 3 portrays the procedure of the systematic mapping study. Section 4 shows the results of this study. Section 5 discusses the results. Section 6 describes the threats to validity. Section 7 concludes the paper along with remarks on future research.

## 2. Background and related work

This section provides the background on agile software development and large-scale agile development and gives an overview of related work, including other secondary studies.

### *2.1. Agile software development*

Software development methods have undergone a dramatic evolution from traditional sequential development approaches to more flexible and adaptive development approaches due to the market's increasing demands and customer requirements volatility [81]. To address these needs, many companies have started to adopt agile software development methods [80]. Various software practitioners introduced several agile methods in recent years, e.g., the Crystal Method, the Dynamic System Development Method, and the Feature-Driven Development Method, that comprise a set of iterative and incremental methods based on specific values and principles defined in the Agile Manifesto [1]. Currently, the most widely adopted agile methods in industry are Scrum and XP [44]. These methods have aroused great interest both in practice and academia [3].

Agile methods have received both acceptance and criticism in the industry [20]. On the one hand, they have proven to be successful in improving quality [93], productivity [39], and customer satisfaction [39]. On the other hand, there is concern that these methods may not be suitable for large-scale environments [39]. Dybå and Dingsøyr [39] note that there is evidence that suggests combining agile and traditional methods, i.e., hybrid methods, in large undertakings and recommends that practitioners should carefully examine and compare project characteristics with the required characteristics of the suitable agile method(s). Previous literature has also reported on the success of hybrid approaches [67], e.g., recent research by Klünder et al. [59] concludes that projects devising hybrid methods have about a 5% better chance of achieving their goals.

According to Digital.ai [31], companies adopt agile methods to accelerate software delivery, manage changing priorities, increase productivity, improve alignment between business and IT, and enhance quality, to name a few. However, adopting agile methods is not easy, as agile methods do not rely solely on the appropriate application of individual tools or practices but rather often demand a holistic way of thinking and mindset. Thus, the adoption of agile methods often requires a change in the entire organizational culture [69]. Developing an agile mindset and changing the company's culture takes time and effort, and if this effort is neglected, the organization can fall back into old habits and fail to reap the full benefits of agility [61].

### *2.2. Large-scale agile development*

The success of agile methods for small, co-located teams has incited companies to increasingly apply agile practices to large-scale projects [32, 36, 84]. However, the large-scale adoption of agile methods has proven to be very challenging [38]. The challenges of adopting agile practices at large-scale are partly related to the organization's size, as the difficulties of adopting agile methods increase with the size of the organization, i.e., in large organizations, products are more complex, and the inter-dependencies between teams are greater than in smaller organizations [7, 38], leading to inertia and slowing down the change process [66]. Another challenge in large organizations is the need for coordination and communication between multiple teams and also between different organizational units that often do not work in an agile manner, requiring additional coordination mechanisms between teams and also organizational units [65]. While agile methods primarily focus on intra-team practices that work well in small organizations, agile methods do not provide sufficient guidance on how agile teams should interact in large environments [68]. Hence, large organizations must adapt the practices to their specific needs. As a result, practices may need to be put in place that require additional formal communication, which might reduce their agility [65]. As often large organizations are globally distributed and agile methods are primarily based on frequent internal and external collaboration and communication [49], the use of agile practices in globally distributed projects can be challenging [46].

To address issues associated with adopting agile practices in large-scale organizations and projects, consultants and software practitioners have proposed several scaling agile frameworks, e.g., SAFe, LeSS, and DAD, which include predefined workflow patterns to deal with concerns related to large numbers of teams, inter-team coordination, and customer involvement [5, 36]. As large organizations are increasingly pressured and expected to become more agile, and scaling frameworks claim to provide off-the-shelf solutions to scaling, companies have begun to adopt these frameworks at an increasing rate in recent years [24]. This trend is also confirmed by the annual non-scientific survey on the State of Agile by Digital.ai [31].

### 2.2.1. Definition of large-scale agile development

"*When can a company be said to be adopting agile methods at scale rather than just at a smaller scale?*" Several researchers have already tried to answer this question and have proposed several definitions on what "*large-scale agile development*" means. These definitions usually include the number of people or agile teams engaged in the effort or associated costs or duration of a project [35]. Berger and Beynon-Davies [18], for example, classify a project as a large-scale agile project if the project costs exceed 10 million GBP. Bjarnason et al. [19] use the project duration of more than two years as an indicator for classifying a project as large-scale. According to Paasivaara et al. [78], a project with more than 40 people or seven agile teams involved can be considered large-scale. To bring a conceptual clarity of what "*large-scale agile development*" means, Dingsøyr et al. [35] have identified several different interpretations and proposed a taxonomy for large-scale agile development that uses the number of collaborating and coordinating teams to define the scale of an agile project. The taxonomy developed by Dingsøyr et al. [35] consists of three categories: (i) small-scale agile projects with one team that can use traditional agile practices, e.g., daily meetings, sprint planning, review, and retrospective meetings, for intra-team coordination, (ii) large-scale agile projects with 2–9 agile teams that use new forums, e.g., a Scrum-of-Scrums (SoS) for cross-team coordination, and (iii) very large-scale agile projects with at least 10 agile teams that require several forums for inter-team coordination, e.g., multiple SoS. According to Dingsøyr et al. [35], a project can be considered a large-scale project if it has at least two coordinating agile teams. Fuchs and Hess [42] extend this definition and state that the term "*large-scale agile development*" has multiple interpretations: (i) the use of agile methods in large teams, (ii) the employment of agile methods in large organizations, (iii) the application of agile methods in large multi-team settings, i.e., "*large agile multi-team settings*", or (iv) the usage of agile practices in organizations as a whole, i.e., "*organizational agility*"[2]. Like Fuchs and Hess [42], we focus on the latter two definitions and understand the large-scale agile adoption of agile methods in large agile multi-team settings with at least two teams or the large-scale adoption of agile methods on the organizational level comprising multiple large agile multi-team settings.

[2]The term "*organizational agility*" should not be confused with the term "*enterprise agility*", which constitutes a research direction for itself. According to our understanding, the term "*enterprise agility*" comprises the adoption of agile methods at the company level. In contrast, the term "*organizational agility*" insinuates the adoption of agile methods in large organizational units of companies, e.g., departments, divisions, or units. Hence, the term "*organizational agility*" can be seen as a subset of the term "*enterprise agility*".

### 2.3. Secondary studies on large-scale agile development

Several secondary studies have systematically analyzed the literature on specific topics related to large-scale agile development. We identified 13 secondary studies (see Table 1), of which 10 were systematic literature reviews, and three were either structured or simple literature reviews. Systematic literature reviews have addressed several topics, e.g., the identification and description of challenges (cf. [32, 95, 100]), success factors (cf. [32, 95]), and typical roles involved in large-scale agile endeavors (cf. [43, 99]). Various researchers also conducted systematic literature reviews and multi-vocal literature reviews on scaling frameworks to identify challenges, benefits, and success factors related to the adoption of scaling frameworks (cf. [40, 53, 84]). We also identified simple literature reviews comparing various scaling frameworks (cf. [5, 77]).

Existing secondary studies cover and analyze specific research topics on large-scale agile development in-depth, e.g., large-scale agile transformations, scaling agile frameworks, or global and distributed software engineering. For instance, Dikert et al. [32] deal with the research topic of large-scale agile transformations and explore the challenges and success factors reported for these transformations. For example, Edison et al. [40] address the research topic of scaling agile frameworks by analyzing and comparing them based on their principles, practices, tools, and metrics, as well as identify gaps in the literature and proposing needs for future research. Shameem et al. [95] deal with the research topic of global and distributed software engineering by identifying key success factors for scaling agile methods in these environments.

While existing studies focus on specific research topics in large-scale agile development and analyze them in-depth, we could not identify any systematic mapping studies that would provide an overview of the overall state of research in the field, structuring the body of knowledge in large-scale agile development. Hence, this paper aims to fill this gap by providing an overview of the research activities in large-scale agile development based on a systematic mapping study. As a systematic mapping study, the scope is broader than any existing systematic literature review, and the goals differ. While there is some overlap between this study and the systematic literature reviews of Dikert et al. [32] and Edison et al. [40], we provide a broader overview of the overall field of large-scale agile development than any single systematic literature review.

## 3. Research process

While systematic literature reviews [21, 56] are a common means for identifying, evaluating, interpreting, and comparing all available research related to a particular research question, a systematic mapping study maps out the existing research rather than answering a detailed research question [22, 82]. Hence, we opted to conduct a systematic mapping study, as it is capable of dealing with broad research areas and provides a systematic and objective procedure for identifying, categorizing, and analyzing the existing literature [22, 56, 82].

Table 1: Secondary studies on large-scale agile development

| Year | Study | No. of studies | Topic |
|---|---|---|---|
| 2014 | [86] | 75 | This study provides an understanding of research problems and themes in large-scale, distributed agile development environments based on IEEE publications between 2005 and 2014. |
| 2015 | [92] | 51 | This paper identifies research issues related to the scalability of agile methods for large-scale projects. Moreover, this study unveils existing methods, approaches, frameworks, and practices that can facilitate the application of agile methods for large-scale agile projects, as well as their limitations. |
| 2016 | [32] | 52 | This article answers the question of how large organizations or projects adopt agile and/or lean methods at scale by focusing on the reported challenges and success factors encountered in large-scale agile transformations. |
| 2017 | [43] | 42 | This work provides an analysis of roles assigned with the responsibility for inter-team coordination of large-scale agile development settings. |
| 2017 | [6] | 60 | This paper presents challenges when developing quality requirements in large-scale distributed agile projects and reveals agile practices that have contributed to the emergence of these challenges. This work also summarizes solutions proposed in the literature to address challenges associated with developing quality requirements in large-scale distributed agile projects. |
| 2017 | [95] | 20 | This study presents key factors that can positively impact agile development activities in large-scale, globally distributed software development environments. |
| 2017 | [99] | 146 | This paper gives an overview of 20 identified scaling agile frameworks and describes the responsibilities of different architects in these frameworks. |
| 2018 | [100] | 76 | This work presents different stakeholders and their recurring challenges in large-scale agile projects. |
| 2018 | [96] | 18 | This article shows an overview of human-related factors that can negatively impact agile practices in large-scale, globally distributed software development environments and proposes a hypothetical model of the identified challenges related to the scaling of agile methods. |
| 2018 | [84] | 88 | This study provides an analysis of the scientific and grey literature describing the challenges and benefits encountered by organizations when adopting SAFe. |
| 2018 | [53] | 12 | This work examines practices, challenges, and success factors for scaling agile methods in large companies, reported in the literature and within a large software company. |
| 2019 | [4] | 58 | This paper shows a set of motivators for the large-scale adoption of agile methods from a management perspective. |
| 2021 | [40] | 191 | This article compares five scaling methods based on each method's principles, practices, tools, and metrics. It also presents the challenges and success factors described in the literature when applying these methods. |

### *3.1. Objectives and research questions*

We used the Goal-Question-Metric paradigm [11] to formulate the objective of this study: to **analyze** peer-reviewed literature **for the purpose of** providing an overview of the state of the art **with respect to** the characterization of the topic, the available research on the topic, salient publications and researchers, well-established research streams, and potential research gaps **from the point of view of** scholars and practitioners **in the context of** large-scale agile development. The overall research question of this study is:

***Research Question:*** *What is the state of the art of the literature pertaining to large-scale agile development?*

To answer this question, we further decomposed it into four specific questions:

***RQ1:*** *How is large-scale agile development characterized in the literature?*

Currently, there is no consensus on the actual meaning of the term "*large-scale agile development*" [90] which is why the lack of conceptual clarity regarding this term inhibits effective collaboration and progress in the research area of large-scale agile development [35]. Thus, this research question strives to explore how the extant literature characterizes this term based on the reported case characteristics.

***RQ2:*** *What are the publication trends and characteristics of existing research on large-scale agile development?*

A valuable instrument for understanding the nature of a research area is the investigation of research trends and the systematic classification of extant studies [22, 82]. Accordingly, this research question intends to map the frequency of publications over time to identify research trends and strives to categorize and aggregate extant studies to structure the research area.

***RQ3:*** *What are the seminal studies in large-scale agile development?*

There is perhaps no better way to understand and explore the intellectual structure of a research field than to identify its seminal works [34, 71]. Therefore, this research question aims to identify the protagonists and salient publications in the research field by employing bibliometric analysis.

***RQ4:*** *Which research streams and promising future research directions exist in large-scale agile development?*

One approach to assess the state of the art and maturity level of a research area is to identify main research streams and reveal potential research gaps [58, 82]. Hence, this research question strives to map the general structure of the research field by identifying central research themes. This research question also aims to outline a research agenda for future research efforts by analyzing existing gaps in the research streams.

### 3.2. Mapping study execution

When conducting this study, we followed the guidelines for performing systematic mapping studies [82] and systematic literature reviews [56]. We decided to combine both approaches [55] since two of our research questions, namely *RQ1* and *RQ4*, could not be answered by mappings alone. The execution procedure of this systematic mapping study consisted of three phases: (i) study search, (ii) study selection, (iii) and data extraction, as described in the following.

#### 3.2.1. Study search

In the first phase, we conducted a two-step study search procedure, which included the definition of a search strategy and the screening of related studies consisting of a preliminary search and main search, as described subsequently.

*Study search strategy.* Defining a proper search strategy is essential to ensure that the literature review results are complete [105]. Various researchers have proposed several techniques to develop appropriate search strategies (cf. [83, 105]). We followed the recommendations by Zhang et al. [105] to elaborate on our search strategy, which we describe in the following.

*Search approach.* Our search comprised two main steps: preliminary search and main search. The purpose of the preliminary search was two-fold. First, we wanted to use the preliminary search to construct and evaluate different search strings for the main search. Second, we used the preliminary search as a "*sanity check*" to identify a set of relevant papers that the actual main search should also retrieve. Following the preliminary search, we performed a database keyword search during the main search to retrieve relevant studies in electronic databases listed in Table 2. Afterward, we merged the search results from the preliminary and main searches and excluded duplicate studies. We then included the resulting collection of potentially relevant papers for the study selection phase.

Table 2: Databases used in the main search

| # | Search engine | Website |
|---|---|---|
| DB1 | IEEE Xplore | `http://ieeexplore.ieee.org/` |
| DB2 | ACM Digital Library | `http://dl.acm.org/` |
| DB3 | Science Direct | `http://www.sciencedirect.com/` |
| DB4 | Web of Science | `https://www.webofknowledge.com/` |
| DB5 | Scopus | `https://www.scopus.com/home.uri` |
| DB6 | AIS eLibrary | `https://aisel.aisnet.org/` |

*Data sources.* According to Brereton et al. [21], many different electronic sources should be searched since no single source can find all relevant primary studies. Therefore, as suggested by Kitchenham and Brereton [55], we selected six electronic databases (see Table 2) as the primary sources for the systematic mapping study for covering as many potentially relevant studies as possible. The selection of the electronic databases was guided by: (i) the fact that two of them, i.e., ACM Digital Library and IEEE Xplore, are the largest scientific databases in the field of software engineering [55, 83], (ii) the fact that three of them offer broad coverage of diverse research fields, i.e., Science Direct, Web of Science, and Scopus [89], (iii) and the fact that one of them, i.e., AIS eLibrary, contains articles from the primary information systems research dissemination outlets [74]. We excluded Google Scholar as the results tend to overlap with ones from the included electronic databases [25].

*Search terms.* To identify all relevant studies, we used a five-step strategy [57] for constructing the search terms:

1. deriving main search terms from the study topic and the formulated research questions based on the PICO (Population, Intervention, Comparison, Outcomes) criteria,
2. identifying synonyms and alternative spellings for the main search terms,
3. checking the keywords in relevant papers,
4. incorporating synonyms and alternative words using the Boolean *OR* operator, and
5. linking the search terms using the Boolean *AND* operator.

We only used the first two components of the PICO approach, i.e., population and intervention, and omitted the outcome and context facets from the search structure since our research questions did not warrant a restriction of the results to a specific outcome or context. Similar to Yang et al. [104], the population facet represents the first search set of the overall search string and contains the terms of agile methods that are popularly used in various systematic literature reviews and surveys on agile software development (cf. [38, 88, 98]). Following Dikert et al. [32], we extended the first search set by explicitly stating that the application of agile methods outside of software engineering, e.g., agile manufacturing, should be excluded. The intervention comprises two search sets. The first set includes terms related to the objective of applying agile methods on a larger scale, namely "*large-scale*" and "*scaling*". These two terms are often used within titles and as keywords in related publications on large-scale agile development (cf. [36, 53]). Inspired by Yang et al. [104], the second intervention search set entails terms of large-scale agile development methods and frameworks. We used the results of Uludağ et al. [99] to obtain a list of these methods and frameworks. Following this strategy, we conducted a series of tests and refinements in the preliminary search. The blending of the search sets resulted in the generic search string for the main search. The final generic search string used was:

*Agile software development* ***AND*** *(Large-scale development* ***OR*** *Scaling agile frameworks)*

Table 3 lists the final list of applied search sets and strings. As each electronic database has a specific syntax for search terms, we adapted our search string to the particular syntax requirements of the search engines.

*Time span.* We cover the period from February 2001, when the

Table 3: Overview of search sets and corresponding terms

| Set | Search term |
|---|---|
| Agile software development | ((agile **OR** agility **OR** extreme programming **OR** XP **OR** feature driven development **OR** FDD **OR** scrum **OR** crystal **OR** pair programming **OR** test-driven development **OR** TDD **OR** leanness **OR** lean software development **OR** lean development **OR** LSD) **AND NOT** manufacturing) |
| | **AND** |
| Large-scale development | ((large-scale **OR** scaling) |
| | **OR** |
| Scaling agile frameworks | (Crystal Family **OR** Dynamic Systems Development Method Agile Project Framework for Scrum **OR** Scrum of Scrums **OR** Enterprise Scrum **OR** Agile Software Solution Framework **OR** Large-Scale Scrum **OR** Scaled Agile Framework **OR** Disciplined Agile **OR** Spotify Model **OR** Mega Framework **OR** Enterprise Agile Delivery and Agile Governance Practice **OR** Recipes for Agile Governance in the Enterprise **OR** Continuous Agile Framework **OR** Scrum at Scale **OR** Enterprise Transition Framework **OR** ScALeD Agile Lean Development **OR** eXponential Simple Continuous Autonomous Learning Ecosystem **OR** Lean Enterprise Agile Framework **OR** Nexus **OR** FAST Agile)) |

Agile Manifesto was proposed, to the end of December 2019, when we started this systematic mapping study.

*Preliminary and main search.* Figure 1 shows the study search process and the individual results obtained in each of both phases of the study search. In the preliminary search, we retrieved 693 studies. After removing duplicate studies, 631 papers were left. The main search returned 2,090 publications. After removing duplicate papers, we ended up with 1,643 papers from the main search. After merging the search results from the preliminary and main search and removing duplicates, we retrieved a total of 2,144 articles that serve as input for the subsequent study selection process (see Section 3.2.2).

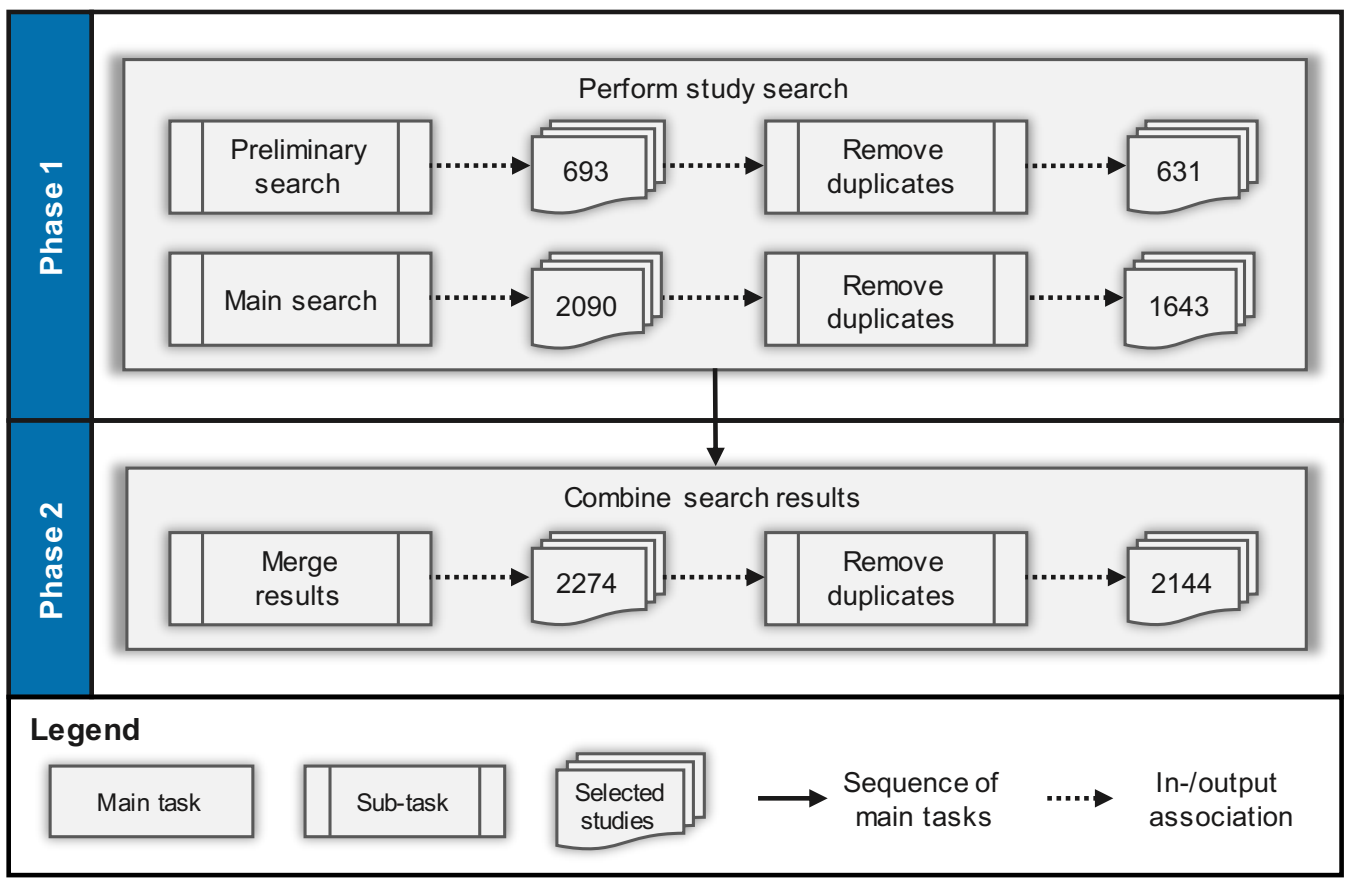

Figure 1: Overview of the study search process

### 3.2.2. *Study selection*

After the study search process, we considered all 2,144 studies for the subsequent study selection consisting of three screening phases: (i) selection of relevant articles based on their metadata (incl. title, keywords, publication year, and publication type), (ii) selection of relevant studies based on their abstract, and (iii) selection of relevant papers based on their full-text. The study selection was based on explicit inclusion and exclusion criteria and was conducted by two researchers in parallel. Following the individual decisions, the researchers harmonized their selection results and resolved conflicts. To ensure that the study selection results were objective, we created a set of well-defined inclusion and exclusion criteria employed in the selection process to filter relevant articles for our study.

*Selection criteria.* We defined selection criteria to reduce the likelihood of bias and to assess the relevance of the studies [56]. Before the study selection process, two researchers discussed and reached an understanding of the inclusion and exclusion criteria (see Table 4). We included an article if it satisfied *all* specified inclusion criteria and discarded it if it met *any* exclusion criterion. We only included peer-reviewed papers (see I3) that describe completed research results (see I4) and cover the large-scale application of agile methods (see I1 and I2), i.e., the application of agile methods in large multi-team settings or the usage of agile practices in organizations as a whole (see Section 2.2.1). Besides articles meeting any of the exclusion criteria (see E1 – E7), we also excluded papers that did not indicate large-scale considerations or described the single team adoption of agile methods, not fulfilling I2.

Table 4: Inclusion and exclusion criteria

| ID | Criteria | Assessment criteria |
|---|---|---|
| I1 | Inclusion | Describe the application of agile methods in software development. |
| I2 | Inclusion | Cover the application of agile methods on a large scale and meet the requirements of being large-scale based on our understanding and definition of large-scale agile development in Section 2.2. |
| I3 | Inclusion | Peer-reviewed, i.e., published in journals, conference or workshop proceedings. |
| I4 | Inclusion | Papers that describe completed research results. |
| E1 | Exclusion | Related to agile manufacturing. |
| E2 | Exclusion | Published in the form of abstracts, book chapters, book and conference reviews, grey literature, magazines, newsletter, short communications, talks, technical reports, and tutorials. |
| E3 | Exclusion | Articles that are not written in English language. |
| E4 | Exclusion | Published before the creation of the Agile Manifesto. |
| E5 | Exclusion | Not available as a full-text. |
| E6 | Exclusion | Experience reports and opinion papers. |
| E7 | Exclusion | Previous version(s) of extended papers. |

*Selection process.* The selection process consisted of three phases (see Figure 2). By the end of the third phase, the first researcher marked 269 studies as relevant, while the second researcher marked 145 studies as related. By the end of the fourth phase, 136 studies were characterized by both researchers as pertinent after resolving conflicts (inclusion rate of 6.34%).

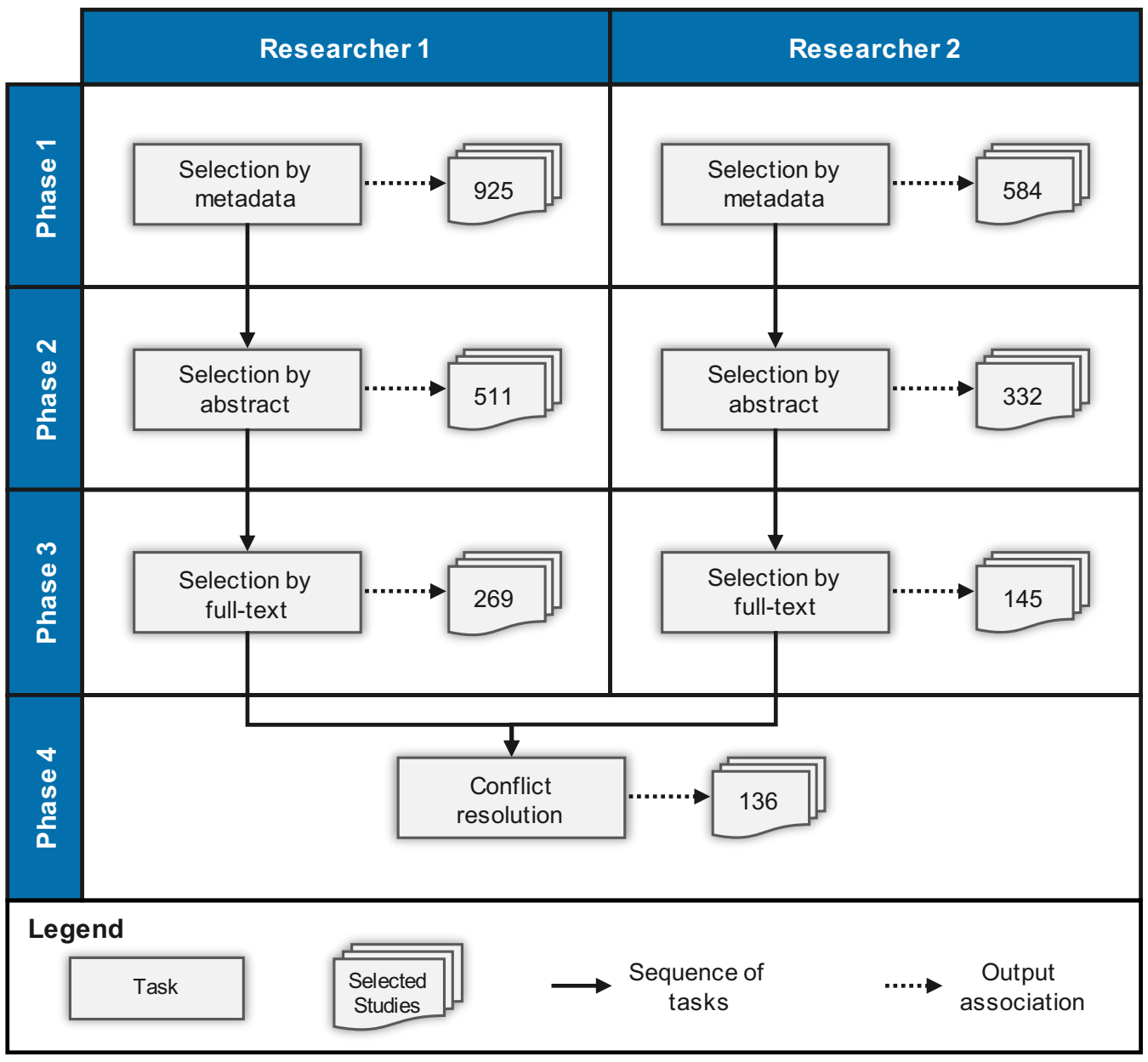

Figure 2: Overview of the study selection process

### 3.2.3. Data extraction

To facilitate data extraction and simplify management of the extracted data, we adopted the approach of categorizing studies into facets [82] and designed a rigorous classification framework based on these facets. Similar to study selection, we used a spreadsheet to record the extracted data. To reduce bias in the data extraction results, two researchers performed the data extraction independently. Before the formal data extraction process, two researchers discussed the definitions of the data items to be extracted to ensure that both researchers had a common understanding. After both scientists completed the data extraction, they discussed their results together and resolved conflicts to reach a consensus on the results. Figure 3 shows the resulting classification framework, which consists of four facets and several subordinate data items.

*Characterization of large-scale agile development (RQ1).* To compile information about the characterization of large-scale agile development, we collected information from the reported case characteristics regarding the four data items: *large-scale agile development category*, *case company characteristics*, *organizational agility characteristics*, and *large agile multi-team setting characteristics*. We extracted data for the *large-scale agile development category* data item to bring more conceptual clarity regarding the actual meaning of the term "*large-scale agile development*". To this end, we used the classification by Fuchs and Hess [42] to categorize the large-scale agile efforts of the reported companies. To get a better picture of companies adopting agile methods at scale, we divided the *case company characteristics* data item into four sub-data items, namely *company name*, *location of company headquarter*, *sector*, and *company size*. To obtain information on the four sub-data items, we combed through the case descriptions. In cases where the name of the company was provided but no other information about the company's location, sector, or size was available, we used the information on the company's website to complete the data. To classify the extracted data related to the *sector* sub-data item, we used the leading Global Industry Classification Standard[3] from MSCI. We used the categorization of Digital.ai [31] for classifying the reported case companies according to their *company size*. Using this categorization also helped us to compare our results more easily with those of Digital.ai [31]. To understand the trend of organizational agility and compile information related to the *organizational agility characteristics* data item, we read the case descriptions of the selected papers to determine the start date of the large-scale agile transformations of the reported case companies. To characterize the reported large agile multi-team settings, we split the *large agile multi-team setting characteristics* data item into three sub-data items, namely *scaling and complexity factors*, *organizational size*, and *applied development and scaling approaches*. We used an integrated approach [28] to extract data related to the *scaling and complexity factors* sub-data item. In doing so, we used the scaling and complexity factors reported by Ambler [7] as a starting list and iteratively expanded it as we identified new factors. For the *organizational size* sub-data item, we read the case descriptions and looked for information on the number of agile teams and people involved in the agile multi-team settings. We used the taxonomy of Dingsøyr et al. [35] to determine whether a setting was small-scale, large-scale, or very large-scale. To extract data related to the *applied development and scaling approaches* sub-data item, we applied an integrated approach [28] and used the lists of Abrahamsson et al. [3] and Uludağ et al. [99] for agile development and scaling approaches and refined the initial lists as new approaches were identified in the studies.

*Publication trends and research characteristics (RQ2).* Figure 3 shows the data items we used to collect data about publication trends and research characteristics. The *publication year*, *country of authorship*, and *publication venue* were retrieved directly from the studies' metadata. Based on the *publication venue*, we derived the *publication type* with a classification scheme consisting of the three categories: *journal*, *conference*, and *workshop*. To classify the *publication domain*, we read the website information of the publication venues and categorized them into one of eight primary publication domains: *enterprise computing*, *information systems*, *IT project management*, *human computer interaction*, *marketing*, *multidisciplinary*, *project management*, and *software engineering*. To classify the research types, we adopted a framework consisting of six categories based on Wieringa et al. [103] (see Table 5). For the *research approach* data item, we used a combination of three taxonomies and definitions of Berg et al. [17], Rodríguez et al. [89], and Unterkalmsteiner et al. [102] to have a complete list of research methods consisting of nine categories (see Table 6). For the *contribution type* data item, we used two existing taxonomies of Shaw [97] and Paternoster et al. [79] to have a complete list of research outcomes consisting of seven categories (see Table

[3] `https://www.msci.com/our-solutions/indexes/gics`, last accessed on: 11-13-2021.

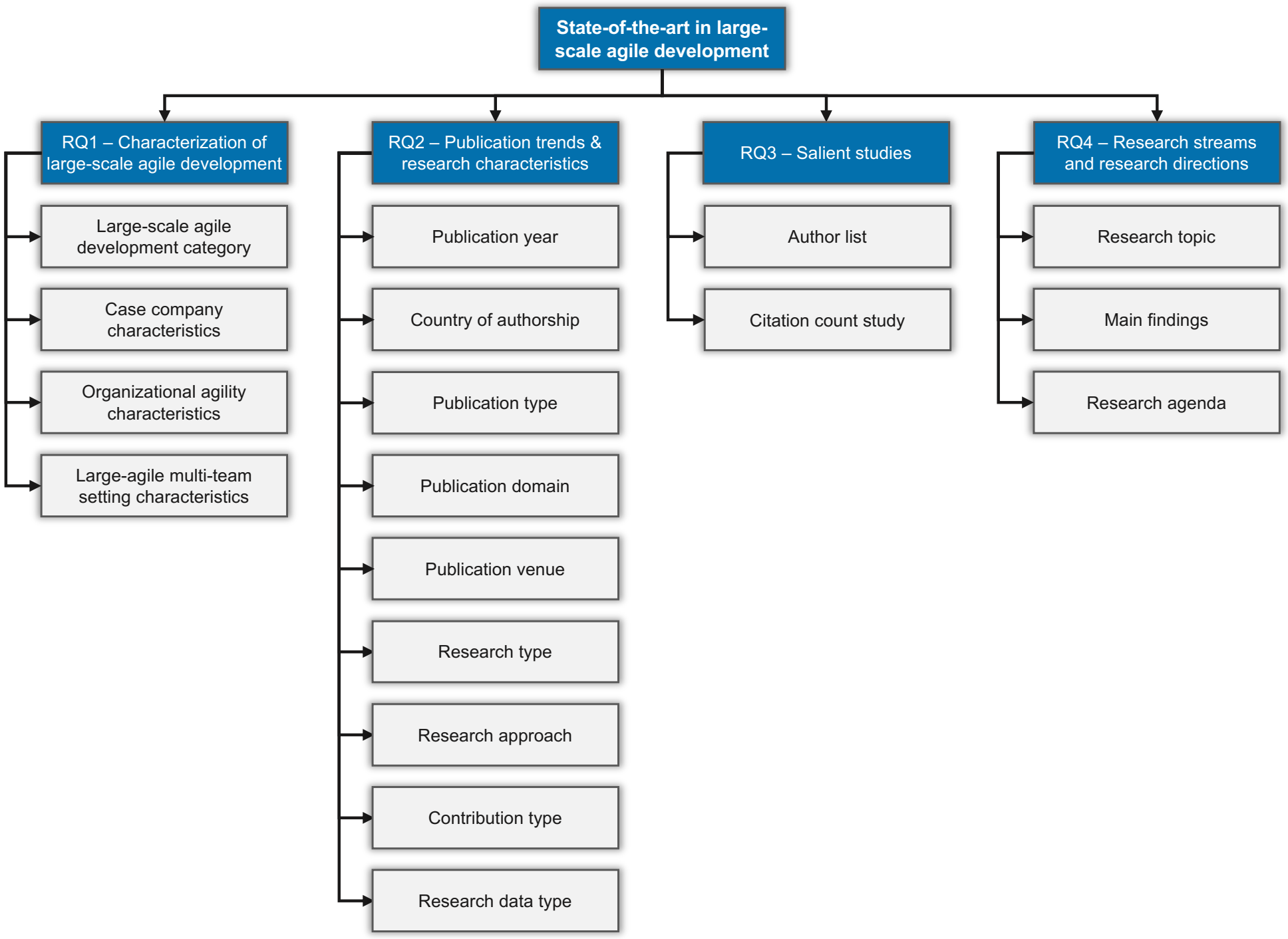

Figure 3: Classification framework

7). The *research data type* consists of three categories indicating whether a study is *primary*, *secondary*, or *tertiary*.

Table 5: Classification scheme for research types based on Wieringa et al. [103]

| Research type | Description |
|---|---|
| Evaluation research | The implementation of existing techniques and solutions are evaluated in practice. |
| Experience reports | Practitioners report their own experiences from one or more real-life projects without discussing the article's underlying research method. |
| Opinion papers | Express the author's personal experience regarding a technique's suitability without relying on related work and research methods. |
| Philosophical papers | These articles sketch a new perspective on looking at existing things by structuring the field using a taxonomy or conceptual framework. |
| Solution proposal | A new or significant extension of an existing technique is shown by demonstrating its benefits and applicability. |
| Validation research | Focus on the investigation of novel techniques that have not yet been implemented in practice. |

*Seminal studies (RQ2).* To compile information about influential studies, we used two data items: *author list* and *citation count study*. We obtained the *author list* from the metadata of the selected papers. We collected data for the *citation count study* data item via Google Scholar.

*Research streams and research directions (RQ4).* To identify central research themes and future research directions, we collected data on research topics and agendas in the selected papers. We followed a systematic process called *keywording* [82] to define the categories of the *research topic* facet. The purpose of the keywording process is to effectively develop a classification framework that fits the selected studies and takes their research focus into account [82]. The keywording process consists of the following three steps:

1. *Identifying keywords and concepts:* Two researchers collected keywords and concepts by reading the full-text of each starting study.
2. *Clustering keywords and concepts:* Two researchers per-

Table 6: Classification scheme for research approaches based on Berg et al. [17], Rodríguez et al. [89], and Unterkalmsteiner et al. [102]

| Research approach | Description |
|---|---|
| Action research | Applies action research to solve a real-world problem while simultaneously scrutinizing the experience of solving the problem. |
| Case study | Uses a case study to provide an in-depth understanding of a real-life situation and contemporary phenomenon or evaluates a theoretical concept by empirically implementing it in a case study. |
| Design and creation | Creates a new IT product, artifact, model, or method. |
| Grounded theory | Uses a systematic process to generate theory from the data obtained based on grounded theory. |
| Mixed methods | Applies more than one research methodology. |
| Systematic literature review/Systematic mapping study | Collects and analyzes primary data to address a specific research question or topic using a systematic literature review or systematic mapping study. |
| Survey | Collects quantitative and/or qualitative data in a standardized, systematic way to find patterns through a questionnaire or interviews. |
| Theoretical | A theoretical study not mentioning grounded theory as the research methodology. |
| Not stated | Does not state the research method, nor can it be derived or interpreted by reading the paper. |

Table 7: Classification scheme for contribution types based on Shaw [97] and Paternoster et al. [79]

| Contribution type | Description |
|---|---|
| Advice/Implication | Discursive and general recommendation based on personal opinions. |
| Framework/Method | Framework or method to facilitate the construction and management of software-intensive systems. |
| Guideline | List of advice or recommendations based on the synthesis of the research results obtained. |
| Lessons learned | Set of outcomes which are analyzed from the research results obtained. |
| Model | Representation of an observed reality using concepts resulting from a conceptualization process. |
| Theory | Construct of cause-effect relationships from determined results. |
| Tool | Technology, program, or application developed to support various aspects of software engineering. |

formed a clustering operation on the collected keywords and concepts into a set of categories and sub-categories.

3. *Refining classification:* Four researchers discussed the preliminary categories and sub-categories. This discussion resulted in the refinement of the classifications to fit them better with the selected studies.

The above-described process ended when no study was left to analyze. During the extraction process, some studies could be classified into more than one research topic. We read the results, discussion, and conclusion sections of the studies to identify their *main findings* and mapped them to the main research topic categories. We used a deductive approach [28] to categorize the *research agenda* of the selected studies based on the final classification of the *research topic* facet. Two researchers read the selected studies' future work sections and mapped the corresponding text fragments to the identified main research topic categories. In addition to the selected studies, we read and mapped relevant data from nine related workshop summaries. They provide a list of important future research topics proposed by researchers and practitioners familiar with large-scale agile development. Following the coding procedure, two researchers merged and aggregated related codes and reformulated the final codes as research questions.

## 4. Study results

In this section, we provide a review of the state of the art of large-scale agile development research and present our answers to the formulated research questions (see Section 3.1). This section is arranged according to the research questions.

### *4.1. Characterization of large-scale agile development*

Although we discussed the term "*large-scale agile development*" in Section 2.2.1 and stated our understanding of it, we were curious to learn what other authors refer to as large-scale agile development. In what follows, we provide an overview of case companies we studied and identify which category of large-scale agile development, i.e., large agile multi-team settings or organizational agility, was reported. We then highlight our key findings for each category.

#### *4.1.1. General overview of the case companies*

We identified 158 case companies, of which 137 remained unnamed in 67 studies[4]. A total of 21 companies were explicitly named in 41 studies. Table 8 shows an overview of the identified companies and the number of studies reporting them. The most frequently identified company was Ericsson (cf. [S1], [S3], [S26]), followed by the Norwegian Public Service Pension Fund (cf. [S28], [S31], [S43]). Other companies repeatedly studied were F-Secure (cf. [S8], [S68]) and Nokia (cf. [S2], [S79]). The remaining 17 companies were mentioned only in one study. As a result, we notice that research on large-scale agile development is mainly empirical and practice-oriented, as almost 80% of the studies examine companies in their respective analyses. We note that a significant portion of the research was conducted with Ericsson, accounting for nearly 17% of all selected studies. We notice that many studies provide superficial case descriptions, making it difficult to generalize and compare results.

Table 8: Overview of the identified case companies

| Company | No. of studies |
|---|---|
| Anonymous | 67 |
| Ericsson | 23 |
| Norwegian Public Service Pension Fund | 8 |
| F-Secure | 2 |
| Nokia | 2 |
| ABB | 1 |
| Apontador | 1 |
| BBC | 1 |
| Caelum | 1 |
| Cisco | 1 |
| Comptel | 1 |
| Dell | 1 |
| Information Mosaic | 1 |
| Intel | 1 |
| Kentico | 1 |
| Paf.com | 1 |
| Rovsing | 1 |
| SAP | 1 |
| SimCorp | 1 |
| Spotify | 1 |
| ThoughtWorks | 1 |
| Universo Online | 1 |

Figure 4 shows the countries where the headquarters of reported companies are located, with circles indicating the number of companies in a given country. We identified the geographic location of 58 companies. Although most companies are located in Europe, large-scale agile development is a relevant practical topic on all continents. Most companies come from Germany, followed by the United States, Norway, and Sweden. Like the State of Agile survey [31], we note a concentration of industry relevance of the topic in Europe and North America. Based on our selection criteria (see Section 3.2.2), we identified fewer companies from North America compared to the State of Agile survey [31] as we excluded many experience reports and opinion articles from North America.

[4]There is some likelihood that some unnamed companies are duplicates. Many case descriptions were too superficial to allow clear identification.

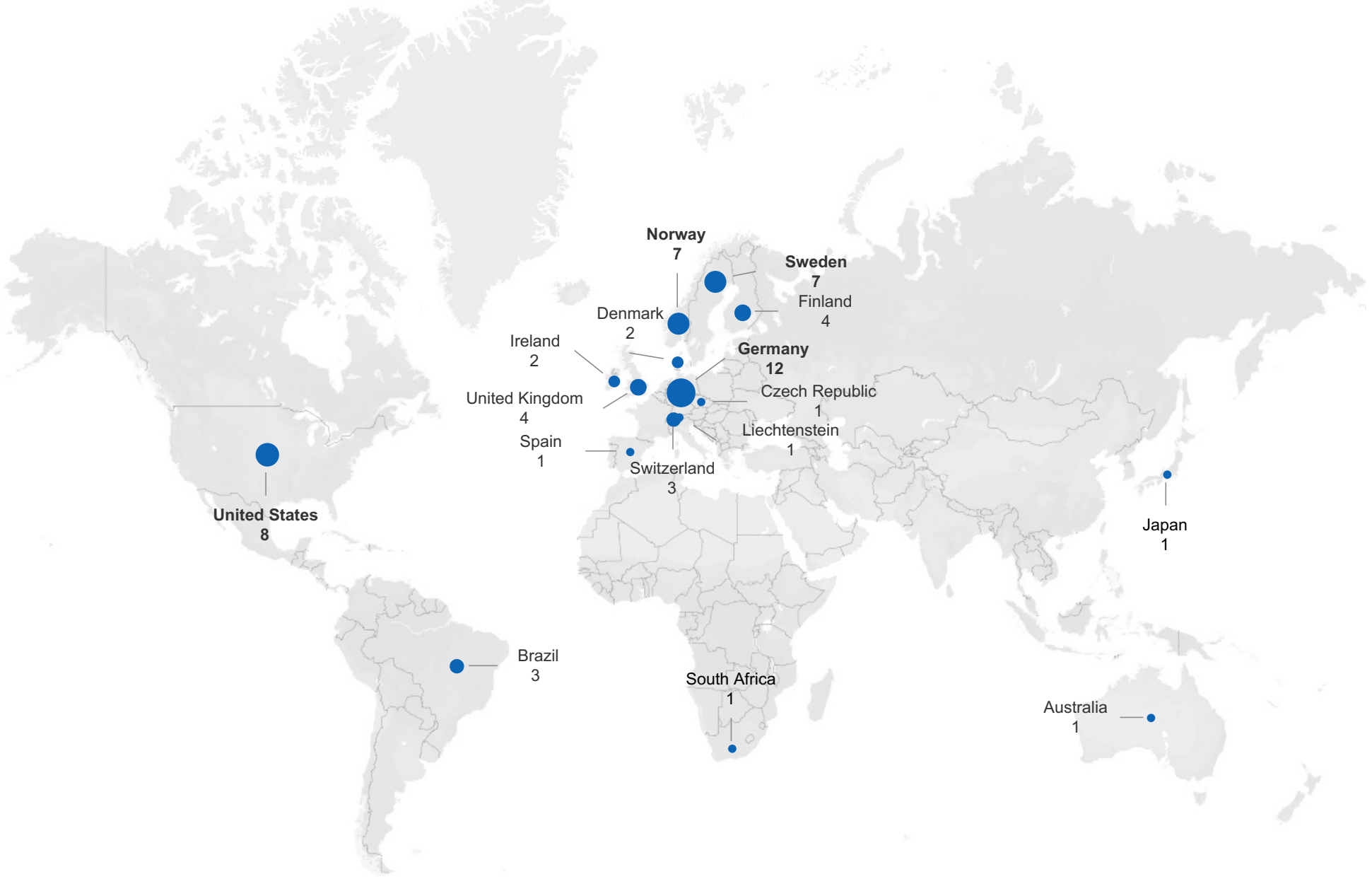

Figure 4: Geographical distribution of the case companies

Figure 5 shows the distribution of the companies across their sectors. While 146 companies come from 10 different sectors, we could not identify the respective sectors for 12 companies. Almost a third of all reported companies come from the information technology sector, followed by the financial and public sectors. Our findings align with the State of Agile survey [31] stating that a large proportion of the companies using (large-scale) agile methods come from the information technology and financial sectors. Notably, 44% of the companies surveyed come from the information technology and finance sectors, while 46.84% of our companies are from these sectors.

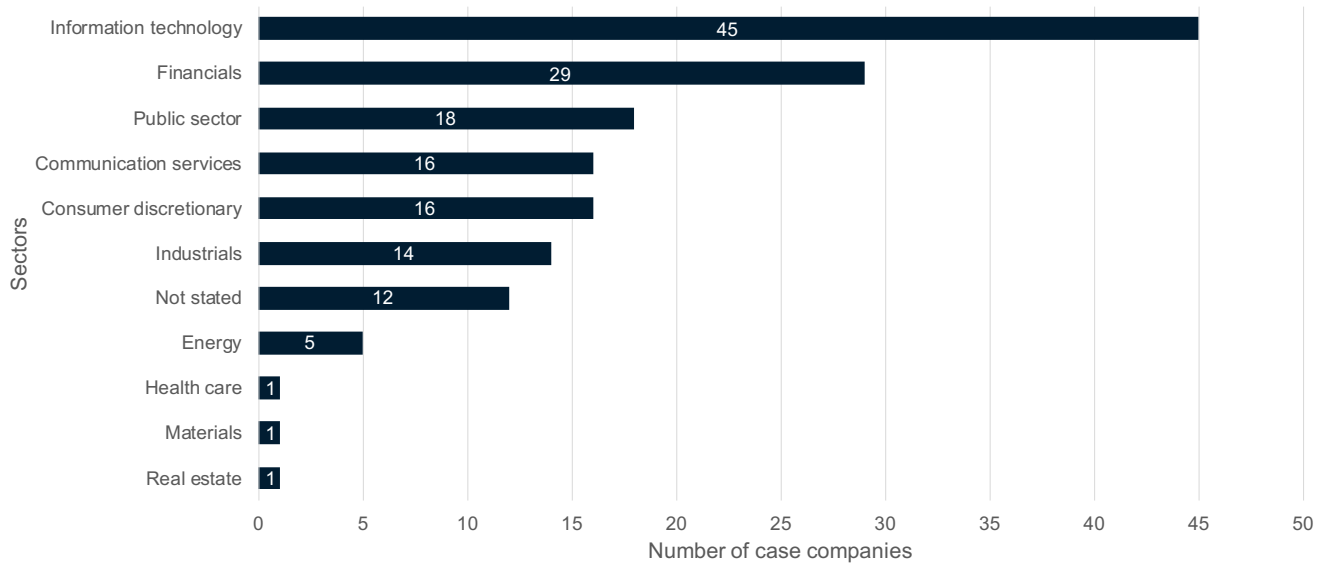

Figure 5: Distribution of the case companies according to their sectors

Figure 6 shows the distribution of the reported companies over their sizes. While we were able to determine the organizational size of 89 companies, we could not reveal the size of 69 companies. Most companies employ $>20,000$ employees, followed by companies with $\leqslant 1,000$ employees. Comparing our results to those of the State of Agile survey [31], we can see some similarities and differences. Like the survey [31], the majority of reported companies are small ($\leqslant 1,000$ employees) or very large ($>20,000$ employees). While these two groups account for 66% of survey respondents, they account for 65.17% of the reported companies in this study. The majority of the companies in our sample are very large, followed by small companies, which is precisely the opposite of the survey [31]. One reason for this could be that the survey [31] covers both the small- and large-scale adoption of agile methods. Our research exclusively considers the large-scale application of agile practices, which may be more relevant for large companies than for small companies with fewer scaling issues.

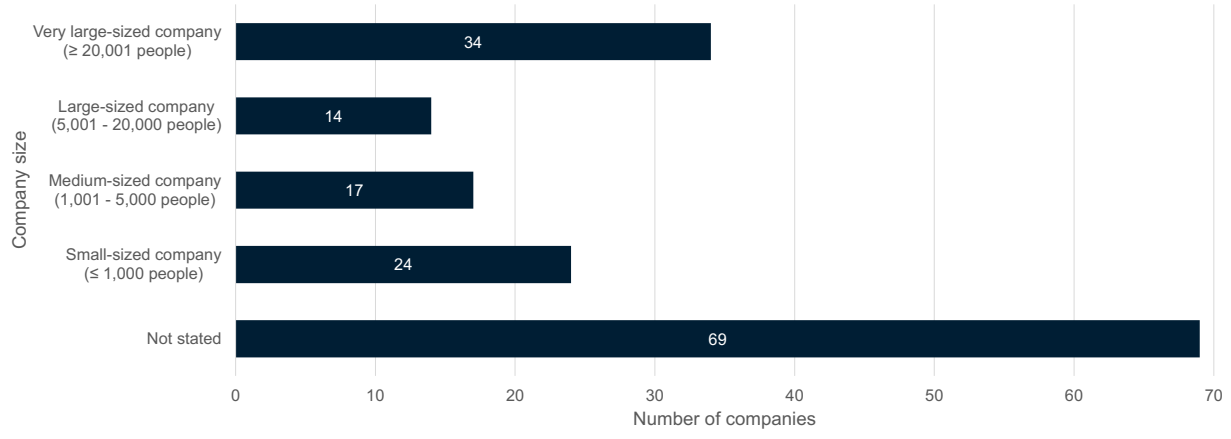

Figure 6: Distribution of the case companies according to their sizes

### 4.1.2. *Understanding of large-scale agile development*

Since there are several definitions of large-scale agile development in the literature (see Section 2.2.1) and there is no consensus on the actual meaning of the term [35, 90], we were curious what is meant by "*large-scale agile development*" in the literature. To this end, we used the classification by Fuchs and Hess [42], i.e., the adoption of agile methods in large multi-team settings ("*large agile multi-team settings*") with at least two agile teams working on a single product or the usage of agile practices in organizations as a whole ("*organizational agility*"), to analyze the large-scale agile efforts of the reported companies. Figure 7 shows the categories that have been reported. Nearly half of all companies applied agile practices in organizations as a whole, while 39.24% of all companies used agile methods in large multi-team settings. An example for organizational agility is provided by Fuchs and Hess [S21] who conceptualize the agile transformation process through the lens of socio-technical systems theory by analyzing the large-scale agile transformations of two companies from the financial and consumer discretionary sectors. An example for large agile multi-team settings is shown by Šāblis and Šmite [S16] who study inter-team coordination mechanisms of a large-scale agile development program of the Norwegian Public Service Pension Fund with eight teams. In four companies, both categories were reported. For instance, Power [S7] explains the notion of "*agile at scale*" by showing an example of Cisco using agile practices in the whole organization and having multiple development efforts with several agile teams working together. For 16 companies (cf. [S107], [S117]), we did not determine the category due to superficial context information.

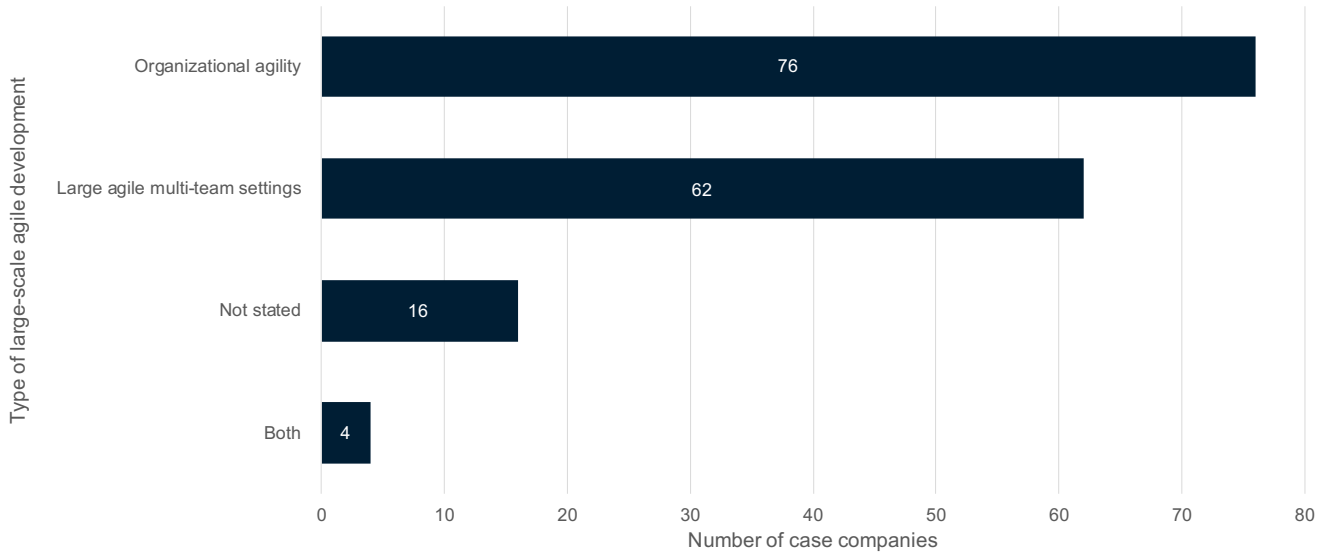

Figure 7: Type of reported large-scale agile development

In the following, we delve into the selected studies and provide more in-depth information in light of both categories of large-scale agile development reported in the companies.

#### 4.1.3. Organizational agility

To understand the trend of organizational agility, we analyzed when the case companies started their large-scale agile transformations. Figure 8 shows the distribution of when case companies started their large-scale agile transformations. We derived information on the agile transformations of 20 case companies. The first two observed transformations towards organizational agility began in 2006 at F-Secure (cf. [S68]) and Paf.com (earlier Eget) (cf. [S88]), when the company-wide adoptions of Scrum were initiated. While nine transformations started between 2006 and 2012, 11 transformations began between 2014 and 2017, indicating a growing number of transformations in recent years. Comparing these numbers with the number of studies published (see Section 4.2.1), we perceive the congruent interest from industry and academia in the topic.

#### 4.1.4. Large agile multi-team settings

We identified 110 agile development efforts[5] with multiple teams. There were also some companies with more than one effort. For instance, we uncovered 11 efforts with 6–40 agile teams at Ericsson, reported in 20 studies (cf. [S1], [S3], [S26]).

[5]There is some likelihood that some large-scale agile development efforts are duplicates, e.g., same settings at different times. Some case descriptions were too superficial to allow unambiguous identification.

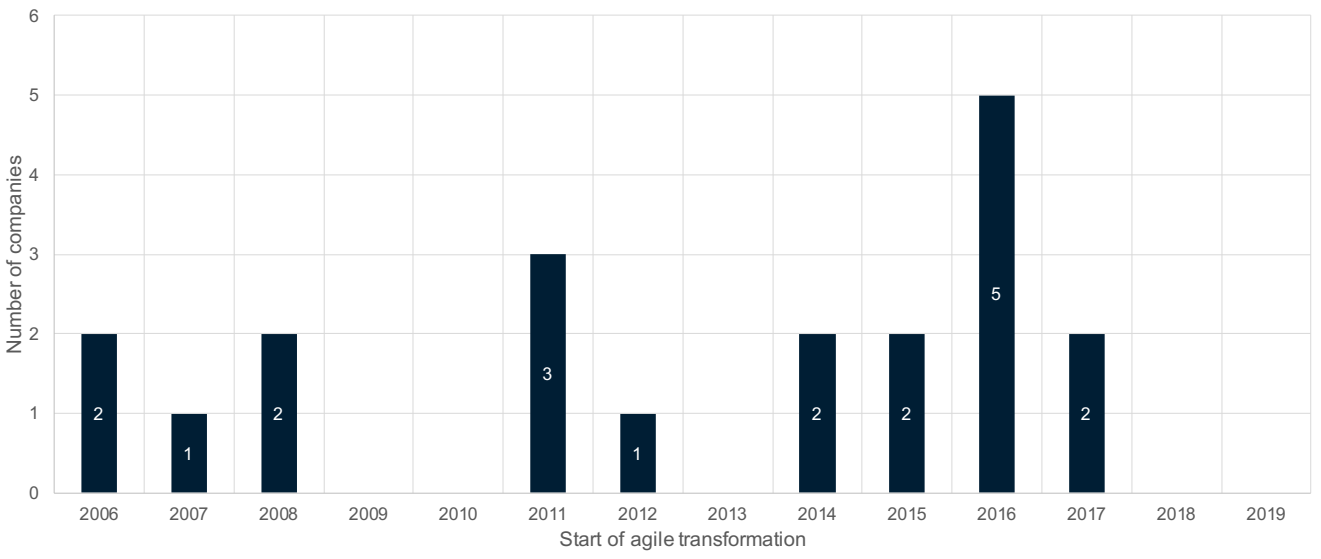

Figure 8: Start of the agile transformations of the case companies

Another example is provided by Bick et al. [S51], who describe the inter-team coordination of five large agile multi-team settings at SAP, involving 4–13 teams.

Based on our ambition to gain a deeper understanding of what the authors of the selected studies mean by the "*scaling of agile methods*", we examined the case descriptions in terms of possible scaling and complexity factors of large agile multi-team settings. Figure 9 provides an overview of the identified factors. We identified six scaling and complexity factors associated with large agile multi-team settings, i.e., organizational size, organizational distribution, product size, number of collaborating companies, number of customers, and budget. Organizational size, i.e., the number of agile teams working together or the number of people in the development effort, was the most frequently observed factor. For instance, Paasivaara and Lassenius [S79] present a case study of scaling Scrum in a large globally distributed software project at Nokia. The study expresses the scale and complexity of the case project by describing that it scaled from two collocated Scrum teams to 20 teams, i.e., the number of agile teams working together, and now employs 170 people, i.e., the number of involved persons in the development effort. Organizational distribution of large agile multi-team settings was the second most frequently identified factor. Here, the studies specify the number of sites a development effort has and provide information regarding the geographic distribution of the development effort. For example, Usman et al. [S37] examine how a large-scale distributed agile project at Ericsson performs effort estimation. The study further delineates the project's organizational distribution by stating that the project is distributed across Sweden, India, Italy, the United States, Poland, and Turkey. The third most frequently cited factor is related to the complexity of the developed system, i.e., product size (measured in lines of code), number of subsystems, and number of requirements and features. For instance, Moe et al. [S67] investigate the intra- and inter-team knowledge sharing in a large-scale distributed agile project at Ericsson, developing a product with 10–12 million lines of code and 30 subsystems. In the case of 14 large agile multi-team settings, the studies did not provide information on scaling and complexity factors. For instance, a multiple case study of five software-intensive companies by Olsson and Bosch [S98] reveals challenges associated with collecting and using customer feedback. Although we can infer from the context of the study that it analyzes large agile multi-team settings of five companies, we did not find fur-ther descriptions of their characteristics. Moreover, the number of companies working together was an additional factor. For example, Uludağ et al. [S104] investigate the collaboration between architects and agile teams of a large-scale agile development program at a German consumer electronics company. The reported large-scale agile program includes 62 persons employed in four companies, i.e., consumer electronics and three suppliers providing external support for their third-party systems. In two large agile multi-team settings, the number of customers/users of the developed software product was cited as a factor. For example, a large-scale agile program at Ericsson creates a complex system used by over 300 operators around the world (cf. [S3], [S38], [S50]). For instance, a large-scale agile endeavor implements a new office automation system for the Norwegian Public Service Pension Fund serving 950,000 customers with various types of services (cf. [S43], [S52]). Budget was reported in one setting, namely at the Norwegian Public Service Pension Fund, which is one of the most extensive IT programs in Norway with a budget of about 140 million EUR (cf. [S28], [S31], [S43]). Comparing our findings with the scaling and complexity factors of Ambler [7], we note some similarities and differences. Like Ambler [7], we believe that organizational size is the primary scaling and complexity factor for characterizing the term "*scaling of agile methods*". Like Ambler [7], we believe that other important factors are organizational distribution and system complexity. Unlike Ambler [7], we identified the number of collaborating companies, the number of customers, and the available budget for development efforts as additional scaling and complexity factors.

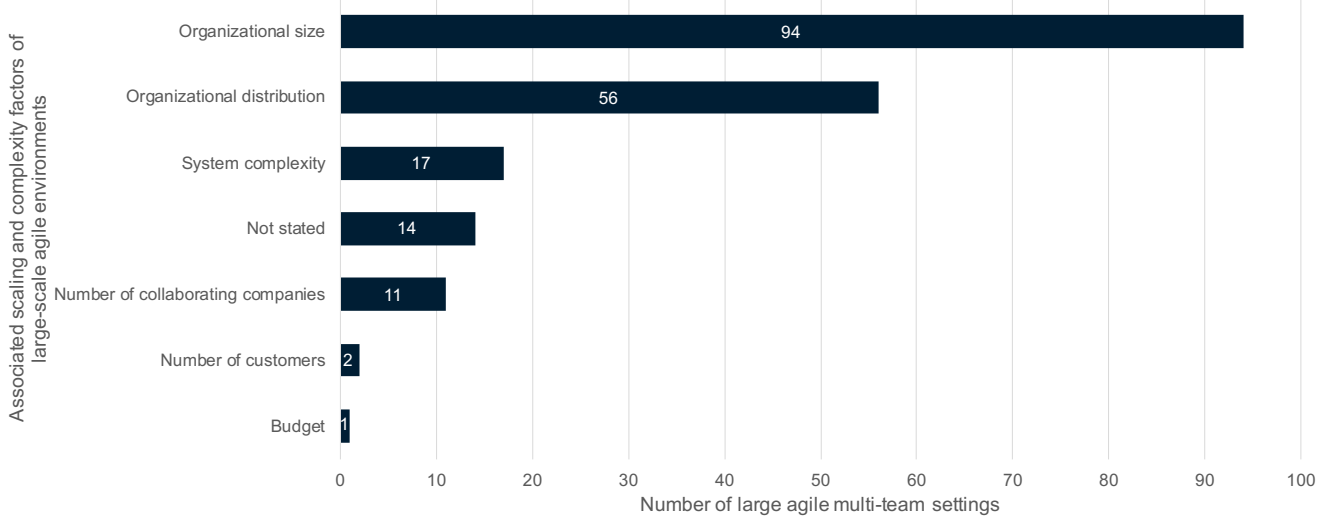

Figure 9: Scaling and complexity factors of large-scale agile environments

Since organizational size was the main reported factor, we were curious to analyze the sizes of the agile multi-team settings. Figure 10 shows the distribution of the number of people and teams involved in the agile multi-team settings. Of the 110 reported agile multi-team settings, we determined the number of people involved for 70 settings and the number of teams for 62 settings. We could not determine the number of people engaged for 40 settings and the number of teams for 48 settings. Most of the reported agile multi-team settings were large and included 11–100 people or 2–9 teams, followed by very large settings that involved 101–8,000 people or 10–40 teams. Based on our understanding of large-scale agile development (see Section 2.2.1) and excluding projects with $<2$ teams, we found no small-scale projects. Our results show some notable differences from the State of Agile survey [31]. While 70% of the settings (excluding settings without numbers) had $\leqslant 100$ employees, 33% of all survey respondents reported that their software organizations had $<100$ employees. While 36% of the survey respondents indicated software organization sizes with 101–1,000 employees, this was the case for 46.67% of agile multi-team settings in this study. A significant difference between this study and the survey is observed for software organization sizes with $>1,001$ employees. While 31% of the survey respondents reported software organization sizes of $>1,001$ people, we identified only one very large agile multi-team setting at Cisco with 8,000 employees (cf. [S7]). This discrepancy may be due to our selection criteria related to experience reports and opinion papers (see Section 3.2.2) led to the exclusion of companies in North America with likely larger sizes.

Finally, we were interested in exploring the applied development and scaling approaches reported in the large agile multi-team settings, visualized in Figure 11. SoS is the most commonly used scaling approach reported, followed by SAFe and LeSS. The Spotify Model was described in three settings, followed by DAD in one setting. Comparing our numbers with those from the State of Agile survey, we see some similarities and differences. While the survey showed SAFe as the most commonly used scaling approach among respondents (35% of their respondents), only 8.18% of the large agile multi-team settings reported using SAFe. For SoS, the numbers are similar as 18.18% of all settings used this scaling approach, while 16% of the survey participants applied it. In addition, 4% of the survey participants reported using LeSS, while 7.27% of the settings in this study used LeSS. While we identified the adoption of DAD in only one setting, 4% of all survey participants adopted it. Scrum was the most frequently cited development approach, followed by Kanban and XP. Re-comparing our results with the survey data confirms that Scrum is by far the most frequently used development approach, i.e., 58% of the survey respondents mentioned Scrum. In contrast, Scrum was used in 72.7% of all large agile multi-team settings reported in this study. While 10.91% of the settings reported using Kanban, that number accounted for 7% of all survey respondents. About half of the settings stated the combined usage of development and scaling approaches. For instance, a large multi-team project at F-Secure, which included $>10$ teams and $>140$ stakeholders, used a combination of Scrum, SoS, and an early version of SAFe (cf. [S8], [S68]). Uludağ et al. [S87], for example, describe a unit with three agile teams who worked with other groups to develop an integrated sales platform for multiple sale distribution channels, using LeSS extended with some XP practices.

### 4.2. *Publication trends and characteristics of existing research on large-scale agile development*

To map publication trends and characteristics of extant research on large-scale agile development, we chose a set of variables focusing on each study's publication and bibliographic data. Below, we detail the key facts drawn from our analysis.

#### 4.2.1. *Distribution of studies over time*

Figure 12 presents the distribution and cumulative sum of selected studies, providing a clear message on an apparent increasing trend in publications on large-scale agile development.

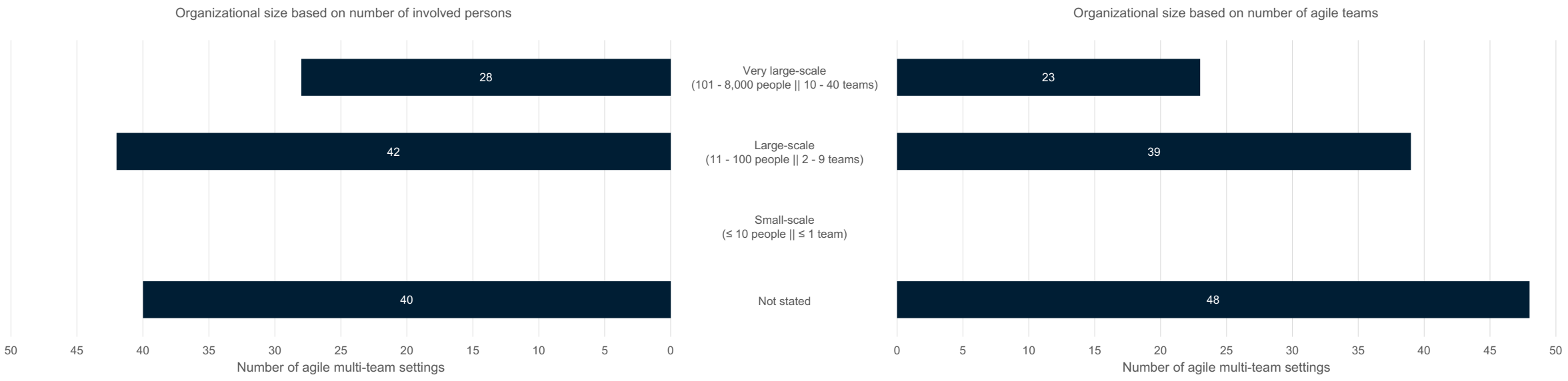

Figure 10: Distribution of organizational sizes of the large agile multi-team settings in terms of number of involved persons and agile teams

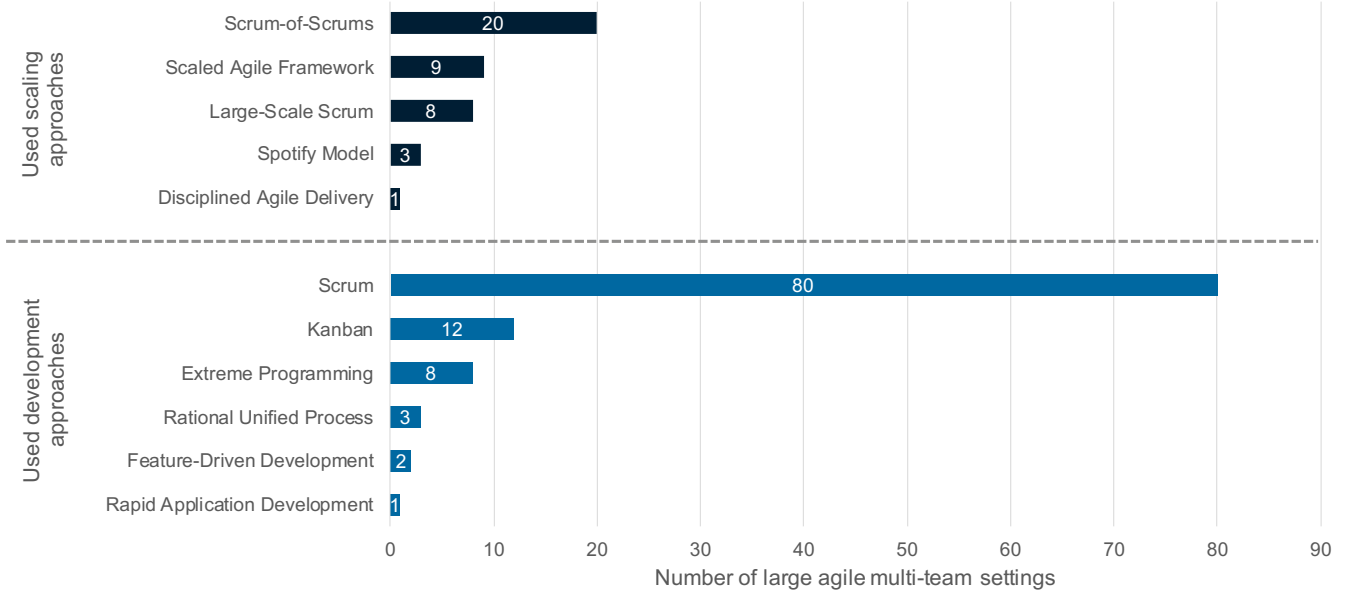

Figure 11: Applied development and scaling approaches in the large agile multi-team settings

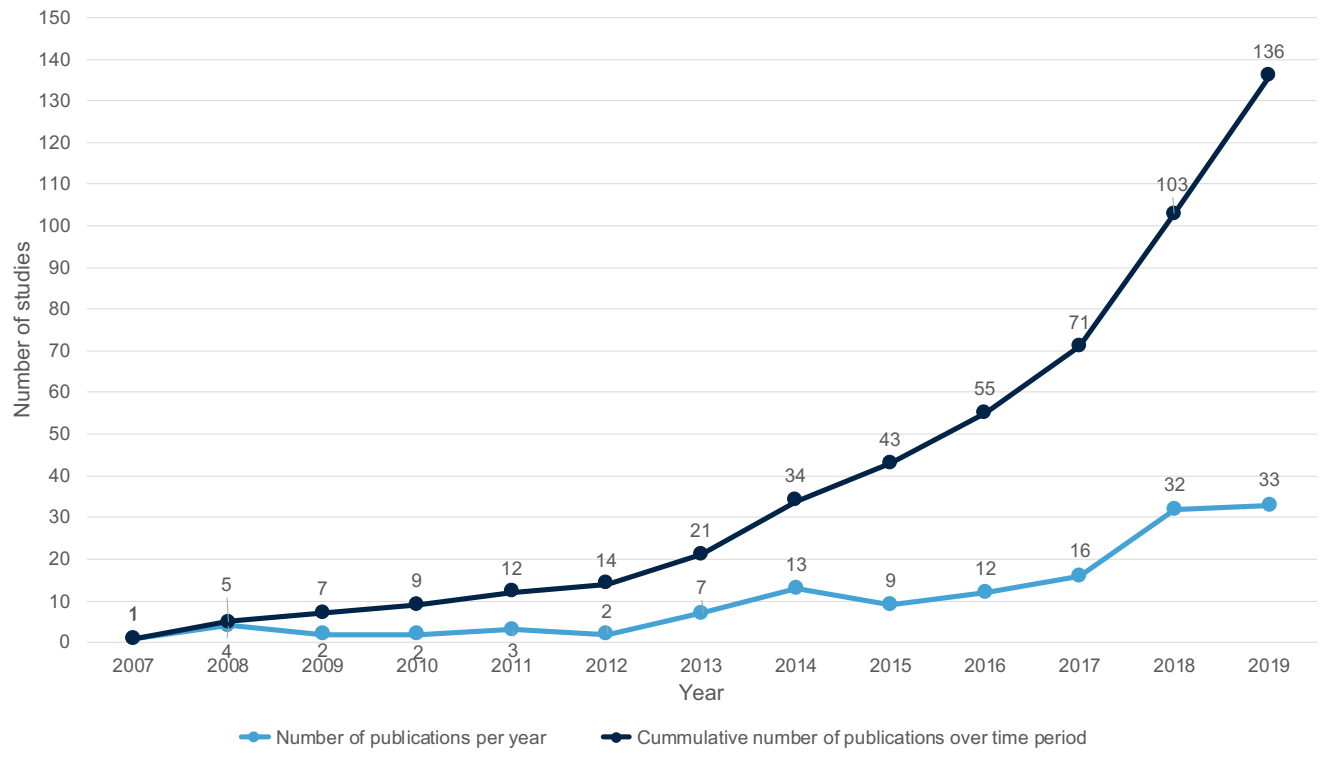

Figure 12: Publications on large-scale agile development from 2007 to 2019

In general, the number of selected studies on this topic has increased from 2007 to 2019, with slight fluctuation. This trend indicates that the large-scale application of agile methods is receiving increasing attention from the research community. Except for 2007, at least two studies were published during the observation period. From 2013 onwards, a growing trend can be observed with at least seven studies published per year, which is a giant leap compared to the years before 2013. Two reasons for this could be that the agile software development community initiated an increasing paradigm shift towards the adoption of agile methods in large projects in 2013 [33, 41] and that the International Workshop on Large-Scale Agile Development was held for the first time in 2013. After 2013, 84.55% of the selected studies were published in the last six years, indicating that the topic of large-scale agile development is relatively infancy when compared to the history of the software engineering discipline and is becoming increasingly attractive from a scientific perspective. The growing trend towards the topic culminated in 2018 and 2019 when the number of publications doubled compared to previous years, accounting for a total of 47.79% of the selected studies. One possible explanation for this is that additional workshops contributing to large-scale agile development research were initiated, e.g., the International Workshop on Autonomous Agile Teams in 2018 and the International Workshop on Agile Transformation in 2019. Based on these observations, we expect this trend to continue.

#### *4.2.2. Most active countries in large-scale agile development research*

Figure 13 shows the countries that are most active in large-scale agile development research. We identified 22 states contributing to large-scale agile development research. The majority of the articles are from Europe, accounting for 88.97% of all publications. The research theme of large-scale agile development received considerable interest in Scandinavia, with 74 papers. Accordingly, the most actives countries in this research area are Finland, Norway, and Sweden. The fourth and fifth most active countries are Germany and the United Kingdom. The remaining 17 states have fewer than 10 publications and contributed only 32 articles. From the available data, we conclude that large-scale agile development is a globally relevant research topic. Although Dingsøyr et al. [34] revealed that the United States and Canada contributed to research on agile software development with 448 out of 1,551 selected studies, we identified only three studies on large-scale agile development. A plausible explanation for this phenomenon is that we identified numerous experience reports and opinion papers from both countries that were excluded by our selection criteria (see Section 3.2.2) (cf. [50], [60], [85]). Comparing the map of most active countries in large-scale agile development research (see Figure 13) with the geographic location of the reported companies (see Figure 4), a considerable overlap can be observed. A logical reason for this observation is that researchers tend to work with companies that are located in closer proximity.

#### *4.2.3. Publication channels*

To examine the pertinent publication channels for large-scale agile development research, we collected data on publication types, venues, and domains for each selected study. Figure 14 shows the proportion of publication types for the 136 selected studies. The predominant publication type is that of conference papers, being almost as much as the combination of journal articles and workshop papers. Such a high number of journal and conference papers may indicate that large-scale agile development is becoming a more mature research area despite its relatively young age. The relatively small number of workshop papers suggests that researchers prefer more scientifically rewarding publication types, e.g., journals and conferences.

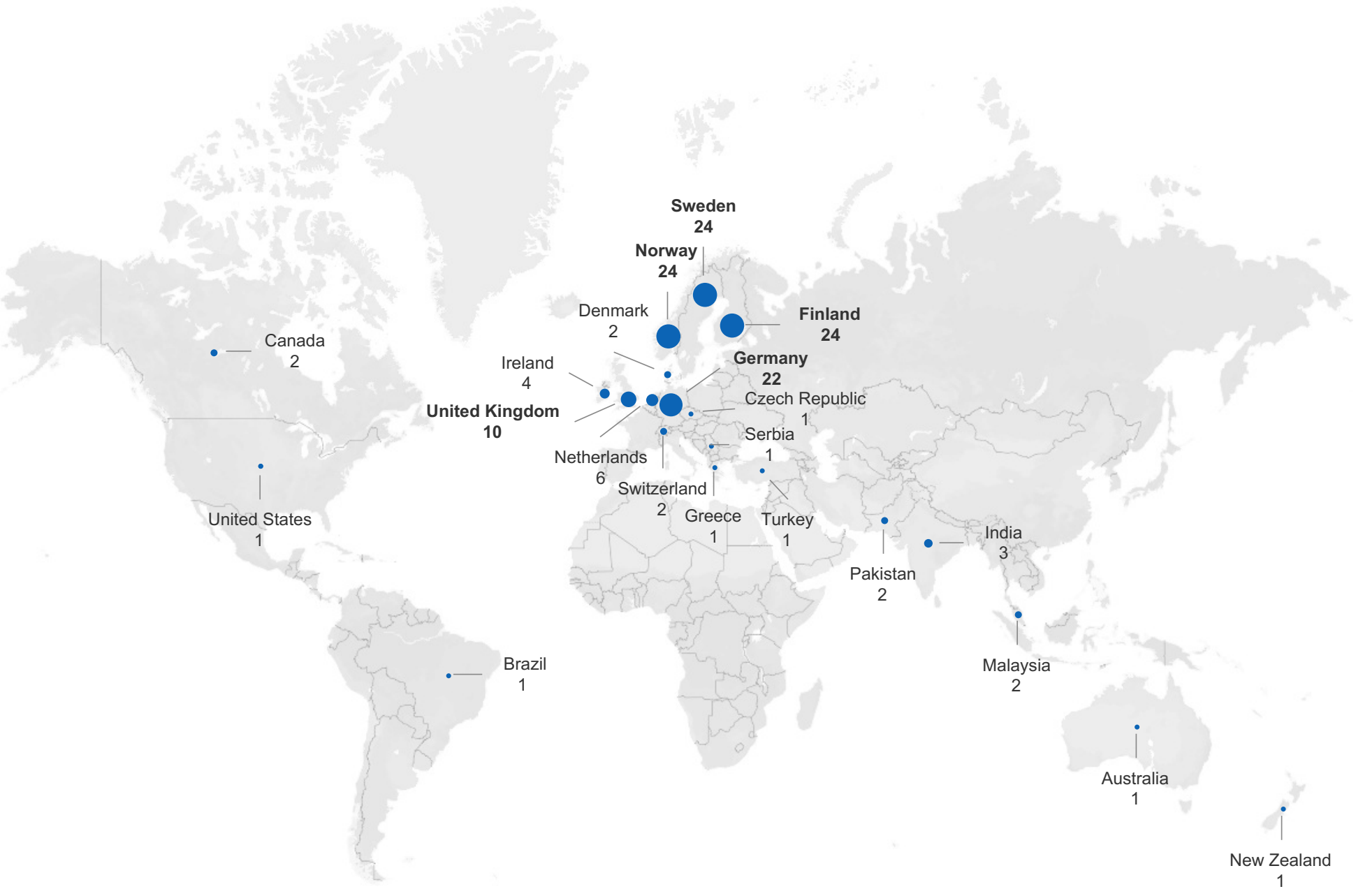

Figure 13: Map of most active countries in large-scale agile development research

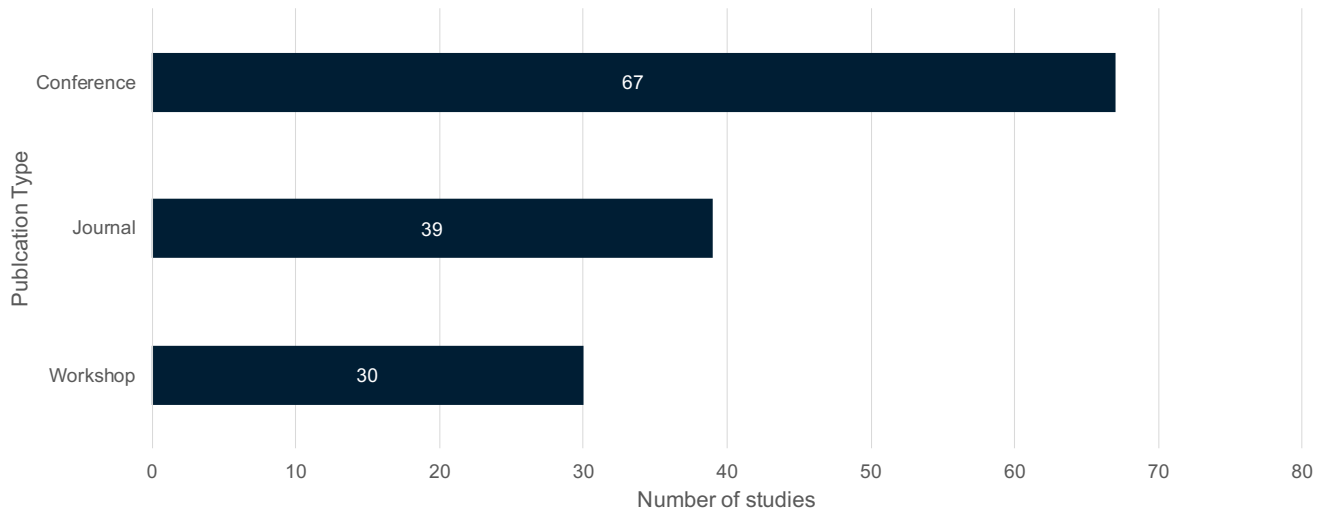

Figure 14: Distribution of publication types

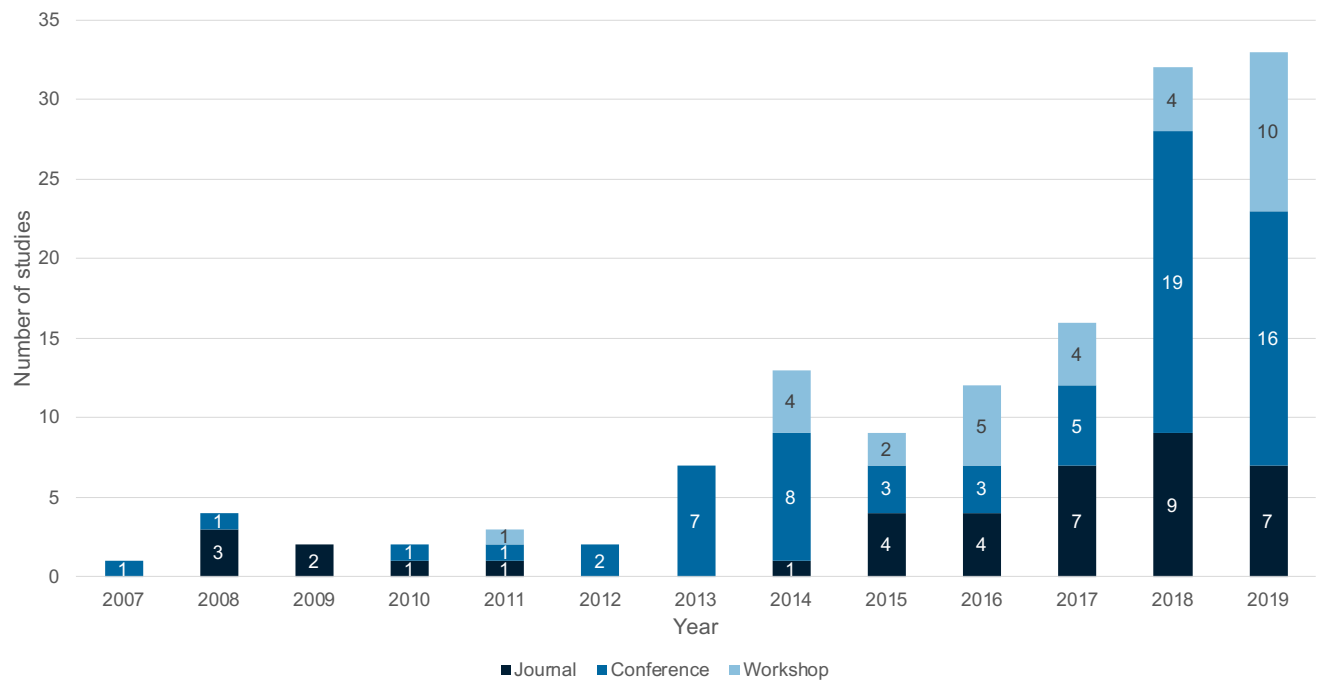

Figure 15: Distribution of publication types over time

Figure 15 shows an annual distribution of the publication type facet. After 2013, an increasing trend can be observed for all publication types, culminating in 2018 and 2019. This observation may further confirm that large-scale agile development is turning into a more mature field, with more fundamental and comprehensive studies published in recent years. Although journals were the predominant publication type between 2007 and 2011, we can observe that conferences and workshops became increasingly prevalent from 2012 onwards, as more venues on the topic were initiated or existing venues included large-scale agile development in their research program.

Figure 16 presents the distribution of selected studies across publication domains. We identified eight publication domains that cover research related to large-scale agile development. Unsurprisingly, most of the publications are covered by publication venues from the software engineering research domain, followed by the information system and project management research areas. Nine articles were published in venues covering various research topics, e.g., IEEE Access embracing all IEEE fields of interest or Procedia Computer Science containing conference proceedings on all computer science topics. Although most of the studies stem from the software engineering field, we recognize a growing interest in large-scale agile development research from further research communities, e.g., information systems, project management, and enterprise computing.

To corroborate this observation, we show in Figure 17 how the number of publications has evolved across the identified publication domains. Large-scale agile development is primarily and expectedly domicile in the software engineering field, with at least one published study per year and a peak in 2019 with 22 publications. Research on large-scale agile development has also aroused an early interest in publication venues related to information systems. After 2013, we observe a grow-ing interest in the topic outside of software engineering. While only two articles were published in venues unrelated to software engineering between 2007 and 2012, further 35 articles were published between 2014 and 2019, representing 94.59% of all studies outside of software engineering. These numbers are coherent with our observations. We notice a general interest in the topic in various academic fields in recent years, as the large-scale adoption of agile methods has far-reaching implications for companies, which in turn is very interesting for many scientists to investigate from an empirical point of view.

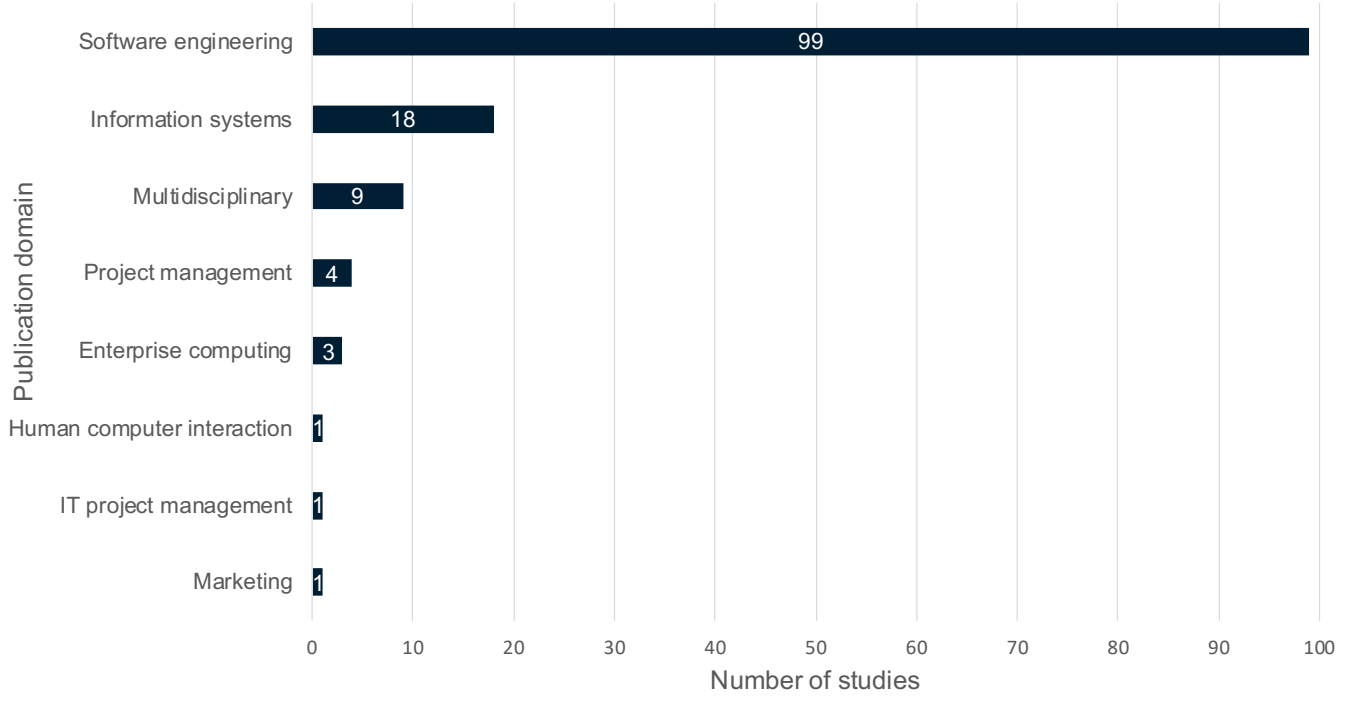

Figure 16: Distribution of publication domains

A look at the specific targeted publications venues shows that research on large-scale agile development is published in various venues. The 136 selected studies are distributed across 46 publication venues (see Appendix A), including 17 journals, 23 conferences, and six workshops, indicating that research on large-scale agile development is receiving wide attention in the research community. Table 9 presents the top-10 venues with the highest number of publications. While eight of the top-10 publication venues belong to the software engineering research field, two conferences are from the information systems research field. The top six venues account for 50% of all publications, while the remaining 40 venues cover half of all studies. The top venues are the International Conference on Agile Software Development, followed by the International Workshop on Large-Scale Agile Development. Both venues have at least twice as many publications as the third-largest contributor, the International Conference on Global Software Engineering. The top contributing journals are IEEE Software and Information and Software Technology. The publication venues' distribution indicates that contributions are published primarily at the International Conference on Agile Software Development and International Workshop on Large-Scale Agile Development, which is unsurprising as they deal exclusively with this topic. Researchers and practitioners can consider these venues as an entry point to explore the state of the art in large-scale agile development research and gain easier access to the community. We note that several studies are present in leading software engineering journals, e.g., IEEE Software, Information and Software Technology, and Empirical Software Engineering. Since six publication venues cover half of all publications, we assume that researchers primarily submit their research to a few specific venues. The publication venues' distribution shows a long tail of 27 venues, each with one published study.

#### *4.2.4. Research types and methods*

Figure 18 shows the distribution of selected studies against their research types. We only identified three research types of the six research types specified by Wieringa et al. [103] (see Table 5), namely evaluation research, solution proposal, and philosophical papers. We note that 110 articles report on evaluating solutions applied in practice, typically through case studies, mixed methods, surveys, or similar empirical methods. Solution proposals were the next most frequently covered research type presenting either novel solutions or significant extensions of existing solutions. Philosophical papers that offer new perspectives on existing things by framing the large-scale agile development field using conceptual frameworks or taxonomies were the third most commonly identified research type. Based on our selection criteria (see Section 3.2.2), we could not identify any experience and opinion papers since they were excluded during our selection process. We could also not identify any validation research studies in which researchers investigate novel techniques that have not yet been implemented in practice. This observation would also have been atypical as most of the selected articles are concerned with identifying, describing, and evaluating the practical application of agile methods on a larger scale. The fact that evaluation research is widely used positively impacts the potential for transferring current research findings to the industry. The prevalence of evaluation research studies is natural in a phenomenon such as large-scale agile development, which is driven primarily by industry instead of resulting from the context of a research lab.

Figure 19 shows the evolution of research types over time. We notice that evaluation research studies dominate large-scale agile development research throughout the period considered, especially before 2013, when 12 out of 14 papers belong to this category. Accordingly, before 2013, only two articles were published from the solution proposal research type. After 2013, an increasing tendency towards all three identified research types can be observed. While between 2014 and 2019, at least six evaluation research papers were published each year, at least three articles per year were published combining philosophical papers and solution proposals. As a result, we observe a promising development in large-scale agile development research, i.e., researchers evaluate existing techniques in practice and develop new solutions based on their observations and frame the observed phenomena using conceptual models.

To investigate research methods applied in large-scale agile development, we visualize the research approaches used in the selected studies in Figure 20. Most studies are empirical and include case studies, mixed methods, ground theory, survey, design and creation, and action research. By far, the most prevalent research method in large-scale agile development is case study research, followed by (systematic) literature review/(systematic) mapping study, mixed methods, and grounded theory. We could not identify the applied research method in seven articles, as it was neither described nor derivable. Only three papers are theoretical, mainly in the form of conceptual models. For instance, Carroll and Conboy [S113] apply normalization process theory to theorize the process of large-scale agile transformations by identifying existing assumptions about these transformations and explaining what factors enable their normalization. Sweetman and Conboy [S127] use the complex adaptive systems lens to critically appraise current thinking on portfolio management in an agile context. We identified only one action research paper, namely the from Pries-Heje and Krohn [S95], that describes the adoption of SAFe at SimCorp and the resulting implementation challenges using a participatory action research project. The results show that the body of knowledge on large-scale agile development is mainly empirical and is still at an exploratory stage. A high percentage of articles are case studies that are sparsely based on grounded theory or theory-building methods.

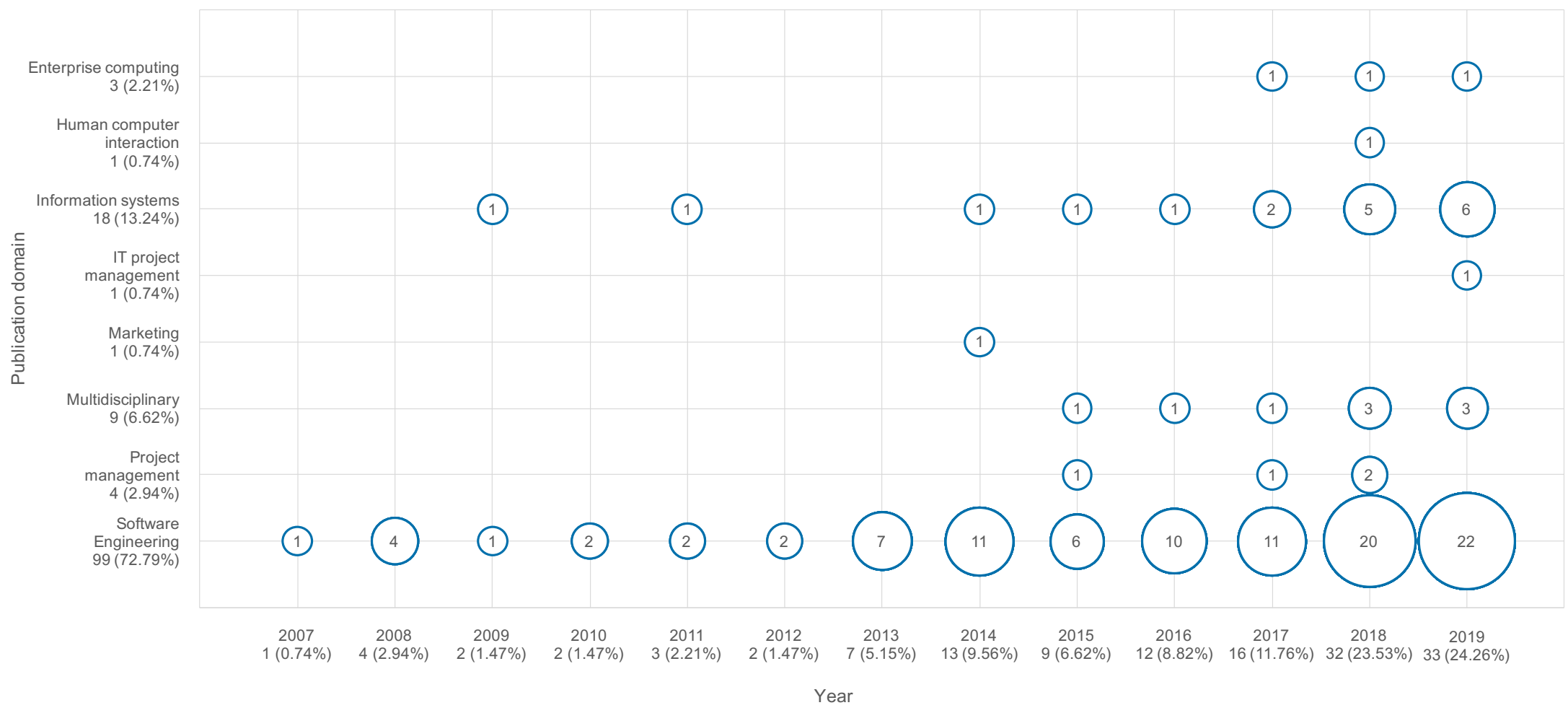

Figure 17: Number of studies per publication domain over time

Table 9: Top-10 publication venues ranked by number of submitted studies

| # | Publication source | Type | Domain | No. | % |
|---|---|---|---|---|---|
| 1 | International Conference on Agile Software Development | Conference | Software engineering | 21 | 15.44 |
| 2 | International Workshop on Large-Scale Agile Development | Workshop | Software engineering | 20 | 14.71 |
| 3 | International Conference on Global Software Engineering | Conference | Software engineering | 9 | 6.62 |
| 4 | IEEE Software | Journal | Software engineering | 6 | 4.41 |
| 5 | Information and Software Technology | Journal | Software engineering | 6 | 4.41 |
| 6 | International Workshop on Autonomous Teams | Workshop | Software engineering | 6 | 4.41 |
| 7 | Empirical Software Engineering | Journal | Software engineering | 5 | 3.68 |
| 8 | Hawaii International Conference on System Sciences | Conference | Information systems | 5 | 3.68 |
| 9 | Americas Conference on Information Systems | Conference | Information systems | 4 | 2.94 |
| 10 | Journal of Systems and Software | Journal | Software engineering | 4 | 2.94 |

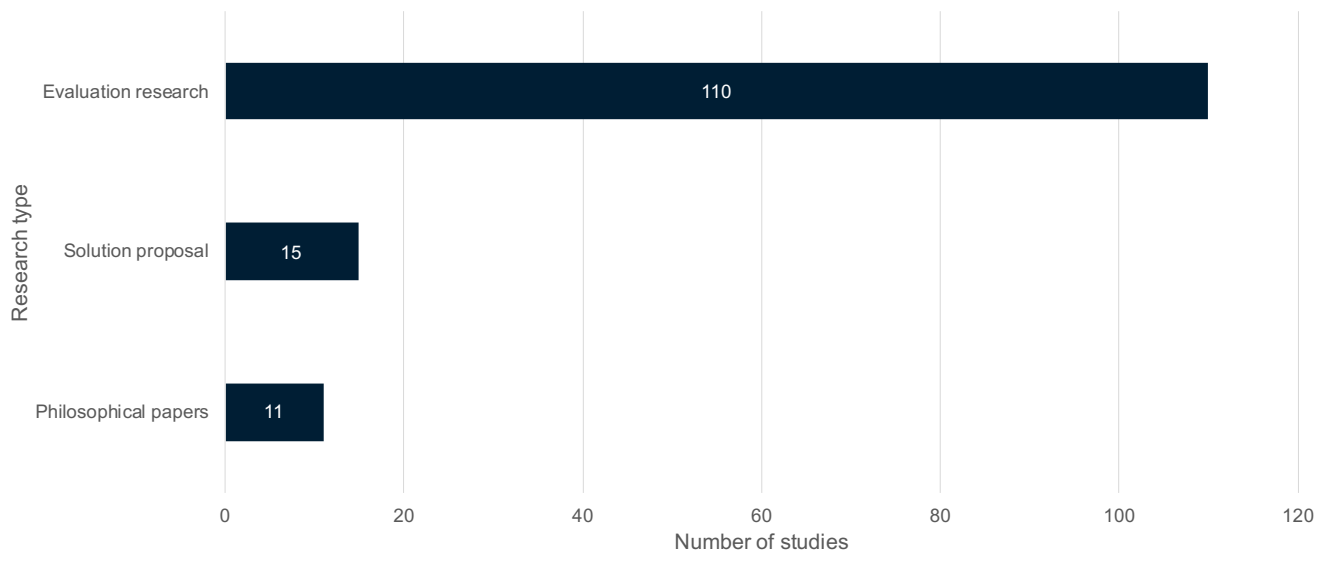

Figure 18: Distribution of research types

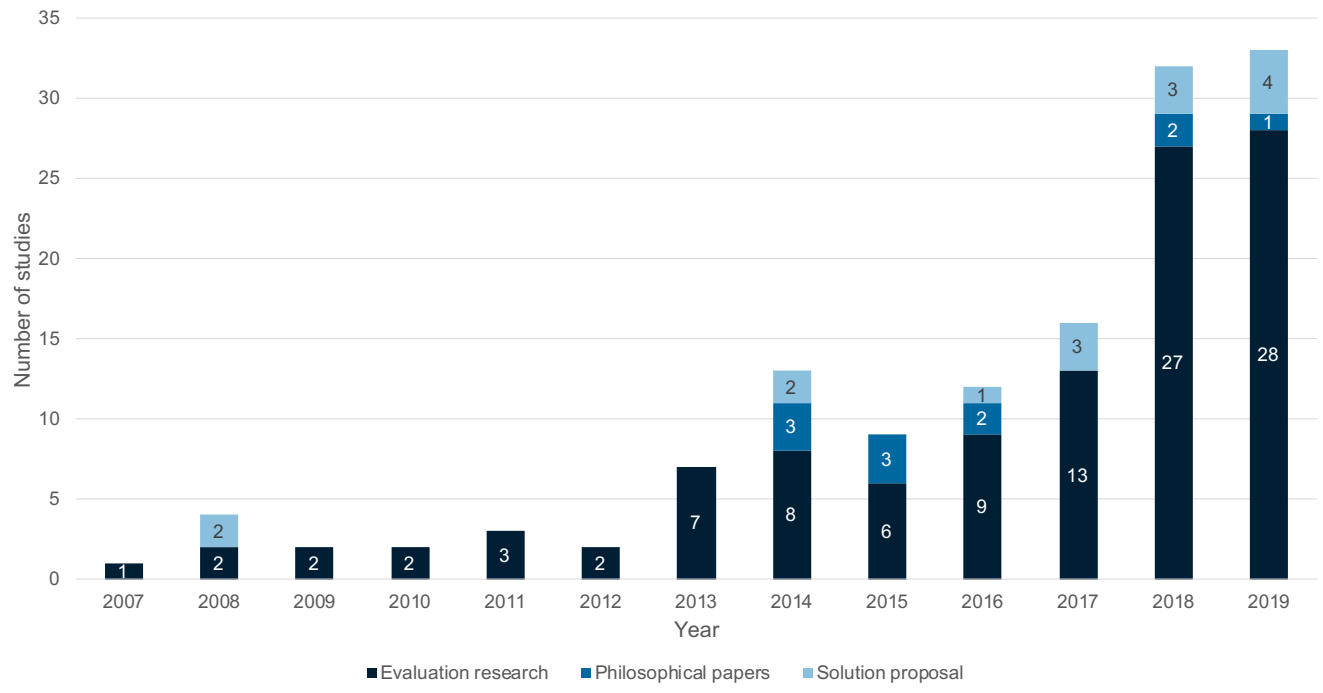

Figure 19: Number of studies per research type over time

To reveal trends, we visualize the annual distribution of applied research approaches over time in Figure 21. Figure 21 confirms the prevalence of case study research from a temporal perspective, i.e., case study research was applied throughout the entire observation period and shows a nearly constant increase over time, peaking in 2018 and 2019. Between 2007 and 2013, 76.19% of the studies were based on case study research. At the same time, only four out of the eight identified research approaches, i.e., case study, design and creation, grounded theory, and survey. As of 2014, the remaining four research meth-ods, i.e., action research, mixed methods, (systematic) literature review/(systematic) mapping study, and theoretical, were applied. The most diverse usage of research approaches was observed in 2019, when seven different research methods were used. The increasing use of various research methods validates our view that large-scale agile development is a multifaceted research field necessitating the investigation of the observation object from different scientific angles. This trend demonstrates the increasing demand from companies for academic support, which fuels the growing research interest of scholars.

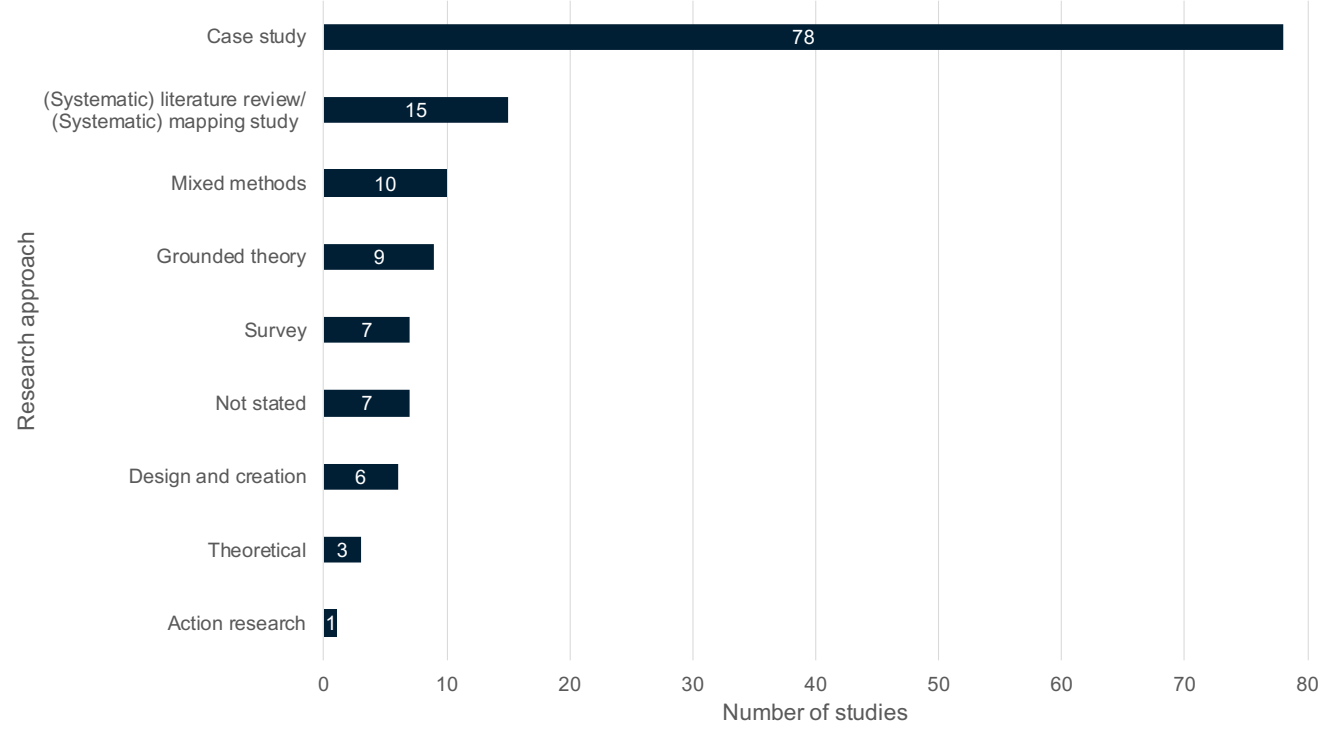

Figure 20: Distribution of applied research approaches

#### *4.2.5. Research outcomes*

Figure 22 shows the distribution of research contributions of the 136 selected studies. A total of 115 papers contribute in the form of lessons learned and guidelines. They are rather observational by analyzing and describing the application of agile practices in large-scale projects. Conversely, the remaining studies contribute models, frameworks/methods, and theories. For instance, Qumer and Henderson-Sellers [S6] propose a new framework called Agile Software Solution Framework to support the adoption and improvement of agile methods in large-scale software projects. Using the human systems dynamics model, Power [S7] presents a new model to determine when scaling agile practices is appropriate for large organizations. Bass [S18] provides another example and uses the Glaserian grounded theory approach to study and explain artifact inventories used in large-scale agile offshore software programs. None of the selected studies contribute to developing new or improved tools. This observation suggests that further research is needed that develop conceptual models and theories to strengthen the theoretical foundations of large-scale agile development research and create rigorously developed frameworks, methods, and tools to assist practitioners.

Figure 23 shows an annual distribution of the contribution type facet. Most articles focused on deriving lessons learned from observations made on large-scale agile endeavors between 2007 and 2019. During this period, at least one lessons learned study was published annually, peaking in 2018 and 2019. Starting in 2014, with at least two published studies per year, a clear tendency towards creating frameworks/methods, models, and theories can be observed. This observation indicates growing empirical evidence owing to the rising interest in the research community and the growing body of knowledge.

To spot possible patterns between the publications domains' preference for made contribution types and applied research approaches, Figure 24 displays the relationship between these three data items. We can see that in almost all domains, except for IT project management, deriving lessons learned from observations made is the most common contribution type. We note that most diverse contributions have been made in the software engineering field, confirming its multidisciplinary nature (cf. [45]) even in the context of large-scale agile development. Concerning the relationship between publication domains and research methods, we notice that case study research is preferred in most domains. Although case study research is the most commonly used approach in the information systems and software engineering domains, we realize that these two domains use a wide variety of research methods.

To better understand the relation between the research outcome and the applied research type and approach of each study, we visualize the relationship between these three data items in Figure 25. The predominance of the selected articles applies case study research to evaluate the usage of agile methods in large-scale projects and derive a set of lessons learned based on the analysis of the research findings. In terms of evaluation research, the remaining articles apply various research methods, mainly (systematic) literature reviews/(systematic) mapping studies, mixed methods, and surveys, to draw some lessons learned. Further 17 papers use case study research, design and creation research, grounded theory, and mixed methods research to make contributions in the form of taxonomies, conceptual frameworks, and theories. For instance, by applying grounded theory, Santos et al. [S44] create a conceptual model that aims to explain the dependence of adequate knowledge sharing across agile teams on the purposeful application of agile practices and organizational conditions and stimuli. Turetken et al. [S19] propose a maturity model to assess the degree of adopting SAFe in companies using design science research. Rolland et al. [S73] employ a mixed methods research design consisting of a literature review and case study to create a conceptual model for examining the underlying assumptions of large-scale agile development. By comparing contribution types and research types and methods, the under-representation of the usage of systematic literature reviews/systematic mapping studies and surveys becomes apparent. Only 16.18% of the studies use (systematic) literature reviews/(systematic) mapping studies and surveys to contribute in terms of lessons learned. These approaches are non-existent in other contribution types. This observation indicates a research gap in creating new conceptual models or adapting existing theoretical models from different research domains to explain the various aspects of large-scale agile development based on quantitative surveys and statistical analyses.

#### *4.2.6. Research data types*

While 121 of the selected papers are primary studies and another 15 articles are secondary studies, we did not find any tertiary studies. Six secondary studies provide ad hoc (systematic) literature reviews on challenges and success factors of applying agile methods in large (and distributed) software projects (cf. [S10], [S22], [S47], [S89], [S135], [S136]). Four secondary studies compare and analyze various scaling agile frameworks based on (systematic) literature reviews (cf. [S9], [S17], [S57], [S108]). Based on a literature review, one secondary study identifies roles responsible for inter-team coordination of agile teams (cf. [S20]). One secondary study reveals a set of motivators for large-scale adoption of agile methods from a management perspective (cf. [S65]). One secondary study provides a multi-vocal literature review describing the benefits and challenges of adopting SAFe (cf. [S132]). One secondary study conceptualizes coordination in large-scale agile development from a multi-team systems perspective (cf. [S30]). One secondary study provides a set of challenges that harm quality requirements in large-scale distributed agile projects and proposes a set of solutions to overcome them (cf. [S134]).

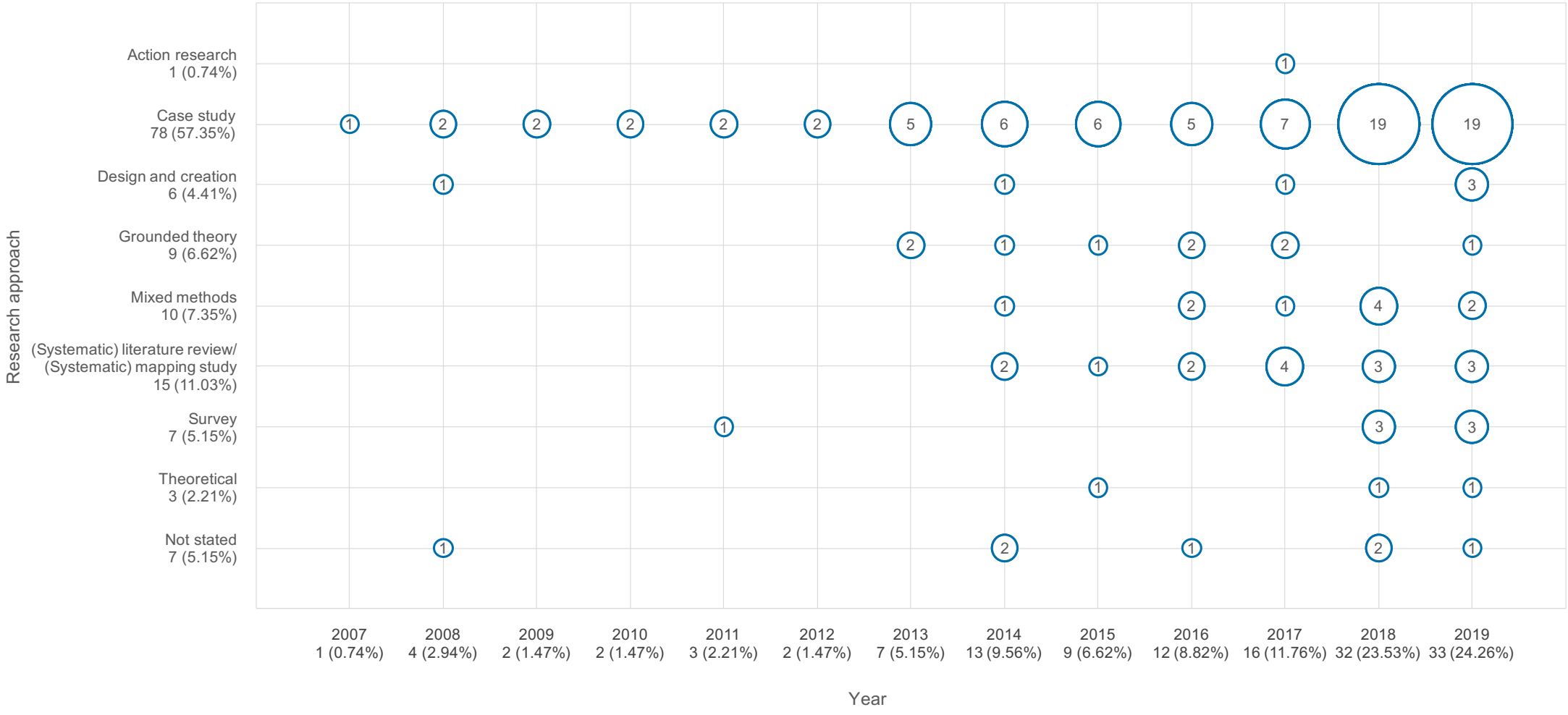

Figure 21: Number of studies per research approach over time

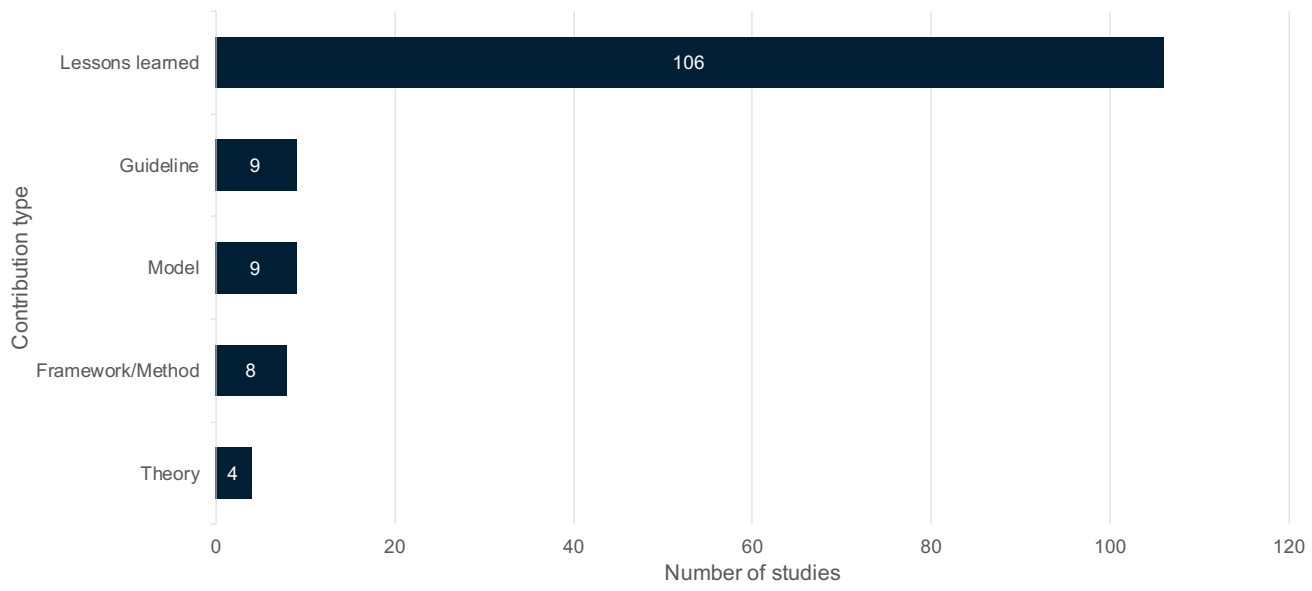

Figure 22: Distribution of research outcomes

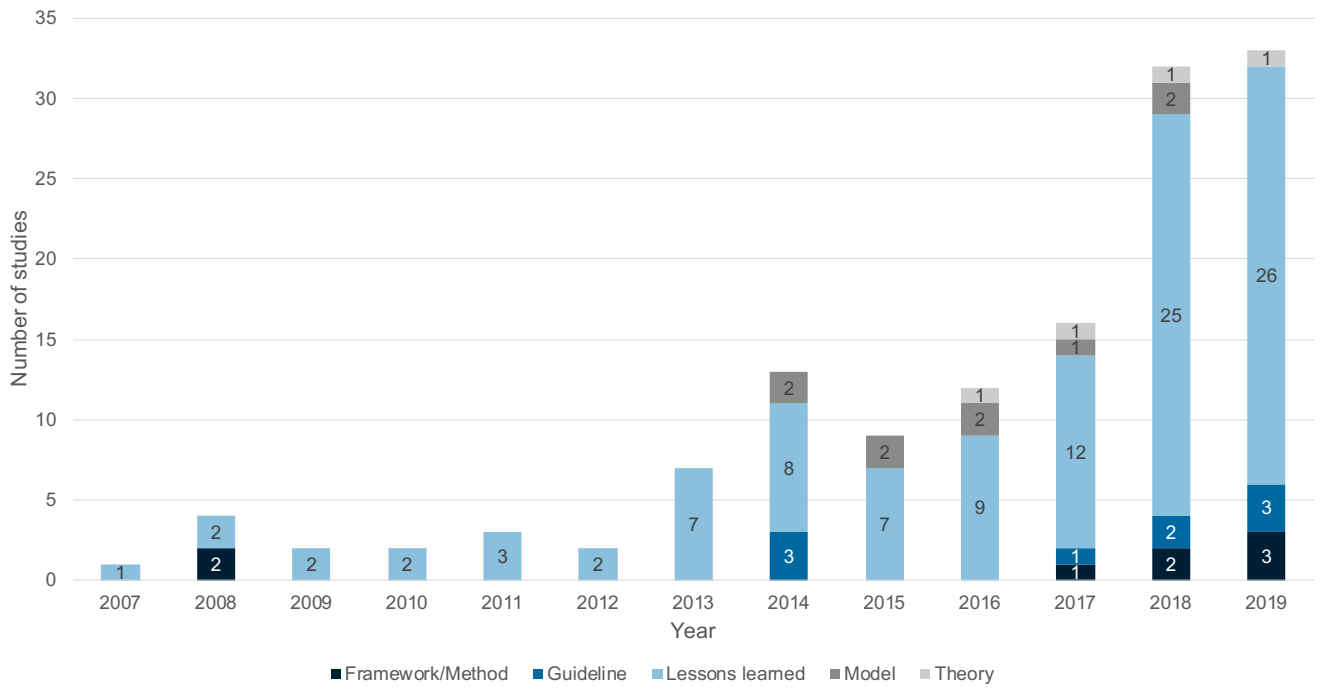

Figure 23: Distribution of research outcomes over time

Figure 26 shows an annual distribution of the research data type facet. Alongside the sharp increase in the number of published studies over the past decade, secondary studies have also increased in recent years. Specifically, all secondary studies were published after 2013. The increase of secondary studies indicates the growing body of knowledge and maturity of the large-scale agile development research field.

### 4.3. *Seminal studies*

As suggested by Dingsøyr et al. [34] and Nerur et al. [71], we identified seminal works as they facilitate understanding and exploration of the intellectual structure of the large-scale agile development research field. Like Herbold et al. [48], we extracted data on citation numbers from Google Scholar to define our criterion for seminal publications and consider the top 10% of the studies with the most citations as influential (see Table 10). The 13 seminal publications:

1. discuss effects, issues, benefits, and success factors related to the large-scale introduction of agile practices in plan-driven organizations (cf. [S1], [S2], [S5], [S22], [S92]),
2. describe experiences in applying agile practices based on Scrum to large, globally distributed software development programs (cf. [S34], [S103]),
3. present the role of communities of practices and the associated challenges and success factors as part of large-scale agile transformations (cf. [S3]),
4. propose a scaling framework for creating new agile software development processes to develop large and complex software applications (cf. [S6]),
5. propose a framework for guiding software process improvement activities concerning agility in large software product development organizations (cf. [S24]),

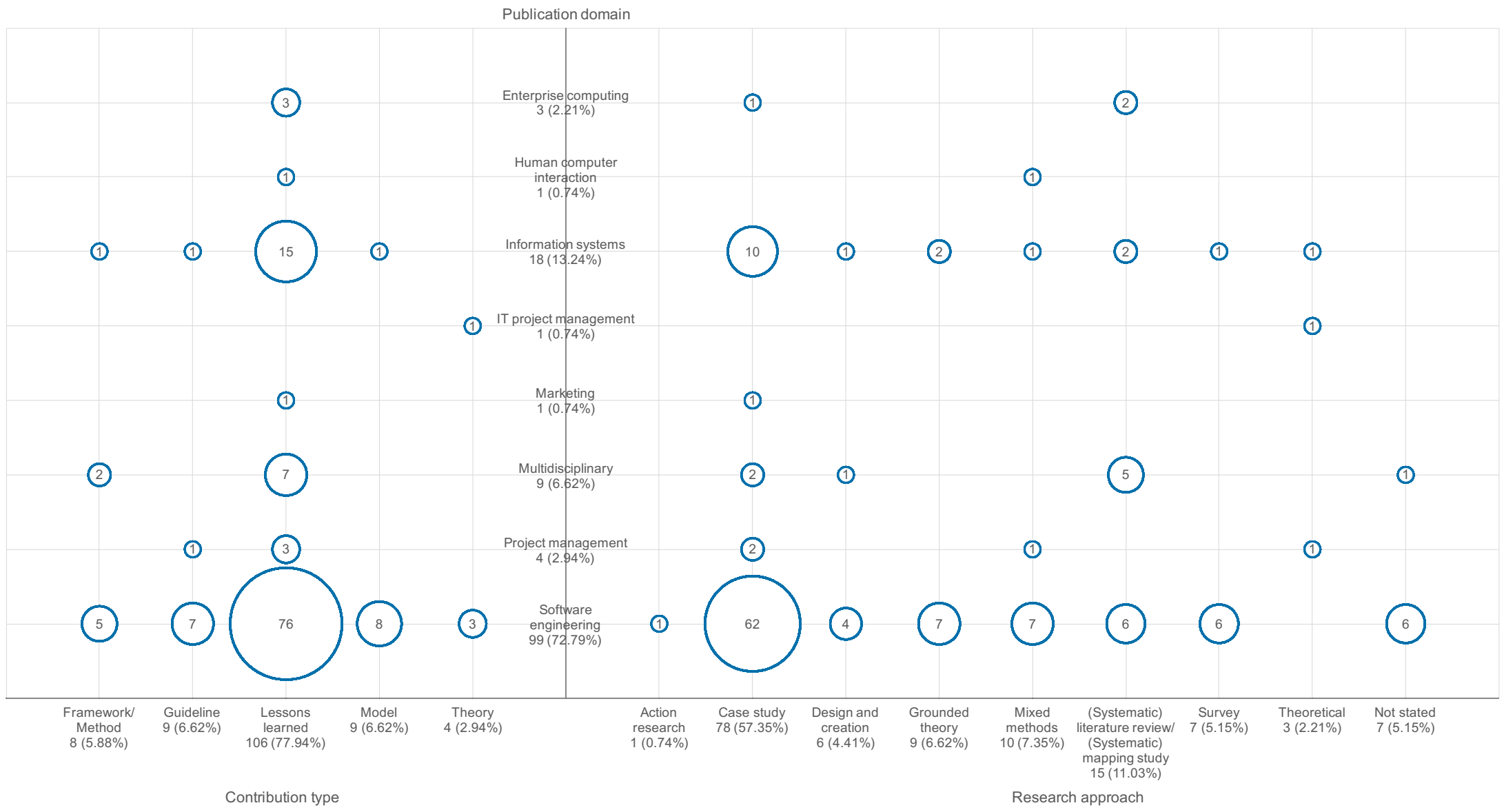

Figure 24: Mapping of publication domains against contribution types and research approaches

6. elucidate the adaption of agile methods in terms of customer involvement, software architecture, and inter-team coordination in a large-scale agile program (cf. [S43]),
7. report the application of the scaling approach SoS and its associated challenges and success factors in large-scale distributed Scrum projects (cf. [S53]), and
8. describe the adoption of portfolio management practices and the associated benefits and side-effects in organizations applying agile methods at large-scale (cf. [S112]).

### *4.4. Research streams and research agenda in large-scale agile development*

As a common goal of systematic mapping studies is to assess the state of the art and maturity level of a research area [58, 82], we present the general structure, central research themes, and research gaps of the large-scale agile development research field. Below we present our key findings related to the identified research streams and gaps before discussing each of them.

#### *4.4.1. General overview*

We explored research clusters to identify pertinent research themes and structured the selected studies according to these themes. Figure 27 provides an overview of the identified research streams and shows the number of studies assigned to each stream. We identified 10 research streams, namely *Agile architecture*, *Agile planning*, *Agile portfolio management*, *Agile practices at scale*, *Communication and coordination*, *Global and distributed software engineering*, *Large-scale agile transformations*, *Scaling agile frameworks*, *Taxonomy*, and *Team autonomy*. While some research streams have enjoyed very little research interest, e.g., *Taxonomy* or *Agile portfolio management*, researchers have devoted a great deal of attention to investigating the scaling of agile practices, i.e., *Agile practices at scale*, studying the coordination of large agile multi-team settings, i.e., *Communication and coordination*, and analyzing the adoption of scaling frameworks, i.e., *Scaling agile frameworks*.

Table 11 shows a tabular representation of the research streams, including their sub-topics[6]. While most research streams exhibit numerous endeavors and clusters dealing with specialized sub-topics, some research streams scarcely show any specific sub-research cluster. For instance, in the research stream *Communication and coordination*, we identified several sub-topics, e.g., *Communication mechanisms, Team performance*, and *Knowledge networks*, while we did not observe any specialized sub-themes in the *Team autonomy* and *Taxonomy* research streams. In almost all research streams (except *Taxonomy*), we observe several studies that derived several lessons learned in the form of challenges and success factors based on their investigations. The complexity and the number of sub-clusters identified per research stream are logically reflected in the number of studies assigned to a stream. While only three papers were assigned to the *Taxonomy* research stream with no specialized sub-topic, nine sub-streams were identified in the *Agile practices at scale* research stream.

Figure 28 visualizes how the number of studies assigned to the research streams has developed over time. The earliest researches on large-scale agile development were conducted between 2007 and 2008 and are related to the *Agile practices at scale*, *Global and distributed software engineering*, and *Scaling agile frameworks* research streams. Tessem and Maurer [S126] represent the first reported paper on large-scale agile development belonging to *Agile practices at scale* research stream and describing the adaptation of Scrum for a project involving 70 people. The initial contributions to the four research streams: *Agile architecture*, *Team autonomy*, *Communication and coordination*, and *Taxonomy* were published between 2013 and 2016, when a great interest in the large-scale agile development research topic emerged. After 2013, there has been a general increase in research interest in all research streams (except *Taxonomy*). The *Scaling agile frameworks* research stream shows a steady research interest between 2016 and 2019, with a peak in 2019, accounting for almost one-third of all published studies in 2019. This observation is congruent with Uludağ et al. [101] stating that many scaling frameworks were created and published between 2011 and 2018, confirming the increasing industry interest in scaling frameworks and sparking a growing interest in analyzing the adoption of these frameworks. Furthermore, 56.25% of all studies published between 2018 and 2019 were related to *Agile practices at scale*, *Communication and coordination*, and *Scaling agile frameworks*, while the remaining seven streams accounted for 43.75% of all studies.

[6] A topic map of Table 11 can be found in Appendix C.

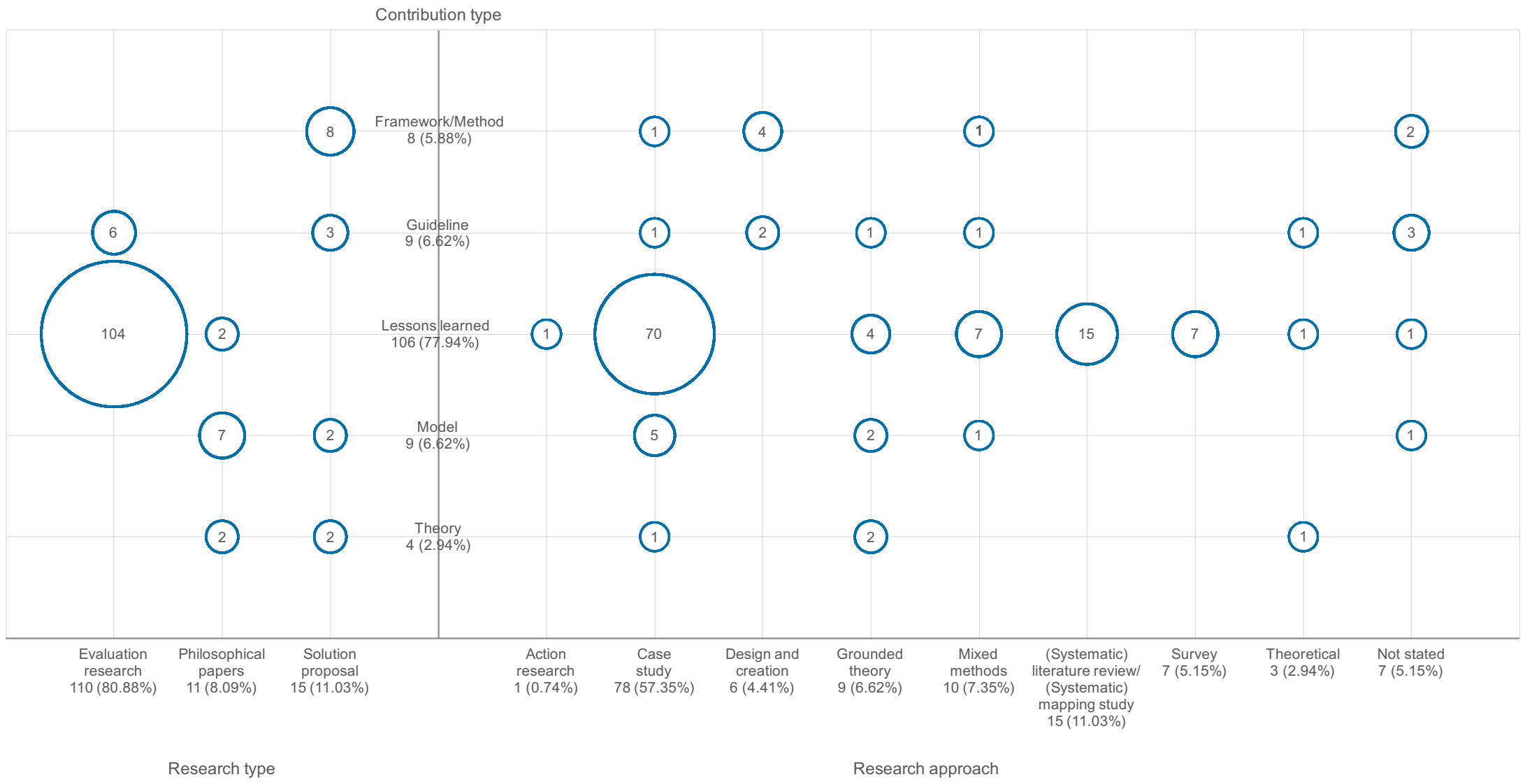

Figure 25: Mapping of contribution types against research types and approaches

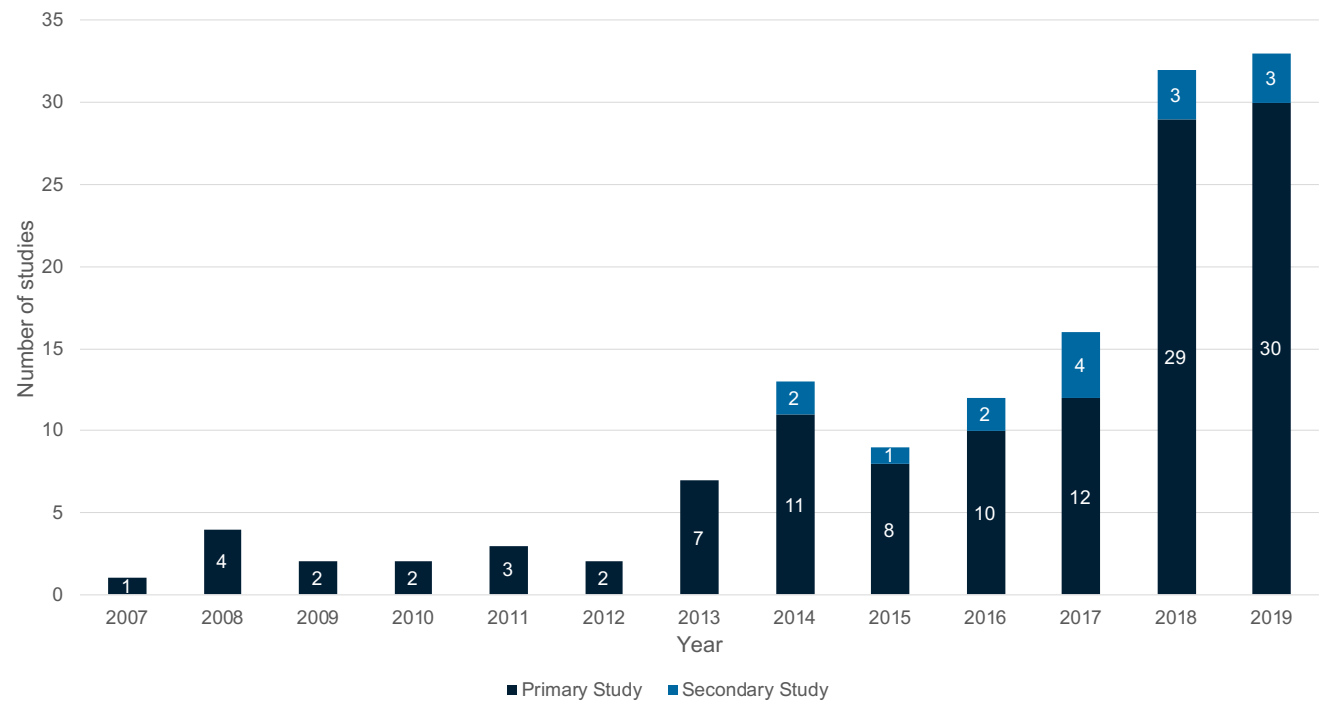

Figure 26: Distribution of used research data types over time

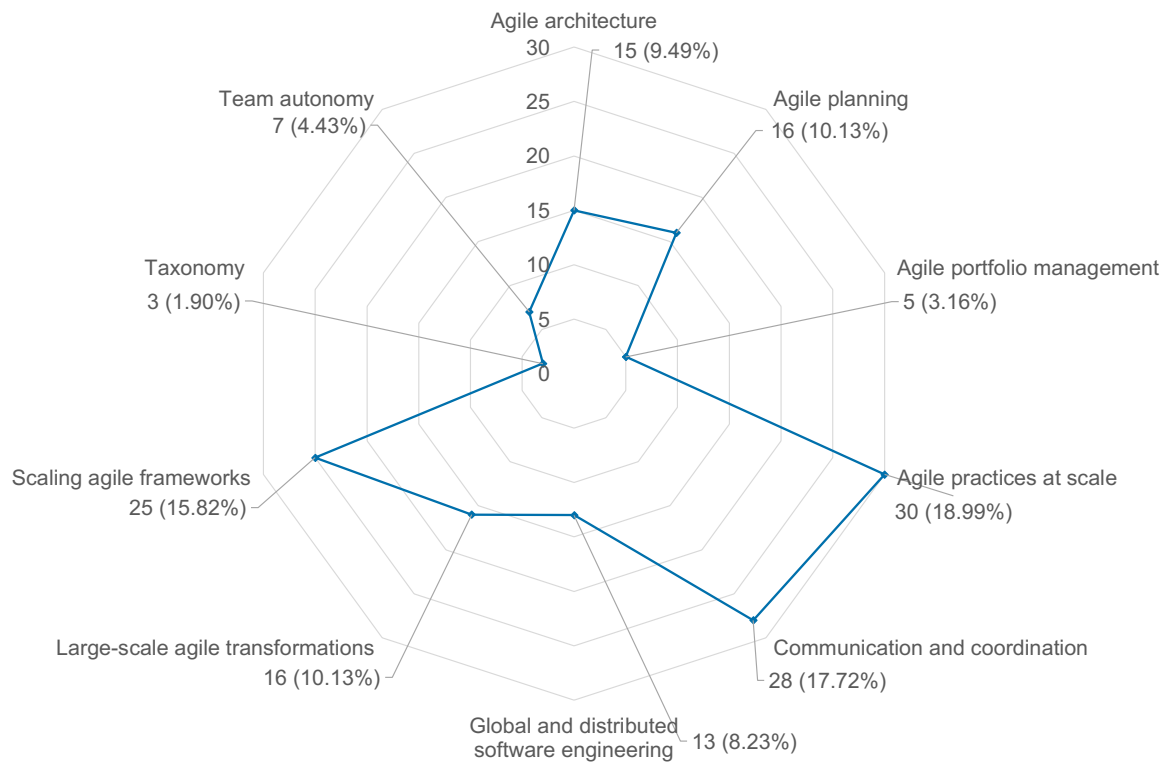

Figure 27: Distribution of the studies in the different research streams

Following Kitchenham et al. [58], we derive a research agenda that scholars can use. Figure 29 shows the number of research questions identified for each research stream. We identified 81 research questions that were mapped to the respective streams (see Table 12). Most research questions were identified in the *Communication and coordination*, *Agile architecture*, and *Agile planning* research streams.

Next, we discuss the research streams and gaps reported in the selected studies.

#### 4.4.2. *Agile architecture*

Agile methods imply that architecture emerges from the system rather than being imposed by some direct structuring force [10]. The practice of emergent design is effective at the team level but insufficient when agile methods are applied at a larger scale, as large-scale agility is enabled by architecture, and vice versa [64], [S106]. The topic of agile architecture is gaining attraction among agilists and researchers (cf. [41, 73, 91]). The *Agile architecture* research stream aims to address the questions of how companies can combine the at first glance contradictory topics of agility and architecture to build complex products and how agile teams can be jointly architecturally aligned at the enterprise level. As the topic of agile architecture is multi-faceted, we identified six sub-topics in the *Agile architecture* stream. The first two sub-topics address agile architecture research at different organizational levels, e.g., at product level, i.e., soft-ware architecture, and enterprise level, i.e., enterprise architecture (cf. [S25], [S81], [S102]). Due to the multitudinousness of the agile architecture topic, we also identified several sub-topics related to software and enterprise architecture sub-topics. Within the software architecture sub-topic, extant literature aims to describe how software reuse can be facilitated in large-scale agile endeavors (cf. [S25]) and how technical debts can be managed in large-scale agile projects (cf. [S64]). Several studies also aim to clarify the roles and responsibilities of software architects in large-scale agile projects (cf. [S43], [S81], [S104]). In the enterprise architecture sub-topic, we identified two additional sub-topics that aim to answer the questions on how enterprise architecture efforts can be effectively adopted in agile environments (cf. [S11], [S102]) and how enterprise architects can support agile teams (cf. [S57], [S106]). The remaining four sub-topics introduce commonly applied architectural toolboxes, i.e., architecture principles, domain-driven design, models, and patterns (cf. [S39], [S87], [S107], [S110]). Notable results of the *Agile architecture* research stream are:

- Observations indicating that an upfront architecture provides agile teams with a stable working environment contributes to team efficiency and that the role of architects is demanding, requiring continuous coordination with many stakeholders [S43].
- An organizational framework, including roles, teams, and practices, to address challenges related to agile architecting in large-scale agile projects [S107].
- A set of architectural tactics for addressing architectural concerns in large-scale agile projects, e.g., the runway-building tactic, which can be used when architecture is needed that is sufficient to meet near-term needs without causing delays or additional work [S110].

Table 10: Top 10% of publications ranked by number of citations according to Google Scholar (data collected on 2019-12-31)

| Study | Title | Authors | Publication type | Year | No. of citations | Citations per year |
|---|---|---|---|---|---|---|
| [S22] | Challenges and success factors for large-scale agile transformations: a systematic literature review | Kim Dikert, Maria Paasivaara, Casper Lassenius | Journal | 2016 | 273 | 68 |
| [S6] | A framework to support the evaluation, adoption and improvement of agile methods in practice | Asif Qumer, Brian Henderson-Sellers | Journal | 2008 | 256 | 21 |
| [S1] | A comparison of issues and advantages in agile and incremental development between state of the art and an industrial case | Kai Petersen, Claes Wohlin | Journal | 2009 | 229 | 21 |
| [S2] | Agile methods rapidly replacing traditional methods at nokia: a survey of opinions on agile transformation | Maarit Laanti, Outi Salo, Pekka Abrahamsson | Journal | 2011 | 227 | 25 |
| [S92] | The effect of moving from a plan-driven to an incremental software development approach with agile practices: an industrial case study | Kai Petersen, Claes Wohlin | Journal | 2010 | 166 | 17 |
| [S112] | Agile portfolio management: an empirical perspective on the practice in use | Christoph J. Stettina, Jeanette Hörz | Journal | 2015 | 126 | 25 |
| [S34] | Distributed agile development: using scrum in a large project | Maria Paasivaara, Sandra Durasiewicz, Casper Lassenius | Conference | 2008 | 119 | 10 |
| [S103] | Using scrum in a globally distributed project: a case study | Maria Paasivaara, Sandra Durasiewicz, Casper Lassenius | Journal | 2008 | 101 | 8 |
| [S3] | Communities of practice in a large distributed agile software development organization – case ericsson | Maria Paasivaara, Casper Lassenius | Journal | 2014 | 98 | 16 |
| [S53] | Inter-team coordination in large- scale globally distributed scrum: do scrum-of-scrums really work? | Maria Paasivaara, Casper Lassenius, Ville T. Heikkilä | Conference | 2012 | 91 | 11 |
| [S5] | A case study on benefits and side-effects of agile practices in large-scale requirements engineering | Elizabeth Bjarnason, Krzysztof Wnuk, Björn Regnell | Workshop | 2011 | 87 | 10 |
| [S24] | Combining agile software projects and large-scale organizational agility | Petri Kettunen, Maarit Laanti | Journal | 2008 | 86 | 7 |
| [S43] | Exploring software development at the very large-scale: a revelatory case study and research agenda for agile method adaptation | Torgeir Dingsøyr, Nils B. Moe, Tor E. Fægri, Eva A. Seim | Journal | 2018 | 83 | 42 |

Observations suggesting that an upfront architecture provides agile teams with a stable working environment that likely contributes to team efficiency, and that the role of architects in large-scale agile development is challenging and requires ongoing coordination and negotiation among many stakeholders

Agile methods have been criticized for their lack of focus on

Table 11: Tabular overview of research streams on large-scale agile development

| Research stream | Influential studies | Topic level 1 / Study | Topic level 2 / Study | Topic level 3 / Study |
|---|---|---|---|---|
| Agile architecture | [S43] | Software architecture | Software reuse | [S25] |
| | | | Role of software architects | [S43], [S57], [S81], [S104] |
| | | | Technical debts | [S64] |
| | | Enterprise architecture | Challenges and success factors | [S11], [S102] |
| | | | Role of enterprise architects | [S57], [S106] |
| | | Architecture principles | [S39], [S41], [S56] | - |
| | | Domain-driven design | [S87] | - |
| | | Models | [S107] | - |
| | | Patterns | [S110] | - |
| Agile planning | [S5], [S43] | Effort estimation | [S37] | - |
| | | Release planning process | [S8], [S27], [S63], [S68] | - |
| | | Challenges and success factors | [S5], [S26], [S69], [S91], [S101], [S125], [S134] | - |
| | | Customer involvement | [S33], [S43], [S98], [S114] | - |
| | | Not specified | [S74] | - |
| Agile portfolio management | [S112] | Challenges and success factors | [S88], [S112], [S127], [S128] | - |
| | | Patterns | [S15] | - |
| Agile practices at scale | [S1], [S3], [S24] | Tailoring | [S14], [S18], [S61], [S80], [S90] | - |
| | | Kanban | [S131] | - |
| | | Retrospectives | [S62] | - |
| | | Communities of practices | [S3], [S38], [S84], [S118] | - |
| | | Rapid application development | [S96] | - |
| | | Challenges and success factors | [S35], [S46], [S47], [S78], [S126], [S135] | - |
| | | Models | [S19], [S24], [S32], [S41], [S87], [S133] | - |
| | | Patterns | [S35], [S46] | - |
| | | Not specified | [S1], [S75], [S94], [S111], [S119] | - |
| Communication and coordination | [S43] | Inter-team coordination | Multiteam systems | [S30], [S51], [S52], [S93] |
| | | | Coordination mechanisms | [S20], [S28], [S31], [S36], [S43], [S54], [S70], [S71], [S85], [S97], [S115] |
| | | | Team performance | [S48], [S58] |
| | | | Challenges and success factors | [S29], [S117] |
| | | | Software ecosystems | [S130] |
| | | Knowledge sharing | Knowledge boundaries | [S76] |
| | | | Knowledge networks | [S16], [S67], [S83], [S104] |
| | | | Not specified | [S44], [S59], [S84] |
| Global and distributed software engineering | [S34], [S103] | Challenges and success factors | [S10], [S34], [S66], [S72], [S89], [S99], [S103], [S136] | - |
| | | Offshoring and outsourcing | [S14], [S18], [S61], [S80] | - |
| | | Not specified | [S37] | - |
| Large-scale agile transformations | [S2], [S22], [S92] | Challenges and success factors | [S2], [S4], [S21], [S22], [S40], [S50], [S60], [S65] , [S92], [S100], [S121], [S122], [S123] | - |
| | | Agile mindset | [S46], [S86] | - |
| | | Theory building | [S113] | - |
| Scaling agile frameworks | [S6], [S53] | Challenges and success factors | [S49] | - |
| | | Large-Scale Scrum | [S42], [S55], [S79] | - |
| | | Scaled Agile Framework | [S13], [S19], [S36], [S71], [S95], [S115], [S116], [S124], [S129], [S132] | - |
| | | Scrum-of-Scrums | [S53] | - |
| | | Spotify Model | [S12], [S45], [S84] | - |
| | | Agile Software Solution Framework | [S6] | - |
| | | Disciplined Agile Delivery | [S40] | - |
| | | Framework comparisons | [S9], [S17], [S57], [S77], [S108] | - |
| Taxonomy | | Not specified | [S7], [S73], [S105] | - |
| Team autonomy | | Challenges and success factors | [S23], [S39], [S109], [S116] | - |
| | | Governance | [S16], [S82], [S120] | - |

Table 12: Tabular overview of open research questions for large-scale agile development

**Agile architecture**

1. How can technical debts be managed and minimized in large-scale agile projects?
2. How is the role of enterprise architects practiced in large-scale agile development?
3. How do architects collaborate with agile teams in large-scale agile development?
4. How can architecture drive large-scale agile transformations?
5. How can the decision-making power between architects and agile teams be balanced?
6. How can software architecture support the coordination of agile teams?
7. How can coordination mechanisms improve architecture sharing at intra- and inter-team level?
8. How can emergent and intentional architecture be balanced?
9. How can the compliance of agile teams with architecture principles automatically determined?
10. What is the effect of applying architecture principles on the outcome of large-scale agile transformations?
11. What are good practices for addressing challenges related to the establishment of architecture principles?
12. Which typical challenges do architects face in large-scale agile development?

**Agile planning**

1. How do organizations that have adopted agile methods implement release planning?
2. How do product owners and customers collaborate with developers in large-scale agile projects?
3. What legal limitations exist in contracts that reduce agility in large scale projects?
4. How can technical dependencies between agile teams be minimized?
5. How can the prioritization between functional and non-functional requirements be balanced in large-scale agile projects?
6. What are good contracting models for organizations with external customers?
7. What are typical requirements engineering challenges in large-scale agile development?
8. Which ceremonies can be used to improve the release planning process?
9. What are factors that impact the accuracy of effort estimations in large scale agile projects?
10. How can customer representatives and agile teams be aligned in large-scale agile projects?
11. How can high-level planning elements be incorporated in agile daily routines of large-scale agile projects?

**Agile portfolio management**

1. What are best practices in agile portfolio management?
2. How is portfolio management interrelated with other governance functions in an agile context?
3. How can traditional portfolio management techniques be applied in an agile environment?
4. How is the new role of project managers practiced in large-scale agile development?
5. How do agile methods affect program and portfolio management?
6. How to enable the strategic alignment and management of agile project portfolios?

**Agile practices at scale**

1. What are challenges, benefits, and success factors of scaling agile practices in organizations?
2. What are recurring concerns and good practices of typical stakeholders in large-scale agile development?
3. What are challenges, benefits, and success factors of establishing communities of practice in large-scale agile projects?
4. How can the on-boarding of new agile team members be facilitated in large-scale agile projects?
5. What is the impact of applying agile practices to the overall performance of the organization?
6. What are appropriate metrics to monitor the progress of agile teams and to support transparency in large-scale agile projects?
7. How can agile practices be scaled in organizations from the public sector?
8. How can continuous improvement at intra- and inter-team level be facilitated?
9. Which issues arise when are retrospectives are organized at inter-team level?
10. How can agile practices be adapted to meet inter-organizational needs?

**Communication and coordination**

1. How can coordination mechanisms be applied effectively in large-scale agile development?
2. How can effective knowledge networks be created in large-scale agile projects?
3. Which tools can be used to support inter-team coordination in large-scale agile projects?
4. How can the number of meetings in large-scale agile projects be reduced?
5. How can meetings in large-scale projects be designed to increase the effectiveness of coordination?
6. How is intra-team coordination affected by increased focus on inter-team coordination or vice versa?
7. Based from a multi-team perspective, how is coordination in large-scale agile development performed?
8. What are effective organizational structures and collaboration models in large projects?
9. Which effect do cultural differences have on inter-team coordination large-scale agile projects?
10. How can focused work and knowledge be balanced in large-scale agile projects?
11. How does co-location of agile teams affect knowledge sharing in large-scale agile projects?
12. Which challenges are caused by inter-team dependencies within large-scale agile projects?
13. How can daily stand-up meetings be organized in a way that they enable inter-team coordination?

**Global and distributed software engineering**

1. What are challenges, benefits, and success factors of applying agile practices in distributed projects?
2. How can virtual agile teams be supported in distributed software development projects?
3. How is knowledge sharing performed in distributed large-scale agile projects?
4. Which human related factors can positively affect globally distributed software projects?
5. How is frequent communication in distributed projects enabled to overcome the challenges of distance?

**Large-scale agile transformations**

1. What are challenges, benefits, and success factors of performing large-scale agile transformations?
2. How can non-agile units be integrated with agile organizational units to support agile transformations?
3. What are reasons and consequences of conducting large-scale agile transformations on the organizations?
4. How can hierarchical and organizational structures be reduced to facilitate large- scale agile transformation?
5. How are agile structures adopted in business units that are not engaged in IT development or delivery?
6. How can local optimization of agile teams be aligned with the enterprise strategy?
7. Which KPIs exist to measure the enterprise agility?
8. How are agile methods adopted at large-scale in highly regulated environments?
9. How do agile teams adopt common values within large-scale agile transformations?

**Scaling agile frameworks**

1. Which scaling agile frameworks are used in organizations and what are their benefits and challenges?
2. How can scaling agile frameworks be selected that are suitable for specific contexts?
3. How is the Scaled Agile Framework adopted in organizations and what are respected challenges and risks when adopting it?
4. How and when should be scaling agile frameworks used in large-scale agile projects?
5. How are scaling agile frameworks tailored to meet the needs of the organizations in which they are adopted?
6. How is the Large-Scale Scrum framework adopted in different types of organizations?
7. Which performance improvements can be observed when adopting scaling agile frameworks?
8. How can agile release trains and value streams be formed in complex organizations?

**Taxonomy**

1. How can agile in the large be conceptualized besides the dimension of number of teams?

**Team autonomy**

1. How can team autonomy in large-scale agile development be increased?
2. How can inter-team coordination and team autonomy be balanced in large-scale agile projects?
3. What are effective intra- and inter-team coordination mechanisms for autonomous agile teams?
4. How can autonomous teams be designed, supported, and coached?
5. What are barriers to team autonomy in large-scale agile development?
6. How do governance structures influence team autonomy in large-scale agile development?

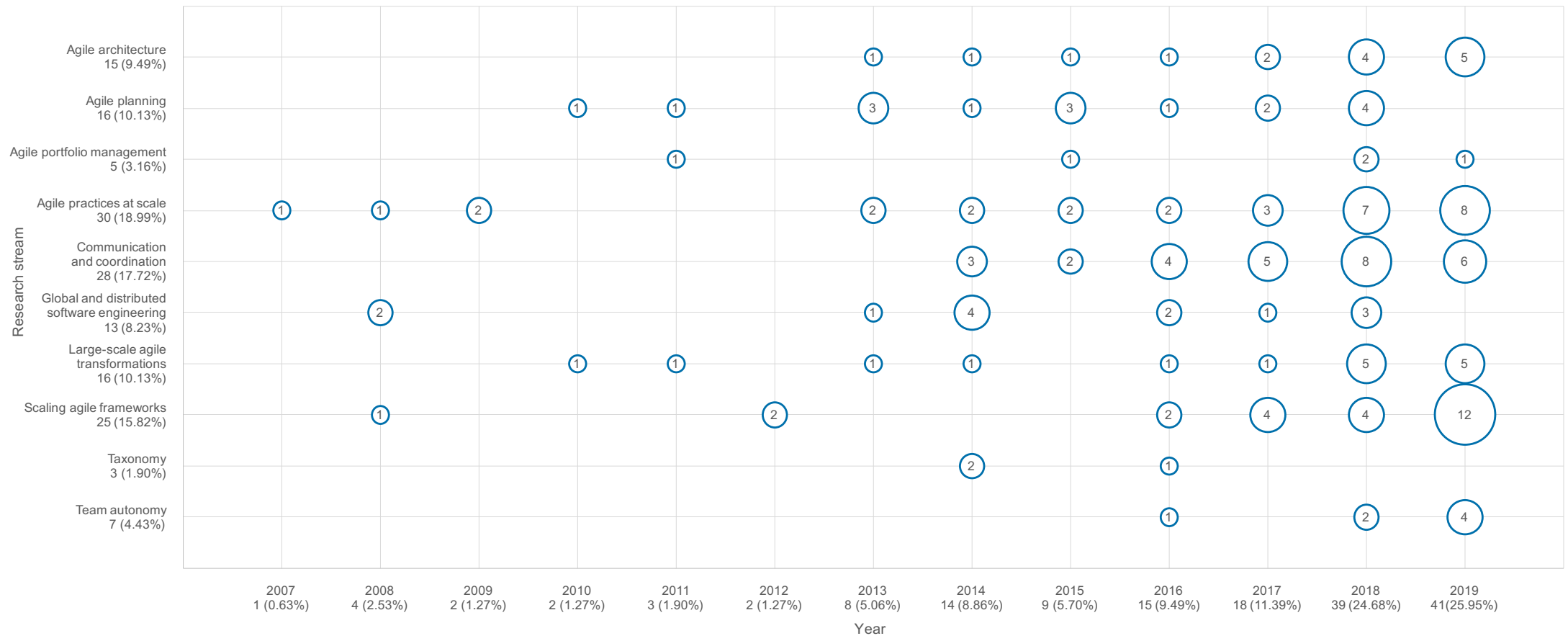

Figure 28: Number of studies per research stream over time

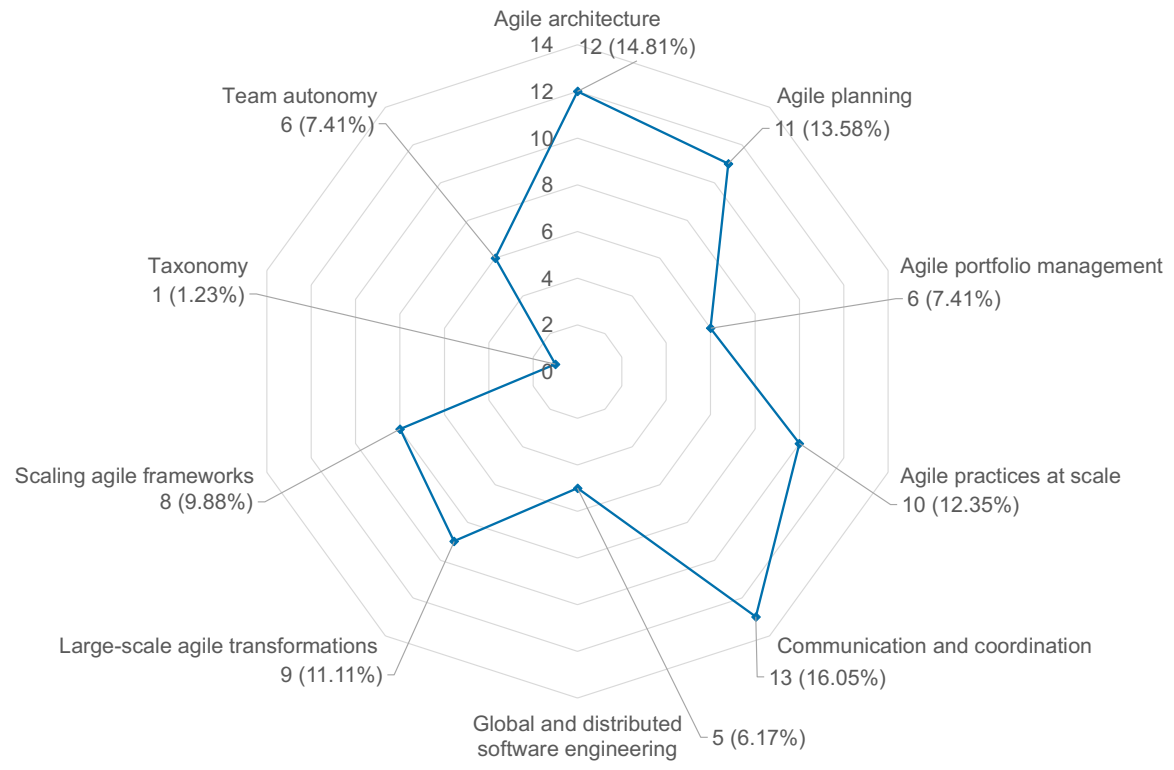

Figure 29: Distribution of research questions in the different research streams

architecture [38], assuming that the best architectures emerge from self-organizing teams [2, 10]. For instance, the incremental design practice of XP asserts that architecture emerges in daily design [10, 13]. Apart from verbal discussions of design decisions, Scrum does not emphasize architecture-related practices as the architecture of a single-project application can always be re-factored and repackaged for a higher level of reuse [10]. There is growing interest from academics and practitioners who wish to combine the two concepts of agility and architecture as some degree of architectural planning and governance becomes increasingly crucial for large-scale agile projects [64]. Due to the existing ambiguity on agile architecture in large-scale agile development, researchers have formulated 12 open research questions on this topic. The first two research questions aim to study the collaboration between architects and agile teams (cf. [W5], [S81], [S87]) and to analyze the tension between architects with decision-making power and self-organizing agile teams (cf. [W5], [S104]). Three studies (cf. [W2], [W6], [S64]) mention the issue of managing technical debts as a research question.

#### *4.4.3. Agile planning*

Many large agile organizations struggle to implement efficient requirements management processes. Moreover, research on agile planning practices in large organizations is scarce [47], [S2]. The *Agile planning* research stream aims to provide scientific evidence on how large-scale agile projects and organizations perform agile planning activities. We identified four sub-topics in this stream. While a lot of research has covered effort estimation in software development projects [52], little research has been conducted on effort estimation in large-scale distributed projects [S22]. Thus, the first sub-topic aims to investigate how large-scale agile projects perform effort estimation (cf. [S37]). Given the importance of release planning to the success of a development project and the lack of solid empirical evidence in the research of release planning in large-scale agile organizations [S68], the second sub-topic aims to describe how these organizations perform release planning. Similar to other research streams, the third sub-topic aims to report on challenges and success factors in adopting agile planning practices (cf. [S5], [S101], [S134]). While agile methods strongly emphasize customer involvement, companies struggle to meet the needs of customers as very large programs encompass many requirements, stakeholders and developers [34], [S33]. Hence, the last sub-topic aims to reveal how large-scale agile projects can actively involve customers (cf. [S33], [S43], [S98]). Notable results pertaining the *Agile planning* research stream are:

- Observations showing that incorporating agile practices into requirements engineering can overcome existing challenges, e.g., communication gaps and over-scoping, but can also introduce new challenges, e.g., balancing agility and stability and ensuring sufficient competencies in cross-functional teams (see [S5]).
- Observations indicating that the scale of a large-scale agile project, e.g., the number of sites, the number of stakeholders, the size and complexity of the legacy code, and the coordination between different types of stakeholders make estimation and planning very challenging (see [S37]).
- A list of ten benefits, e.g., improved communication, transparency, and dependency management, and nine challenges, e.g., lack of preparation and prioritization of requirements, unrealistic schedules, and inadequate architectural planning, when using a release iteration planning method in large-scale agile projects (see [S68]).

We identified 11 research questions for future studies. Although release planning is a critical success factor in agile software projects [26], there is little research on large-scale agile release planning [S27]. Researchers call for studying how companies that have adopted agile methods implement release planning (cf. [W1], [S8], [S27]). They suggest analyzing how companies can use agile ceremonies in the release planning process and how companies can incorporate high-level planning elements into daily routines of large-scale agile projects (cf. [S26], [S51]). As large-scale agile programs often have a large number of stakeholders and users, and their needs have to be communicated to a large number of developers, customer engagement can be a major challenge [S43]. Thus, scientists suggest further research on customer-developer collaboration in large-scale agile projects (cf. [W1], [W2]). They also call for research on identifying appropriate contracting models for companies with external customers (cf. [W6]) and how companies can facilitate alignment between customer representatives and agile teams in large-scale agile projects (cf. [S43]). Researchers should examine legal constraints in contracts limiting the agility of large-scale projects (cf. [W1], [W2]).

#### 4.4.4. Agile portfolio management

While the concept of portfolio management is not new [S88] and is established in the traditional project management literature, the iterative nature of agile methods poses new challenges to the current management practice [S112], e.g., the need for portfolio management to be able to respond quickly to changes while connecting agile teams to strategy [S15]. The *Agile portfolio management* research stream deals with the question of how companies can adapt their traditional portfolio management approaches to agile environments. This stream can be divided into two sub-topics, namely identifying challenges and success factors of agile portfolio management approaches (cf. [S88], [S112], [S127]) and creating patterns for agile portfolio management (cf. [S15]). Stettina and Hörz [S112] provide an overview of portfolio management practices of 14 large European companies affected by agile methods and describe the challenges, implications, and benefits of agile methods on portfolio management practices. Horlach et al. [S15] propose four design goals for an effective agile portfolio management and six design principles for achieving these goals. Interesting findings of the *Agile portfolio management* research stream include:

- Observations indicating that implementing a structured portfolio management process can help reduce time-to-market, improve project visibility, and coordinate the work of agile teams (see [S88]).
- Observations suggesting that most challenges associated with agile portfolio management are process-related and that an alignment with the customer and involvement of the business is beneficial (see [S112]).
- Formulation of 16 propositions for the effective management of agile projects portfolios, e.g., that portfolio managers need to find the right balance between control and autonomy in agile projects, and that agile project portfolios require mechanisms for simple, fast, and collective decision-making (see [S127]).

Within the *Agile portfolio management* research stream, we identified six open research questions. As existing best practices for portfolio management are more suited for stable environments [S15], two studies (cf. [W6], [S15]) suggest researchers identifying best practices for agile portfolio management. Hobbs and Petit [S111] suggest further research on the impact of adopting large-scale agile methods on portfolio management. Rautiainen et al. [S88] recommend exploring how companies can apply traditional portfolio management techniques in agile environments.

#### 4.4.5. Agile practices at scale

Adopting and scaling agile practices outside of their ideal context requires rethinking the original underlying assumptions of agile practices [S73]. The *Agile practices at scale* research stream aims to explore how companies tailor genuine agile practices to fit large-scale projects. We identified eight sub-topics in this research stream. The first sub-topic mainly analyzes how companies tailor Scrum-related roles and artifacts in large projects (cf. [S61], [S80], [S90]). The next four sub-topics describe the application of different agile and lean practices in large agile projects, namely Kanban (cf. [S131]), Retrospectives (cf. [S62]), Community of Practices (cf. [S3], [S84], [S118]), and Rapid Application Development (cf. [S96]). The sixth sub-topic, like other research streams, deals with identifying challenges and success factors for applying agile practices in large-scale projects (cf. [S47], [S78], [S135]). The seventh sub-topic mainly generalizes and conceptualizes observations from case studies in the form of models related to applying agile practices in large-scale projects (cf. [S32], [S41], [S133]). Inspired by existing pattern languages on agile software development, the eighth sub-topic aims to provide best practices for large-scale agile endeavors (cf. [S35], [S46]). Notable findings of the *Agile practices at scale* research stream are:

- Observations showing that adopting agile practices in large-scale projects brings several benefits, e.g., increased control and transparency over the project and better learning and understanding through direct communication, but also raises new issues, e.g., administrative overhead and coordination, and limited focus on architecture (see [S1]).
- Observations pointing that organizational culture, previous agile experience, and management support are key success factors in scaling agile practices and that resistance to change, an overly aggressive time-frame for adoption, and integration with existing non-agile business processes are critical challenges in scaling agile methods (see [S78]).
- A study showing that medium and large software projects using agile methods perform better on average than projects using non-agile methods and that agile methods are more likely to succeed compared to traditional methods as project size increases (see [S119]).

The use of agile methods on large-scale projects brings unprecedented challenges [S43]. We identified 10 related research questions, of which the unveiling the challenges, benefits, and success factors of scaling agile practices in organizations was the most frequently stated (cf. [S2], [S78], [S125]). Although several pattern languages documenting best practices for agile projects have been published in agile software development research (cf. [15, 16, 27]), the body of knowledge about best practices in large-scale agile development is still emerging. Several studies suggest identifying best practices for large-scale agile endeavors (cf. [S1], [S35], [S78]). Given the importance of communities of practices for knowledge sharing, inter-team coordination, and technical work [63], [S3], several researchers recommend future studies to investigate the challenges, benefits, and success factors of establishing communities of practices in large-scale agile projects (cf. [W3], [W6], [S3]). Since onboarding newcomers in large-scale agile projects might be challenging [W3], three studies propose as future work to investigate how companies can facilitate the onboarding process of new agile team members in these projects (cf. [W3], [W6], [W9]). Several studies suggest identifying metrics to quantify the impact of applying agile practices on overall organizational performance and to monitor the progress of agile teams in large-scale agile projects (cf. [W2], [S26], [S63]).

#### 4.4.6. Communication and coordination

Coordinating work is critical when managing large projects with multiple teams [S97]. This circumstance is also true for large-scale agile projects, as work is performed by many developers and development teams simultaneously, and the frequent and iterative delivery of results requires work and knowledge coordination at different levels [S43, S54]. Achieving effective coordination in large-scale agile projects is difficult due to the complexity of these projects [S54]. The *Communication and coordination* research stream tackles the issue of how agile teams can effectively coordinate and communicate with each other. The *Communication and coordination* stream covers multiple topic levels consisting of two sub-topics, namely inter-team coordination and knowledge sharing, and another seven subordinate sub-topics. Since coordination between teams is critical in managing large-scale endeavors [S31], the inter-team coordination sub-topic mainly aims to identify and describe different coordination mechanisms and coordination modes for aligning agile teams (cf. [S28], [S31] [S54]). Within the inter-team coordination sub-topic, we identified five additional sub-topics that aim (i) to advance the conceptual understanding of inter-team coordination through the lens of multi-team systems (cf. [S30], [S52], [S93]), (ii) describe different mechanisms and modes for coordinating agile teams (cf. [S31], [S85], [S97]), (iii) reveal factors that influence the performance of agile teams (cf. [S48], [S58]), (iv) identify challenges and success factors related to the coordination of agile teams (cf. [S29], [S117]), and (v) describe the adoption of agile methods beyond organizational boundaries (cf. [S130]). Since resource availability in a team's knowledge network and the effective knowledge coordination between agile teams become paramount in large-scale agile projects [S83], the knowledge sharing sub-topic mainly aims to describe how agile teams can share their knowledge and expertise with other teams. We identified two sub-topics in the knowledge sharing sub-topic that aim to answer the questions on how companies tailor agile practices to enable scaling across different knowledge boundaries (cf. [S76]) and how companies can build effective knowledge networks in large-scale agile projects (cf. [S16], [S67], [S83]). Notable results of the *Communication and coordination* research stream are:

- Observations suggesting that the combination of traditional planning at an inter-team level and agile development at a team level leads to ineffective coordination in large-scale agile development, as a lack of dependency awareness is hampered by misaligned planning activities at the team and inter-team levels (see [S29]).
- Observations indicating that group mode coordination is central for achieving effective inter-team coordination in large-scale agile projects, using scheduled and unscheduled meetings, e.g., product owner meetings and spontaneous discussions in the open work area (see [S83]).
- Observations pointing that networking is essential in solving complex tasks in large-scale agile projects and that several mechanisms can support the creation of knowledge networks, e.g., introducing formal technical experts and facilitating communities of practice (see [S97]).

Due to the high number of articles in the *Communication and coordination* research stream, we identified 13 research questions that were most frequently for a given stream. Because the coordination of many agile teams is a key challenge [S28], several researchers emphasize further research to investigate how companies can design and apply coordination mechanisms for increasing the effectiveness of coordination (cf. [S18], [S28], [S70]). They also call for research to analyze how companies can reduce the number of meetings in large-scale agile projects (cf. [W5], [S70]) and identify tools that companies can use for inter-team coordination (cf. [W6], [W8], [S70]).

#### 4.4.7. Global and distributed software engineering

Given the benefits to both customer and vendor companies in terms of low cost, early product delivery, and high-quality products, companies are increasingly deploying virtual teams that operating across geographical boundaries to develop software [72, 75]. The competitive advantages and business profits of agile methods motivate companies to adopt agile methods in large, globally distributed projects [S89]. Agile methods are more difficult to scale in distributed projects due to additional challenges, e.g., communication and coordination issues, cultural differences, and temporal issues [S22], [S103]. The *Global and distributed software engineering* research stream deals with how companies can use agile methods in large, globally distributed projects. Studies mainly focus on two sub-topics, namely identifying challenges and success factors in adopting agile methods in large, globally distributed projects (cf. [S72], [S89], [S103]) and the tailoring of agile methods to meet the needs of these projects (cf. [S18], [S61], [S80]). As an example for the first sub-topic, Shameem et al. [S89] describe critical factors that positively impact the adoption of agile methods in large, globally distributed projects based on a systematic literature review. Related to the second sub-topic, Bass [S14] uses 46 interviews with eight companies to inves-tigate how companies adapt the role of the product owner to the needs of large, geographically distributed software projects. Noteworthy findings related to the *Global and distributed software engineering* research stream include:

- A list of 11 challenges identified in large, globally distributed agile projects, of which 6 were evaluated as critical: lack of requirements analysis, customer involvement, communication, roles and responsibilities, management commitments, and knowledge sharing (see [S10]).
- A classification model for large, globally distributed agile projects contrasting solution focused development and execution focused development (see [S61]).
- Observations indicating that adopting agile practices based on Scrum in a large, globally distributed agile project helps mitigate the biggest problem of large, globally distributed projects, namely communication, by creating frequent possibilities to communicate across distributed sites, e.g., through Daily Scrum meetings (see [S103]).

We identified five open research questions in the *Global and distributed software engineering* stream. As many companies scale agile methods to distributed environments due to fast development rates and low development costs [S72], identifying challenges, benefits, and success factors when scaling agile to distributed organizations was the most frequently cited research question (cf. [W1], [S72], [S103]).

#### 4.4.8. Large-scale agile transformations

Companies are striving to become agile to respond to dynamic environments and sustain their survival. As a result, many companies are extensively introducing agile methods leading to large-scale agile transformations [S4], [S21], [S22]. These transformations entail new managerial challenges, e.g., lack of top management engagement [S21], skepticism towards the new way of working [S22] or establishing an enterprise-wide agile culture and mindset [S46]. The *Large-scale agile transformations* research stream aims to shed light on how companies undergo these transformations to meet the imperatives of agile companies. We identified three sub-topics in this research stream. Like other streams, the first sub-topic deals with identifying a set of observed success factors and challenges when companies undertake large-scale agile transformations (cf. [S2], [S4], [S100]. As these transformations represent large episodic change processes that have a large impact on the employees [S21] and changing the peoples' mindset is more difficult than teaching new practices [S46], the second sub-topic discusses various ways in which companies can establish an agile mindset and define common values among employees (cf. [S46], [S86]). The third sub-topic takes a more theoretical stance and aims to theorize the process behind large-scale agile transformations (cf. [S113]). Fuchs and Hess [S21] provide an example for the first sub-topic and captures the interplay of challenges, coping, and scaling actions in the execution of large-scale agile transformations through socio-technical systems theory. Relating to the second sub-topic, Paasivaara et al. [S86] report how Ericsson established value workshops to align different sites and teams as part of its large-scale agile transformation. Carroll and Conboy [S113] provide an example for the third sub-topic, applying normalization process theory to examine the normalization of these transformations. Notable results of the *Large-scale agile transformations* research stream are:

- A survey indicating that while a large-scale agile transformation poses new challenges, e.g., agile deployment and adaptation of agile methods to fit the organization and requirements management and planning in a flexible and iterative manner, it also brings several benefits, e.g., higher satisfaction, a feeling of effectiveness, and increased transparency and autonomy, which is why most respondents do not want to return to the old way of working (see [S2]).
- A list of 35 challenges, e.g., agile difficult to implement and integrating non-development functions, and 29 success factors, e.g., management support and choosing and customizing the agile model, were reported for large-scale agile transformations (see [S22]).
- Observations stating introducing agile practices in an organization that has been working in a plan-driven way leads to several improvements, e.g., increased release frequency and better reflection of the customers' needs, but also raises several challenges, e.g., support for coordinating a high number of teams and integration of release projects in the overall development process (see [S92]).

As the *Large-scale agile transformations* research stream has flourished in recent years, especially in the last two years, several studies formulated important research questions as avenues for future research. The most frequently asked research question aims to identify challenges, benefits, and success factors in conducting large-scale agile transformations (cf. [S24], [S45], [S100]). As "*agile breaks everything*" and adopting agile practices may have far-reaching implications for companies [62], [S4], several researchers call for examining the reasons for performing large-scale agile transformations and their impact on companies (cf. [S5], [S45], [S65]). Since companies often encounter challenges when agile units need to collaborate with non-agile departments [62], some studies propose to investigate how companies have overcome these challenges from a longitudinal perspective (cf. [W7], [S4], [S45]). From an integrational perspective, three studies suggest further research on how companies can adopt agile structures in business units that are not engaged in software development or delivery (cf. [W4], [W6], [S45]). Hierarchical control and bureaucracy mechanisms can act as barriers for the successful performance of large-scale agile transformations [S122], researchers should explore how companies can dissolve their hierarchical structures to facilitate these transformations (cf. [W4], [W6], [W7]).

#### 4.4.9. Scaling agile frameworks

Some custodians of existing agile methods and practitioners have created several scaling frameworks that claim to provide off-the-shelf solutions for solving issues related to the large-scale adoption of agile methods [101], [S43], [S113]. As there is a increasing interest in adopting scaling frameworks from a practical perspective [31], there is also a growing academic interest in studying the adoption of these frameworks within the *Scaling agile frameworks* research stream. We identified eight sub-topics in this stream. As empirical evidence on the adoption of scaling frameworks is still very much in its infancy [S49], the first sub-topic aims to highlight challenges and recommendations for companies seeking to adopt scaling frameworks (cf. [S49]). The other six sub-topics mainly analyze how popular scaling frameworks are adopted (cf. [S12], [S36], [S40]). The last sub-topic aims to provide a comparison of these frameworks (cf. [S9], [S17], [S77]). Conboy and Carroll [S49] provide an example for the first sub-topic and unveils nine challenges associated with implementing scaling frameworks and a set of recommendations to address these challenges adequately. Relating to the six sub-topics analyzing the adoption of popular frameworks, Petit et al. [S12] present a case study of how a large company adapted Spotify practices to promote the effectiveness of team autonomy in a mission-critical project. Lal and Clear [S40] provide another example, presenting a case study on how a global software vendor transitioned to an agile company by adopting DAD. Related to the last sub-topic, Alqudah and Razali [S9] compare six scaling frameworks, e.g., DAD, LeSS, and SAFe, based on various criteria, e.g., team size and available training. Noteworthy results of the *Scaling agile frameworks* research stream are:

- A maturity model consisting of levels, principles, and practices for defining a roadmap for SAFe adoption and assessing the level of adoption (see [S19]).
- A list of nine challenges, e.g., comparing scaling frameworks and top-down vs. bottom-up implementation, and a set of recommendations to overcome these challenges, e.g., using a small number of metrics to compare frameworks and determining whether a framework promotes a top-down or bottom-up implementation approach, when implementing scaling frameworks (see [S49]).
- A survey indicating that the SAFe adoption leads to several benefits, e.g., improved transparency, cooperation, and cadence, but also entails several obstacles, e.g., old mindset and culture, SAFe has not correctly fitted to the organization, and missing fluency when using SAFe (see [S129]).

The selected studies within the *Scaling agile frameworks* research stream named eight research questions for further investigation by researchers. The most frequently stated research question on scaling agile frameworks deals with the observation of adopting specific scaling frameworks in companies and the associated benefits and challenges (cf. [W6], [S40], [S60]). As scaling frameworks are often not selected systematically but merely based on the popularity of the framework or recommendation of consultants, future work should identify contextual factors and comparison criteria that companies can use to select scaling frameworks systematically (cf. [W3], [S79], [S92]). Since many companies also adapt scaling frameworks to their organizations, two studies (cf. [S22], [S60]) call for future research on how companies tailor scaling frameworks to meet their needs. There is also a call for research on the two most widely adopted scaling frameworks, namely SAFe and LeSS, to study their adoption in companies (cf. [S19], [S55], [S124]).

#### *4.4.10. Taxonomy*

The *Taxonomy* research stream deals with providing more conceptual clarity related to large-scale agile development, as well as the scale and implications for scalability of agile methods. Power [S7] explores when scaling approaches are appropriate in large organizations and discusses three contexts for agility at scale. Rolland et al. [S73] take a more theoretical stance to clarify the meaning of large-scale agile development by examining its underlying assumptions in existing studies. Dingsøyr et al. [S105] provide a taxonomy for characterizing large-scale agile projects based on the number of agile teams. Noteworthy results of the *Taxonomy* research stream are:

- Three contexts for agility at scale: (i) being agile in a team inside a large organization, (ii) using agile approaches in a large development effort inside a large organization, (iii) and the large organization being itself agile (see [S7]).
- A comparison of 10 prevailing assumptions in existing studies with a set of alternative assumptions better suited to the characteristics of large-scale agile projects, e.g., collective code ownership vs. transferring, translating, and transforming knowledge across boundaries (see [S73]).
- A taxonomy of scale, consisting of three categories: (i) small-scale (1 team), (ii) large-scale (2–9 teams), and very large-scale ($\geqslant$10 teams) (see [S105]).

We identified one research question in the *Taxonomy* stream. Two articles (cf. [S73], [S105]) suggest further research verifying the current taxonomy of scaling, which is currently merely based on the number of teams involved in large-scale agile projects [S105], and identifying other adequate scaling dimensions for classifying the scale of agile development projects.

#### *4.4.11. Team autonomy*

In large-scale agile development, the effective functioning of team autonomy is challenged as a certain amount of autonomy has to be sacrificed to reach consensus on common standards and align work with other teams [S109], [S116]. The *Team autonomy* research stream is primarily concerned with how complex organizations affect team autonomy and how they can strike a balance between self-organizing teams focused on their own goals and those of the broader organization. We identified two sub-topics in *Team autonomy* stream related to identifying challenges and success factors in establishing team autonomy in large-scale agile endeavors (cf. [S23], [S109], [S116]) and in the context of the interplay between governing and autonomizing agile teams (cf. [S16], [S82], [S120]). For instance, Moe et al. [S23] analyze large-scale projects and presents barriers that reduce the extent of team autonomy. Concerning the second sub-topic, Šāblis and Šmite [S16] investigate the interplay between governance and team autonomy by identifying governance roles that support teams. Interesting findings related to the *Team autonomy* research stream include:

- Observations indicating that there are two barriers to team autonomy in large-scale agile development: overarching goals that are often set by management without team involvement and organizational dependencies that result in teams having to manage additional tasks (see [S23]).

- Observations suggesting that reliance on resources outside of the large-scale agile program, e.g., shared IT resources, creates external dependencies and reduces the autonomy of agile teams (see [S109]).
- A model for assessing the autonomy of agile teams in large-scale agile development (see [S120]).

We identified six research questions in the *Team autonomy* research stream. The most frequently cited research questions relate to how to improve team autonomy in large-scale agile endeavors (cf. [W4], [S16], [S29]) and how to balance coordination between agile teams and their autonomy in large-scale agile projects (cf. [W8], [S16], [S43]), as the need for coordination and control can restrict a team's autonomy [S16].

## 5. Discussion

In this section, we discuss our general observations on the start-of-the-art on large-scale agile development.

**Reflection on the research of the past 13 years.** Since the first publication in 2007, researchers worldwide have lavished attention to study the application of agile methods in large projects and organizations. While the number of published studies started to accelerate in 2013, the academic interest has been steadily growing. The maturation of the research field can be seen both in the increasing number of published studies, as well as in increasing number of articles in top journals. While the 52 included papers from the most prominent systematic literature review on large-scale agile development by Dikert et al. [S22] comprised mostly experience reports indicating a lack of sound academic research, the 136 included studies in this study covered only research papers excluding personal opinion and experience papers from practitioners and scientists signaling a tremendous shift of the scientific foundation of large-scale agile development. We believe that this academic traction is sparked by the omnipresent industrial relevance of the topic demanding scientific assistance by researchers (cf. [37], [S22]). While almost 60% of the papers represent case studies mainly being exploratory and less theoretical, they do not pay enough attention to establish theoretical underpinnings as similar to studies on agile software development [34]. Analogous to agile software development and agreeing with Dingsøyr et al. [34], we believe that the area of large-scale agile development can mature as a research discipline only if adequate efforts are made to provide a solid theoretical scaffold.

**Current structure of the research landscape.** Research on large-scale agile development has been conducted mainly empirically and qualitatively to describe and explain how companies adopt large-scale agile methods. Consequently, extant research has been dominated by evaluation research assessing the large-scale adoption of agile methods and deriving a set of lessons learned instead of creating solution artifacts in the form of models, frameworks/methods, and tools. Like Batra [12], we can further observe an apparent lack of quantitative studies using surveys as data collection instruments to provide quantitative investigations and assessments, e.g., quantitatively assessing the strengths and weaknesses of scaling frameworks or performance of agile teams in large agile multi-team settings.

**Outlook for future research undertakings.** We encountered some intriguing observations, which is why we believe that large-scale agile development will continue to be a relevant topic in practice as well as in academia. An analysis of the past State of Agile surveys conducted by Digital.ai (cf. [29–31]) reveals an ever-increasing adoption of agile methods in organizations, including the adoption of scaling agile frameworks. According to Digital.ai [31], this trend will likely continue and also intensify, especially concerning the more significant expansion and scaling of agile methods beyond software development, namely across the whole company. Parallel to this development, the number of scaling agile frameworks proposed by software practitioners is still flourishing and likely to further grow as the creators of these frameworks feel committed to develop their frameworks further [101]. The number of published studies in the last years (see Figure 4.2) and the increasing interest of various publication venues (see Appendix A) from different research domains (see Figure 17) indicate that the growing industrial interest in large-scale agile development is backed by a growing scientific interest across diverse research communities. Although researchers made considerable efforts to close research gaps in various research streams of large-scale agile development, a high number of open research questions (see Table 12) are still waiting to be addressed by researchers.

**Contemplation of the current phenomenon.** Although the early advice from the agile community was that scaling agile methods to larger projects and organizations is "*probably the last thing anyone would want to do*" [87], and the advice from several fields is to reduce the size of software projects as much as possible [9], why are companies still trying to adopt agile methods outside of their sweet spot in larger projects and organizations? One plausible explanation is that solutions often demand too much work for a single team, or that new solutions are so complex or so dependent on existing systems that it is considered inefficient or impractical to split the development efforts into small projects, making agile methods a way to reduce risk at scale while also facilitating innovation [37]. Despite the challenges of large-scale adoption of agile methods, we observe that the idea permeates almost every continent and industry sector. We revealed more than 150 companies distributed over the globe across various sectors make the use of agile methods in larger projects and organizations (see Section 4.1.1). Our findings and the survey results of Digital.ai [31] indicate that, regardless of their organizational size, companies are adopting agile methods at scale. However, our results show that most of the adopting companies have more than 5,000 employees, accounting for almost 70% of all identified case companies with stated size, indicating that this phenomenon is probably more relevant for large companies than for small companies.

**New emerging research themes.** Early research on large-scale agile development started with contributions related to the *Agile practices at scale*, *Global and distributed software engineering*, and *Scaling agile frameworks* research streams (see Section 4.4.1). Themes recently receiving more focus include *team autonomy* and *large-scale agile transformations*. For instance, as large-scale agile transformations can be characterized as episodic organizational change processes [S21], there is a need to conduct longitudinal case studies that accompany these transformations to investigate their long-lasting effects.

## 6. Threats to validity

Although we employed a rigorous study design and paid particular attention to the selection and analysis of published studies, our study has some limitations. The results of this systematic mapping study may be affected by various threats to validity, which are (i) study search incompleteness, (ii) study selection bias, (iii) study distribution imbalance, and (iv) data extraction inaccuracy, which we will discuss in the following.

### 6.1. Incompleteness of study search

There may be some relevant publications that we did not retrieve by our search, which may affect the completeness of our study. To mitigate this risk, we searched the most common electronic databases in which a large number of journals as well as conference and workshop proceedings in the fields of software engineering and information systems are indexed. We also performed a preliminary search before the main search to improve the correctness and completeness of the search results. These measures reduced the probability of missing relevant studies.

### 6.2. Bias on study selection

The selection of relevant studies largely depends on the personal knowledge of the researchers, which may lead to a bias in the results of the study selection. To mitigate this bias, we created a set of selection criteria (see Section 3.2.2). As the researchers of this study may have different understandings of the selection criteria, we conducted a preliminary search before the main search to ensure that the researchers had a consistent understanding of the selection criteria. Two reviewers also performed the study selection process in parallel and independently and then discussed and resolved any conflicts between their results to mitigate personal bias in study selection. A potential bias may arise from excluding the grey literature, e.g., white papers, technical reports, or editorials. This potential bias is inherent to our study design and did not significantly impact our study. We wanted to focus exclusively on state of the art presented in high-quality scientific papers that have undergone rigorous peer-reviewed publication processes.

### 6.3. Imbalance of study distribution

Around one-third of the selected publications come from the proceedings of the International Conference on Agile Software Development and International Workshop on Large-Scale Agile Development (see Appendix A). These studies may, to some extent, carry the bias of conference and workshop organizers and committee members. However, we did not address this type of bias as there is no effective way to determine whether such bias exists. Hence, we were not able to mitigate this kind of bias. Moreover, conferences and workshops, by definition, allow the publication of immature results that may distort the level of evidence of the selected studies.

### 6.4. Inaccuracy of data extraction

Data extraction bias may negatively affect the accuracy of data extraction results, affecting the classification results of the selected publications. Two researchers specified a list of extracted data items to mitigate this risk and reduce possible misunderstandings on the data items to be extracted. Two researchers also performed a pilot data extraction process before the formal data extraction. Further, two researchers conducted the main data extraction process in parallel and independently. Two researchers discussed and resolved conflicts arising from the data extraction results in several workshop sessions.

## 7. Conclusions and future work

Large-scale agile development is on the verge of becoming a mature research area, as many publications on large-scale agile development have appeared in scientific conferences and journals, leading to a growing body of knowledge. However, until now, no systematic mapping study has been published that systematically identifies, analyzes, and classifies the state of research. This mapping study aims to fill this gap and provide an overview of the research activities in large-scale agile development. Based on this objective, we selected 136 studies as a result of the systematic mapping process.

Our findings show that the adoption of agile methods has indeed inspired large companies around the world and in various sectors to apply these methods to larger projects and organizations. Our results suggest that this industrial interest has sparked significant academic interest in the topic of large-scale agile development, as a total of 136 articles were published in 46 publication venues by more than 200 authors worldwide between 2007 and 2019. In addition, our results show that research on large-scale agile development is mainly empirical and observational rather than solution-oriented, as most studies contribute in the form of lessons learned through case studies. Our results reveal that 10 research streams have emerged over time that focus on different aspects of large-scale agile development, e.g., agile architecture, scaling agile frameworks or team autonomy. Our findings show that the identified research streams raise many open research questions.

We hope that this mapping study will serve as a starting point for new research efforts that address previously neglected or emerging research topics and assist practitioners in the field of large-scale agile development. Based on our findings, we suggest that future research endeavors should pay greater attention to building a solid theoretical scaffold for the observed phenomena in large-scale agile development and should create rigorously developed frameworks, methods, and tools to meet practitioners' needs. Moreover, we recommend researchers to provide more conceptual clarity on the meaning of large-scale agile development, which has not received much attention but plays a crucial role in advancing the research field. We encourage researchers to use the compiled research questions to address the still open research gaps.

## Appendix A. Distribution of selected studies by publication channels

| # | Publication source | Type | Domain | No. | % |
|---|---|---|---|---|---|
| 1 | International Conference on Agile Software Development | Conference | Software engineering | 21 | 15.44 |
| 2 | International Workshop on Large-Scale Agile Development | Workshop | Software engineering | 20 | 14.71 |
| 3 | International Conference on Global Software Engineering | Conference | Software engineering | 9 | 6.62 |
| 4 | IEEE Software | Journal | Software engineering | 6 | 4.41 |
| 5 | Information and Software Technology | Journal | Software engineering | 6 | 4.41 |
| 6 | International Workshop on Autonomous Teams | Workshop | Software engineering | 6 | 4.41 |
| 7 | Empirical Software Engineering | Journal | Software engineering | 5 | 3.68 |
| 8 | Hawaii International Conference on System Sciences | Conference | Information systems | 5 | 3.68 |
| 9 | Americas Conference on Information Systems | Conference | Information systems | 4 | 2.94 |
| 10 | Journal of Systems and Software | Journal | Software engineering | 4 | 2.94 |
| 11 | Euromicro Conference on Software Engineering and Advanced Applications | Conference | Software engineering | 3 | 2.21 |
| 12 | International Conference on Enterprise Distributed Object Computing | Conference | Enterprise computing | 3 | 2.21 |
| 13 | International Symposium on Empirical Software Engineering and Measurement | Conference | Software engineering | 3 | 2.21 |
| 14 | Journal of Software: Evolution and Process | Journal | Software engineering | 3 | 2.21 |
| 15 | Project Management Journal | Journal | Project management | 3 | 2.21 |
| 16 | European Conference on Information Systems | Conference | Information systems | 2 | 1.47 |
| 17 | International Conference on Information Systems | Conference | Information systems | 2 | 1.47 |
| 18 | International Conference on Product-Focused Software Process Improvement | Conference | Software engineering | 2 | 1.47 |
| 19 | Software Process: Improvement and Practice | Journal | Software engineering | 2 | 1.47 |
| 20 | Bled eConference | Conference | Multidisciplinary | 1 | 0.74 |
| 21 | CIRP Design Conference | Conference | Multidisciplinary | 1 | 0.74 |
| 22 | European Conference on Pattern Languages of Programs | Conference | Multidisciplinary | 1 | 0.74 |
| 23 | Human Computer Interaction | Journal | Human computer interaction | 1 | 0.74 |
| 24 | IEEE Access | Journal | Multidisciplinary | 1 | 0.74 |
| 25 | IEEE Transactions on Software engineering | Journal | Software engineering | 1 | 0.74 |
| 26 | Information Systems Journal | Journal | Information systems | 1 | 0.74 |
| 27 | International Conference on Advanced Information Systems Engineering | Conference | Information systems | 1 | 0.74 |
| 28 | International Conference on Computing Communication and Automation | Conference | Multidisciplinary | 1 | 0.74 |
| 29 | International Conference on Information Systems Development | Conference | Information systems | 1 | 0.74 |
| 30 | International Conference on Pattern Languages of Programs | Conference | Multidisciplinary | 1 | 0.74 |
| 31 | International Conference on Perspectives in Business Informatics Research | Conference | Information systems | 1 | 0.74 |
| 32 | International Conference on Strategic Innovative Marketing | Conference | Marketing | 1 | 0.74 |
| 33 | International Journal of Information Systems and Project Management | Journal | Information systems | 1 | 0.74 |
| 34 | International Journal of Advanced Computer Science and Applications | Journal | Multidisciplinary | 1 | 0.74 |
| 35 | International Journal of Project Management | Journal | Project management | 1 | 0.74 |
| 36 | International Journal on Advanced Science, Engineering and Information Technology | Journal | Multidisciplinary | 1 | 0.74 |
| 37 | International Requirements Engineering Conference | Conference | Software engineering | 1 | 0.74 |
| 38 | International Research Workshop on IT Project Management | Workshop | IT project management | 1 | 0.74 |
| 39 | International Systems and Software Product Line Conference | Conference | Software engineering | 1 | 0.74 |
| 40 | International Working Conference on Requirements Engineering | Conference | Software engineering | 1 | 0.74 |
| 41 | International Workshop on Evidential Assessment of Software Technologies | Workshop | Software engineering | 1 | 0.74 |
| 42 | International Workshop on Requirements Engineering in Agile Development | Workshop | Software engineering | 1 | 0.74 |
| 43 | Malaysian Software Engineering Conference | Conference | Software engineering | 1 | 0.74 |
| 44 | Procedia Computer Science | Journal | Multidisciplinary | 1 | 0.74 |
| 45 | Software Quality Professional | Journal | Software engineering | 1 | 0.74 |
| 46 | Workshop on Agile Requirements Engineering | Workshop | Software engineering | 1 | 0.74 |
| | Total | | | 136 | 100 |

## Appendix B. Systematic map overview

| Study | Publication type | Research type | Research approach | Contribution type | Research data type |
|---|---|---|---|---|---|
| [S1] | Journal | Evaluation research | Case study | Lessons learned | Primary study |
| [S2] | Journal | Evaluation research | Survey | Lessons learned | Primary study |
| [S3] | Journal | Evaluation research | Case study | Lessons learned | Primary study |
| [S4] | Conference | Evaluation research | Case study | Lessons learned | Primary study |
| [S5] | Workshop | Evaluation research | Case study | Lessons learned | Primary study |
| [S6] | Journal | Solution proposal | Design & creation | Framework/Method | Primary study |
| [S7] | Workshop | Philosophical papers | Case study | Model | Primary study |
| [S8] | Conference | Evaluation research | Case study | Lessons learned | Primary study |
| [S9] | Journal | Evaluation research | (Systematic) literature review | Lessons learned | Secondary study |
| [S10] | Conference | Evaluation research | (Systematic) literature review | Lessons learned | Secondary study |
| [S11] | Conference | Philosophical papers | Case study | Model | Primary study |
| [S12] | Workshop | Evaluation research | Case study | Lessons learned | Primary study |
| [S13] | Conference | Evaluation research | Case study | Lessons learned | Primary study |
| [S14] | Conference | Evaluation research | Grounded theory | Lessons learned | Primary study |
| [S15] | Conference | Solution proposal | Design & creation | Guideline | Primary study |
| [S16] | Workshop | Evaluation research | Mixed methods | Lessons learned | Primary study |
| [S17] | Workshop | Evaluation research | (Systematic) literature review | Lessons learned | Primary study |
| [S18] | Journal | Philosophical papers | Grounded theory | Theory | Primary study |
| [S19] | Journal | Solution proposal | Design & creation | Framework/Method | Primary study |
| [S20] | Workshop | Evaluation research | (Systematic) literature review | Lessons learned | Secondary study |
| [S21] | Conference | Philosophical papers | Case study | Model | Primary study |
| [S22] | Journal | Evaluation research | (Systematic) literature review | Lessons learned | Secondary study |
| [S23] | Conference | Evaluation research | Case study | Lessons learned | Primary study |
| [S24] | Journal | Solution proposal | Not stated | Framework/Method | Primary study |
| [S25] | Conference | Evaluation research | Case study | Lessons learned | Primary study |
| [S26] | Conference | Evaluation research | Case study | Lessons learned | Primary study |
| [S27] | Conference | Evaluation research | Case study | Lessons learned | Primary study |
| [S28] | Journal | Evaluation research | Case study | Lessons learned | Primary study |
| [S29] | Journal | Evaluation research | Grounded theory | Guideline | Primary study |
| [S30] | Conference | Philosophical papers | (Systematic) literature review | Lessons learned | Primary study |
| [S31] | Journal | Evaluation research | Case study | Lessons learned | Primary study |
| [S32] | Conference | Solution proposal | Case study | Model | Primary study |
| [S33] | Conference | Evaluation research | Case study | Lessons learned | Primary study |
| [S34] | Conference | Evaluation research | Case study | Lessons learned | Primary study |
| [S35] | Conference | Solution proposal | Design & creation | Framework/Method | Primary study |
| [S36] | Conference | Evaluation research | Case study | Lessons learned | Primary study |
| [S37] | Journal | Evaluation research | Case study | Lessons learned | Primary study |
| [S38] | Journal | Evaluation research | Case study | Lessons learned | Primary study |
| [S39] | Workshop | Evaluation research | Case study | Lessons learned | Primary study |
| [S40] | Conference | Evaluation research | Case study | Lessons learned | Primary study |
| [S41] | Conference | Solution proposal | Mixed methods | Framework/Method | Primary study |
| [S42] | Conference | Evaluation research | Case study | Lessons learned | Primary study |
| [S43] | Journal | Evaluation research | Case study | Lessons learned | Primary study |
| [S44] | Journal | Philosophical papers | Grounded theory | Model | Primary study |
| [S45] | Conference | Evaluation research | Case study | Lessons learned | Primary study |
| [S46] | Conference | Solution proposal | Design & creation | Framework/Method | Primary study |
| [S47] | Conference | Evaluation research | (Systematic) literature review | Lessons learned | Secondary study |
| [S48] | Conference | Evaluation research | Survey | Lessons learned | Primary study |
| [S49] | Journal | Evaluation research | Not stated | Guideline | Primary study |
| [S50] | Conference | Evaluation research | Case study | Lessons learned | Primary study |
| [S51] | Workshop | Evaluation research | Case study | Lessons learned | Primary study |
| [S52] | Conference | Evaluation research | Case study | Lessons learned | Primary study |
| [S53] | Conference | Evaluation research | Case study | Lessons learned | Primary study |
| [S54] | Workshop | Evaluation research | Case study | Lessons learned | Primary study |
| [S55] | Conference | Evaluation research | Case study | Lessons learned | Primary study |
| [S56] | Conference | Evaluation research | Case study | Lessons learned | Primary study |
| [S57] | Conference | Evaluation research | (Systematic) literature review | Lessons learned | Secondary study |
| [S58] | Conference | Evaluation research | Survey | Lessons learned | Primary study |
| [S59] | Workshop | Evaluation research | Not stated | Lessons learned | Primary study |
| [S60] | Journal | Evaluation research | Case study | Lessons learned | Primary study |
| [S61] | Workshop | Solution proposal | Grounded Theory | Model | Primary study |
| [S62] | Conference | Evaluation research | Case study | Lessons learned | Primary study |
| [S63] | Journal | Evaluation research | Case study | Lessons learned | Primary study |
| [S64] | Workshop | Evaluation research | Case study | Lessons learned | Primary study |
| [S65] | Journal | Evaluation research | (Systematic) literature review | Lessons learned | Secondary study |
| [S66] | Conference | Evaluation research | Case study | Lessons learned | Primary study |
| [S67] | Conference | Evaluation research | Mixed methods | Lessons learned | Primary study |
| [S68] | Journal | Evaluation research | Case study | Lessons learned | Primary study |

| Study | Publication type | Research type | Research approach | Contribution type | Research data type |
|---|---|---|---|---|---|
| [S69] | Conference | Evaluation research | Grounded theory | Lessons learned | Primary study |
| [S70] | Workshop | Solution proposal | Case study | Theory | Primary study |
| [S71] | Conference | Evaluation research | Case study | Lessons learned | Primary study |
| [S72] | Journal | Evaluation research | Mixed methods | Lessons learned | Primary study |
| [S73] | Conference | Philosophical papers | Mixed methods | Model | Primary study |
| [S74] | Workshop | Evaluation research | Case study | Lessons learned | Primary study |
| [S75] | Journal | Evaluation research | Survey | Lessons learned | Primary study |
| [S76] | Workshop | Evaluation research | Case study | Lessons learned | Primary study |
| [S77] | Workshop | Evaluation research | Not stated | Guideline | Primary study |
| [S78] | Journal | Evaluation research | Mixed methods | Lessons learned | Primary study |
| [S79] | Conference | Evaluation research | Case study | Lessons learned | Primary study |
| [S80] | Conference | Evaluation research | Grounded theory | Lessons learned | Primary study |
| [S81] | Journal | Evaluation research | Case study | Lessons learned | Primary study |
| [S82] | Journal | Evaluation research | Mixed methods | Lessons learned | Primary study |
| [S83] | Journal | Evaluation research | Case study | Lessons learned | Primary study |
| [S84] | Journal | Evaluation research | Mixed methods | Guideline | Primary study |
| [S85] | Conference | Evaluation research | Case study | Lessons learned | Primary study |
| [S86] | Workshop | Evaluation research | Case study | Lessons learned | Primary study |
| [S87] | Conference | Solution proposal | Case study | Framework/Method | Primary study |
| [S88] | Conference | Evaluation research | Case study | Lessons learned | Primary study |
| [S89] | Workshop | Evaluation research | (Systematic) literature review | Lessons learned | Secondary study |
| [S90] | Journal | Evaluation research | Grounded theory | Lessons learned | Primary study |
| [S91] | Conference | Evaluation research | Case study | Guideline | Primary study |
| [S92] | Journal | Evaluation research | Case study | Lessons learned | Primary study |
| [S93] | Conference | Philosophical papers | Theoretical | Lessons learned | Primary study |
| [S94] | Conference | Evaluation research | Case study | Lessons learned | Primary study |
| [S95] | Workshop | Evaluation research | Action research | Lessons learned | Primary study |
| [S96] | Journal | Evaluation research | Case study | Lessons learned | Primary study |
| [S97] | Journal | Evaluation research | Case study | Lessons learned | Primary study |
| [S98] | Workshop | Philosophical papers | Case study | Model | Primary study |
| [S99] | Conference | Evaluation research | Case study | Lessons learned | Primary study |
| [S100] | Journal | Solution proposal | Grounded Theory | Theory | Primary study |
| [S101] | Conference | Evaluation research | Case study | Lessons learned | Primary study |
| [S102] | Conference | Evaluation research | Case study | Lessons learned | Primary study |
| [S103] | Journal | Evaluation research | Case study | Lessons learned | Primary study |
| [S104] | Conference | Evaluation research | Case study | Lessons learned | Primary study |
| [S105] | Conference | Philosophical papers | Not stated | Model | Primary study |
| [S106] | Conference | Evaluation research | Case study | Lessons learned | Primary study |
| [S107] | Conference | Evaluation research | Case study | Lessons learned | Primary study |
| [S108] | Journal | Evaluation research | (Systematic) literature review | Lessons learned | Secondary study |
| [S109] | Workshop | Evaluation research | Case study | Lessons learned | Primary study |
| [S110] | Workshop | Solution proposal | Not stated | Guideline | Primary study |
| [S111] | Journal | Evaluation research | Mixed methods | Lessons learned | Primary study |
| [S112] | Journal | Evaluation research | Case study | Lessons learned | Primary study |
| [S113] | Workshop | Philosophical papers | Theoretical | Theory | Primary study |
| [S114] | Workshop | Evaluation research | Case study | Lessons learned | Primary study |
| [S115] | Workshop | Evaluation research | Case study | Lessons learned | Primary study |
| [S116] | Workshop | Evaluation research | Case study | Lessons learned | Primary study |
| [S117] | Workshop | Solution proposal | Design & creation | Guideline | Primary study |
| [S118] | Conference | Evaluation research | Case study | Lessons learned | Primary study |
| [S119] | Conference | Evaluation research | Survey | Lessons learned | Primary study |
| [S120] | Workshop | Evaluation research | Case study | Lessons learned | Primary study |
| [S121] | Conference | Evaluation research | Case study | Lessons learned | Primary study |
| [S122] | Conference | Evaluation research | Case study | Lessons learned | Primary study |
| [S123] | Workshop | Evaluation research | Survey | Lessons learned | Primary study |
| [S124] | Conference | Evaluation research | Case study | Lessons learned | Primary study |
| [S125] | Conference | Evaluation research | Mixed methods | Lessons learned | Primary study |
| [S126] | Conference | Evaluation research | Case study | lessons learned | Primary study |
| [S127] | Journal | Evaluation research | Theoretical | Guideline | Primary study |
| [S128] | Conference | Evaluation research | Case study | Lessons learned | Primary study |
| [S129] | Workshop | Evaluation research | Survey | Lessons learned | Primary study |
| [S130] | Conference | Evaluation research | Case study | Lessons learned | Primary study |
| [S131] | Conference | Evaluation research | Case study | Lessons learned | Primary study |
| [S132] | Conference | Evaluation research | (Systematic) literature review | Lessons learned | Secondary study |
| [S133] | Conference | Evaluation research | Not stated | Framework/Method | Primary study |
| [S134] | Conference | Evaluation research | (Systematic) literature review | Lessons learned | Secondary study |
| [S135] | Journal | Evaluation research | (Systematic) literature review | Lessons learned | Secondary study |
| [S136] | Conference | Evaluation research | (Systematic) literature review | Lessons learned | Secondary study |

## Appendix C. Overview of research streams on large-scale agile development

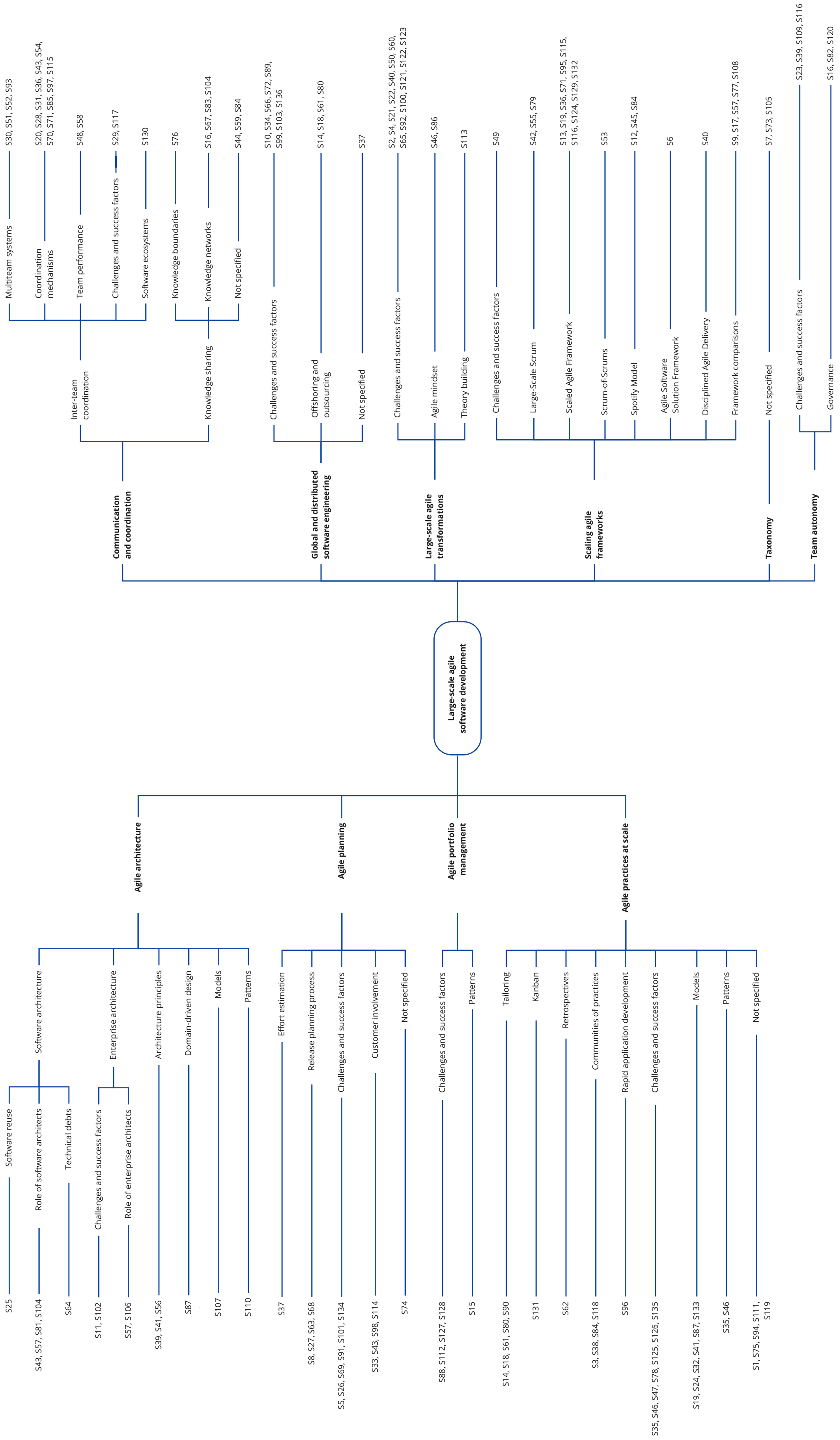

## References

[1] Pekka Abrahamsson, Juhani Warsta, Mikko T Siponen, and Jussi Ronkainen. New directions on agile methods: A comparative analysis. In *Proceedings of the 25th International Conference on Software Engineering*, pages 244–254. IEEE, May 2003.

[2] Pekka Abrahamsson, Muhammad Ali Babar, and Philippe Kruchten. Agility and architecture: Can they coexist? *IEEE Software*, 27(2), 2010.

[3] Pekka Abrahamsson, Outi Salo, Jussi Ronkainen, and Juhani Warsta. Agile software development methods: Review and analysis. *arXiv preprint arXiv:1709.08439*, 2017.

[4] Muhammad Faisal Abrar, Muhammad Sohail Khan, Sikandar Ali, Umar Ali, Muhammad Faran Majeed, Amjad Ali, Bahrul Amin, and Nasir Rasheed. Motivators for large-scale agile adoption from management perspective: A systematic literature review. *IEEE Access*, 7:22660–22674, 2019.

[5] Mashal Alqudah and Rozilawati Razali. A review of scaling agile methods in large software development. *International Journal on Advanced Science, Engineering and Information Technology*, 6(6):828–837, 2016.

[6] Wasim Alsaqaf, Maya Daneva, and Roel Wieringa. Quality requirements in large-scale distributed agile projects – a systematic literature review. In *Requirements Engineering: Foundation for Software Quality*, pages 219–234, Cham, February 2017. Springer.

[7] Scott W. Ambler. Agile software development at scale. In *IFIP Central and East European Conference on Software Engineering Techniques*, pages 1–12. Springer, 2007.

[8] Scott W. Ambler and Mark Lines. *Disciplined agile delivery: A practitioner's guide to agile software delivery in the enterprise*. IBM Press, 2012.

[9] Stephen J Andriole. The death of big software. *Communications of the ACM*, 60(12):29–32, 2017.

[10] Muhammad Ali Babar. An exploratory study of architectural practices and challenges in using agile software development approaches. In *Software Architecture, 2009 & European Conference on Software Architecture. WICSA/ECSA 2009. Joint Working IEEE/IFIP Conference on*, pages 81–90. IEEE, 2009.

[11] Victor Basili, Gianluigi Caldiera, and Dieter Rombach. The goal question metric approach. *Encyclopedia of Software Engineering*, pages 528–532, 1994.

[12] Dinesh Batra. Research challenges and opportunities in conducting quantitative studies on large-scale agile methodology. *Journal of Database Management (JDM)*, 31(2):64–73, 2020.

[13] Kent Beck. *Extreme programming explained: embrace change*. addison-wesley professional, 2000.

[14] Kent Beck. *Extreme programming explained: embrace change*. Addison-Wesley Professional, 2000.

[15] Mike Beedle, Martine Devos, Yonat Sharon, Ken Schwaber, and Jeff Sutherland. Scrum: An extension pattern language for hyperproductive software development. *Pattern Languages of Program Design*, 4:637–651, 1999.

[16] Mike Beedle, James O. Coplien, Jeff Sutherland, Jens C. Østergaard, Ademar Aguiar, and Ken Schwaber. Essential scrum patterns. In *14th European Conference on Pattern Languages of Programs*, pages 1–17, Irsee, 2010. The Hillside Group.

[17] Vebjørn Berg, Jørgen Birkeland, Anh Nguyen-Duc, Ilias O. Pappas, and Letizia Jaccheri. Software startup engineering: A systematic mapping study. *Journal of Systems and Software*, 144:255–274, 2018.

[18] Hilary Berger and Paul Beynon-Davies. The utility of rapid application development in large-scale, complex projects. *Information Systems Journal*, 19(6):549–570, 2009.

[19] Elizabeth Bjarnason, Krzysztof Wnuk, and Björn Regnell. A case study on benefits and side-effects of agile practices in large-scale requirements engineering. In *Proceedings of the 1st Workshop on Agile Requirements Engineering (AREW)*, pages 1–5. Association for Computing Machinery (ACM), July 2011.

[20] Barry Boehm. Get ready for agile methods, with care. *Computer*, 35(1): 64–69, 2002.

[21] O. Pearl Brereton, Barbara A. Kitchenham, David Budgen, Mark Turner, and Mohamed Khalil. Lessons from applying the systematic literature review process within the software engineering domain. *Journal of Systems and Software*, 80(4):571–583, 2007.

[22] David Budgen, Mark Turner, O. Pearl Brereton, and Barbara A. Kitchenham. Using mapping studies in software engineering. In *Psychology of Programming Interest Group*, volume 8, pages 195–204, 2008.

[23] The LeSS Company B.V. Large Scale Scrum. `https://less.works/case-studies`, 2022. [Online; accessed 18-APR-2022].

[24] Noel Carroll and Kieran Conboy. Applying normalization process theory to explain large-scale agile transformations. In *Proceedings of the 14th International Research Workshop on IT Project Management*, January 2019.

[25] Lianipng Chen, Muhammad Ali Babar, and He Zhang. Towards an evidence-based understanding of electronic data sources. In *Proceedings of the 14th International Conference on Evaluation and Assessment in Software Engineering (EASE)*, pages 1–4. BCS Learning & Development Ltd, April 2010.

[26] Tsun Chow and Dac-Buu Cao. A survey study of critical success factors in agile software projects. *Journal of systems and software*, 81(6):961–971, 2008.

[27] James O. Coplien and Neil B. Harrison. *Organizational Patterns of Agile Software Development*. Addison-Wesley, Boston, 2004.

[28] Daniela S. Cruzes and Tore Dybå. Recommended steps for thematic synthesis in software engineering. In *Proceedings of the 2011 International Symposium on Empirical Software Engineering and Measurement*, pages 275–284. Institute of Electrical and Electronics Engineers (IEEE), September 2011.

[29] Digital.ai. 12th State of Agile Survey. `https://stateofagile.com/#ufh-i-613553652-12th-annual-state-of-agile-report/7027494`, 2018. [Online; accessed 21-JUL-2021].

[30] Digital.ai. 13th State of Agile Survey. `https://stateofagile.com/#ufh-i-613553418-13th-annual-state-of-agile-report/7027494`, 2019. [Online; accessed 21-JUL-2021].

[31] Digital.ai. 14th State of Agile Survey. `https://stateofagile.com/#ufh-i-615706098-14th-annual-state-of-agile-report/7027494`, 2020. [Online; accessed 21-JUL-2021].

[32] Kim Dikert, Maria Paasivaara, and Casper Lassenius. Challenges and success factors for large-scale agile transformations: A systematic literature review. *Journal of Systems and Software*, 119:87–108, 2016.

[33] Torgeir Dingsøyr and Nils Brede Moe. Research challenges in large-scale agile software development. *ACM SIGSOFT Software Engineering Notes*, 38(5):38–39, 2013.

[34] Torgeir Dingsøyr, Sridhar Nerur, Venugopal Balijepally, and Nils Brede Moe. A decade of agile methodologies: Towards explaining agile software development. *Journal of Systems and Software*, 85(6):1213 – 1221, 2012. ISSN 0164-1212. Special Issue: Agile Development.

[35] Torgeir Dingsøyr, Tor Erlend Fægri, and Juha Itkonen. What is large in large-scale? a taxonomy of scale for agile software development. In *Proceedings of the 15th International Conference on Product-Focused Software Process Improvement (PROFES)*, pages 273–276. Springer, December 2014.

[36] Torgeir Dingsøyr, Nils Brede Moe, Tor Erlend Fægri, and Eva Amdahl Seim. Exploring software development at the very large-scale: a revelatory case study and research agenda for agile method adaptation. *Empirical Software Engineering*, 23(1):490–520, 2018.

[37] Torgeir Dingsøyr, Davide Falessi, and Ken Power. Agile development at scale: the next frontier. *IEEE Software*, 36(2):30–38, 2019.

[38] Tore Dybå and Torgeir Dingsøyr. Empirical studies of agile software development: A systematic review. *Information and Software Technology*, 50(9):833 – 859, 2008. ISSN 0950-5849.

[39] Tore Dybå and Torgeir Dingsøyr. What do we know about agile software development? *IEEE Software*, 26(5):6–9, 2009.

[40] Henry Edison, Xiaofeng Wang, and Kieran Conboy. Comparing methods for large-scale agile software development: A systematic literature review. *IEEE Transactions on Software Engineering*, 2021.

[41] Sallyann Freudenberg and Helen Sharp. The top 10 burning research questions from practitioners. *Ieee Software*, 27(5):8–9, 2010.

[42] Christoph Fuchs and Thomas Hess. Becoming agile in the digital transformation: The process of a large-scale agile transformation. In *Proceedings of the 39th International Conference on Information Systems (ICIS)*, December 2018.

[43] Tomas Gustavsson. Assigned roles for inter-team coordination in large-scale agile development: A literature review. In *Proceedings of the 5th International Workshop on Large-Scale Agile Development (XP)*, pages 1–5. Association for Computing Machinery (ACM), May 2017.

[44] Amani Mahdi Mohammed Hamed and Hisham Abushama. Popular agile approaches in software development: Review and analysis. In *Proceedings of the 2013 International Conference on Computing, Electrical and Electronic Engineering (ICCEEE)*, pages 160–166. Institute of Electrical and Electronics Engineers (IEEE), August 2013.

[45] Jo E. Hannay, Dag I. K. Sjoberg, and Tore Dybå. A systematic review of theory use in software engineering experiments. *IEEE Transactions on Software Engineering*, 33(2):87–107, 2007.

[46] Geir K. Hanssen, Darja Smite, and Nils Brede Moe. Signs of agile trends in global software engineering research: A tertiary study. In *2011 IEEE Sixth International Conference on Global Software Engineering Workshop*, pages 17–23, 2011.

[47] Ville Heikkilä, Daniela Damian, Casper Lassenius, and Maria Paasivaara. A mapping study on requirements engineering in agile software development. In *2015 41st Euromicro conference on software engineering and advanced applications*, pages 199–207. IEEE, 2015.

[48] Steffen Herbold, Aynur Amirfallah, Fabian Trautsch, and Jens Grabowski. A systematic mapping study of developer social network research. *Journal of Systems and Software*, 171:110802, 2021. ISSN 0164-1212.

[49] Jim Highsmith and Alistair Cockburn. Agile software development: the business of innovation. *Computer*, 34(9):120–127, 2001.

[50] Peter Hodgkins and Luke Hohmann. Agile program management: Lessons learned from the verisign managed security services team. In *Agile 2007 (AGILE 2007)*, pages 194–199. Institute of Electrical and Electronics Engineers (IEEE), August 2007.

[51] Scaled Agile Inc. Scaled Agile Framework. `http://www.scaledagileframework.com/case-studies/`, 2022. [Online; accessed 18-APR-2022].

[52] Magne Jorgensen and Martin Shepperd. A systematic review of software development cost estimation studies. *IEEE Transactions on software engineering*, 33(1):33–53, 2006.

[53] Martin Kalenda, Petr Hyna, and Bruno Rossi. Scaling agile in large organizations: Practices, challenges, and success factors. *Journal of Software: Evolution and Process*, 30(10):e1954, 2018.

[54] Petri Kettunen. Extending software project agility with new product development enterprise agility. *Software Process: Improvement and Practice*, 12(6):541–548, November 2007.

[55] Barbara A. Kitchenham and O. Pearl Brereton. A systematic review of systematic review process research in software engineering. *Information and Software Technology*, 55(12):2049–2075, 2013.

[56] Barbara A. Kitchenham and Stuart Charters. Guidelines for performing systematic literature reviews in software engineering. Technical report, Keele University and University of Durham, EBSE, 2007.

[57] Barbara A. Kitchenham, Emilia Mendes, and Guilherme H. Travassos. Cross versus within-company cost estimation studies: A systematic review. *IEEE Transactions on Software Engineering*, 33(5):316–329, 2007.

[58] Barbara A. Kitchenham, David Budgen, and O. Pearl Brereton. Using mapping studies as the basis for further research - a participant-observer case study. *Information and Software Technology*, 53(6):638–651, 2011.

[59] Jil Klünder, Regina Hebig, Paolo Tell, Marco Kuhrmann, Joyce Nakatumba-Nabende, Rogardt Heldal, Stephan Krusche, Masud Fazal-Baqaie, Michael Felderer, Marcela Fabiana Genero Bocco, et al. Catching up with method and process practice: An industry-informed baseline for researchers. In *2019 IEEE/ACM 41st International Conference on Software Engineering: Software Engineering in Practice (ICSE-SEIP)*, pages 255–264. IEEE, 2019.

[60] Harry Koehnemann and Mark Coats. Experiences applying agile practices to large systems. In *Agile Conference*, pages 295–300, 2009.

[61] Dina Koutsikouri, Sabine Madsen, and Nataliya Berbyuk Lindström. Agile transformation: How employees experience and cope with transformative change. In *International Conference on Agile Software Development*, pages 155–163. Springer, Cham, 2020.

[62] Daryl Kulak and Hong Li. *The journey to enterprise agility: Systems thinking and organizational legacy*. Springer, 2017.

[63] Craig Larman. *Practices for scaling lean & Agile development: large, multisite, and offshore product development with large-scale scrum*. Pearson Education India, 2010.

[64] Dean Leffingwell, Ryan Martens, and Mauricio Zamora. Principles of agile architecture. *Leffingwell, LLC. and Rally Software Development Corp*, 2008.

[65] Mikael Lindvall, Dirk Muthig, Aldo Dagnino, Christina Wallin, Michael Stupperich, David Kiefer, John May, and Tuomo Kahkonen. Agile software development in large organizations. *Computer*, 37(12):26–34, 2004.

[66] Jeffrey A Livermore. Factors that significantly impact the implementation of an agile software development methodology. *Journal of Software*, 3(4): 31–36, 2008.

[67] Aniket Mahanti. Challenges in enterprise adoption of agile methods-a survey. *Journal of Computing and Information technology*, 14(3):197–206, 2006.

[68] Chuck Maples. Enterprise agile transformation: The two-year wall. In *Proceedings of the 2009 Agile Conference*, pages 90–95. IEEE, August 2009.

[69] Subhas Chandra Misra, Vinod Kumar, and Uma Kumar. Identifying some critical changes required in adopting agile practices in traditional software development projects. *International Journal of Quality & Reliability Management*, 2010.

[70] Nils Brede Moe and Torgeir Dingsøyr. Emerging research themes and updated research agenda for large-scale agile development: a summary of the 5th international workshop at xp2017. In *Proceedings of the XP2017 Scientific Workshops*, pages 1–4. Association for Computing Machinery (ACM), May 2017.

[71] Sridhar P Nerur, Abdul A Rasheed, and Vivek Natarajan. The intellectual structure of the strategic management field: An author co-citation analysis. *Strategic Management Journal*, 29(3):319–336, 2008.

[72] Mahmood Niazi, Sajjad Mahmood, Mohammad Alshayeb, Mohammed Rehan Riaz, Kanaan Faisal, Narciso Cerpa, Siffat Ullah Khan, and Ita Richardson. Challenges of project management in global software development: A client-vendor analysis. *Information and Software Technology*, 80:1–19, 2016.

[73] Robert L Nord, Ipek Ozkaya, and Philippe Kruchten. Agile in distress: Architecture to the rescue. In *International Conference on Agile Software Development*, pages 43–57. Springer, 2014.

[74] BJ Briony Oates. Evidence-based information systems: A decade later. In *Proceedings of the 19th European Conference on Information Systems (ECIS)*. Association for Information Systems, June 2011.

[75] Helena Holmström Olsson, Eoin Ó. Conchúir, Pär J. Ågerfalk, and Brian Fitzgerald. Global software development challenges: A case study on temporal, geographical and socio-cultural distance. In *2006 IEEE International Conference on Global Software Engineering (ICGSE'06)*, pages 3–11. IEEE, 2006.

[76] Wanda J. Orlikowski. Improvising organizational transformation over time: A situated change perspective. *Information Systems Research*, 7(1): 63–92, 1996.

[77] Necmettin Ozkan and Ayca Tarhan. A review of scaling approaches to agile software development models. *Software Quality Professional*, 21(4): 11–20, 2019.

[78] Maria Paasivaara, Sandra Durasiewicz, and Casper Lassenius. Using scrum in a globally distributed project: a case study. *Software Process: Improvement and Practice*, 13(6):527–544, 2008.

[79] Nicolò Paternoster, Carmine Giardino, Michael Unterkalmsteiner, Tony Gorschek, and Pekka Abrahamsson. Software development in startup companies: A systematic mapping study. *Information and Software Technology*, 56(10):1200–1218, 2014.

[80] Kai Petersen and Claes Wohlin. A comparison of issues and advantages in agile and incremental development between state of the art and an industrial case. *Journal of Systems and Software*, 82(9):1479–1490, 2009.

[81] Kai Petersen and Claes Wohlin. The effect of moving from a plan-driven to an incremental software development approach with agile practices. *Empirical Software Engineering*, 15(6):654–693, 2010.

[82] Kai Petersen, Robert Feldt, Shahid Mujtaba, and Michael Mattsson. Systematic mapping studies in software engineering. In *Proceedings of the 12th International Conference on Evaluation and Assessment in Software Engineering (EASE)*, pages 68–77, Swindon, UK, June 2008. BCS Learning & Development Ltd.

[83] Kai Petersen, Sairam Vakkalanka, and Ludwik Kuzniarz. Guidelines for conducting systematic mapping studies in software engineering: An update. *Information and Software Technology*, 64:1–18, 2015.

[84] Abheeshta Putta, Maria Paasivaara, and Casper Lassenius. Benefits and challenges of adopting the scaled agile framework (safe): Preliminary results from a multivocal literature review. In *Product-Focused Software Process Improvement*, pages 334–351, Cham, November 2018. Springer.

[85] Mark Rajpal. Lessons learned from a failed attempt at distributed ag-

ile. In *Proceedings of the 17th International Conferences XP 2016: Agile Processes in Software Engineering and Extreme Programming*, pages 235–243, Cham, May 2016. Springer.

[86] Abbas Moshref Razavi and Rodina Ahmad. Agile development in large and distributed environments: A systematic literature review on organizational, managerial and cultural aspects. In *Proceedings of the 2014 8th. Malaysian Software Engineering Conference (MySEC)*, pages 216–221. Institute of Electrical and Electronics Engineers (IEEE), September 2014.

[87] Donald J Reifer, Frank Maurer, and Hakan Erdogmus. Scaling agile methods. *IEEE software*, 20(4):12–14, 2003.

[88] Pilar Rodríguez, Jouni Markkula, Markku Oivo, and Kimmo Turula. Survey on agile and lean usage in finnish software industry. In *Proceedings of the ACM-IEEE International Symposium on Empirical Software Engineering and Measurement (ESEM)*, pages 139–148, New York, NY, USA, September 2012. Association for Computing Machinery (ACM). ISBN 9781450310567.

[89] Pilar Rodríguez, Alireza Haghighatkhah, Lucy Ellen Lwakatare, Susanna Teppola, Tanja Suomalainen, Juho Eskeli, Teemu Karvonen, Pasi Kuvaja, June M. Verner, and Markku Oivo. Continuous deployment of software intensive products and services: A systematic mapping study. *Journal of Systems and Software*, 123:263–291, 2017.

[90] Knut H. Rolland, Brian Fitzgerald, Torgeir Dingsøyr, and Klaas-Jan Stol. Problematizing agile in the large: Alternative assumptions for large-scale agile development. In *Proceedings of the 37th International Conference on Information Systems (ICIS)*, December 2016.

[91] Dominik Rost, Balthasar Weitzel, Matthias Naab, Torsten Lenhart, and Hartmut Schmitt. Distilling best practices for agile development from architecture methodology. In *European Conference on Software Architecture*, pages 259–267. Springer, 2015.

[92] Hina Saeeda, Hannan Khalid, Mukhtar Ahmed, Abu Sameer, and Fahim Arif. Systematic literature review of agile scalability for large scale projects. *International Journal of Advanced Computer Science and Applications (IJACSA)*, 6(9):63–75, 2015.

[93] Christoph Tobias Schmidt, Srinivasa Ganesha Venkatesha, and Juergen Heymann. Empirical insights into the perceived benefits of agile software engineering practices: A case study from sap. In *Companion Proceedings of the 36th International Conference on Software Engineering*, pages 84–92. Association for Computing Machinery (ACM), May 2014.

[94] Ken Schwaber and Mike Beedle. *Agile software development with Scrum*, volume 1. Prentice Hall Upper Saddle River, 2002.

[95] Mohammad Shameem, Chiranjeev Kumar, Bibhas Chandra, and Arif Ali Khan. Systematic review of success factors for scaling agile methods in global software development environment: A client-vendor perspective. In *Proceedings of the 24th Asia-Pacific Software Engineering Conference Workshops (APSEC)*, pages 17–24. Institute of Electrical and Electronics Engineers (IEEE), December 2017.

[96] Mohammad Shameem, Bibhas Chandra, Rakesh Ranjan Kumar, and Chiranjeev Kumar. A systematic literature review to identify human related challenges in globally distributed agile software development: towards a hypothetical model for scaling agile methodologies. In *Proceedings of the 2018 4th International Conference on Computing Communication and Automation (ICCCA)*, pages 1–7. Institute of Electrical and Electronics Engineers (IEEE), December 2018.

[97] Mary Shaw. Writing good software engineering research papers. In *Proceedings of the 25th International Conference on Software Engineering, 2003.*, pages 726–736. Institute of Electrical and Electronics Engineers (IEEE), May 2003.

[98] Stavros Stavru. A critical examination of recent industrial surveys on agile method usage. *Journal of Systems and Software*, 94:87 – 97, 2014. ISSN 0164-1212.

[99] Ömer Uludağ, Martin Kleehaus, Xian Xu, and Florian Matthes. Investigating the role of architects in scaling agile frameworks. In *Proceedings of the 21st IEEE International Enterprise Distributed Object Computing Conference (EDOC)*, pages 123–132. Institute of Electrical and Electronics Engineers (IEEE), October 2017.

[100] Ömer Uludağ, Martin Kleehaus, Christoph Caprano, and Florian Matthes. Identifying and structuring challenges in large-scale agile development based on a structured literature review. In *Proceedings of the 22nd IEEE International Enterprise Distributed Object Computing Conference (EDOC)*, pages 191–197. Institute of Electrical and Electronics Engineers (IEEE), October 2018.

[101] Ömer Uludağ, Abheeshta Putta, Maria Paasivaara, and Florian Matthes. Evolution of the agile scaling frameworks. In *Proceedings of the 22nd International Conference on Agile Software Development*, Cham, June 2021. Springer.

[102] Michael Unterkalmsteiner, Tony Gorschek, A. K. M. Moinul Islam, Chow Kian Cheng, Rahadian Bayu Permadi, and Robert Feldt. Evaluation and measurement of software process improvement—a systematic literature review. *IEEE Transactions on Software Engineering*, 38(2):398–424, 2011.

[103] Roel Wieringa, Neil Maiden, Nancy Mead, and Colette Rolland. Requirements engineering paper classification and evaluation criteria: A proposal and a discussion. *Requirements engineering*, 11(1):102–107, 2006.

[104] Chen Yang, Peng Liang, and Paris Avgeriou. A systematic mapping study on the combination of software architecture and agile development. *Journal of Systems and Software*, 111:157–184, 2016.

[105] He Zhang, Muhammad Ali Babar, and Paolo Tell. Identifying relevant studies in software engineering. *Information and Software Technology*, 53 (6):625–637, 2011.

## Selected studies

[S1] Kai Petersen and Claes Wohlin. A comparison of issues and advantages in agile and incremental development between state of the art and an industrial case. *Journal of Systems and Software*, 82(9):1479 – 1490, 2009.

[S2] Maarit Laanti, Outi Salo, and Pekka Abrahamsson. Agile methods rapidly replacing traditional methods at nokia: A survey of opinions on agile transformation. *Information and Software Technology*, 53(3): 276 – 290, 2011.

[S3] Maria Paasivaara and Casper Lassenius. Communities of practice in a large distributed agile software development organization – case ericsson. *Information and Software Technology*, 56(12):1556 – 1577, 2014.

[S4] Daniel Gerster, Christian Dremel, and Prashant Kelker. "agile meets non-agile": Implications of adopting agile practices at enterprises. In *Proceedings of the 24th Americas Conference on Information Systems (AMCIS)*, pages 836 – 845.

[S5] Elizabeth Bjarnason, Krzysztof Wnuk, and Björn Regnell. A case study on benefits and side-effects of agile practices in large-scale requirements engineering. In *Proceedings of the 1st Workshop on Agile Requirements Engineering (AREW)*, page 3. Association for Computing Machinery (ACM), July 2011.

[S6] Asif Qumer and Brian Henderson-Sellers. A framework to support the evaluation, adoption and improvement of agile methods in practice. *Journal of Systems and Software*, 81(11):1899–1919, 2008.

[S7] Ken Power. A model for understanding when scaling agile is appropriate in large organizations. In *Proceedings of the 2nd International Workshop on Large-Scale Agile Development (XP)*, pages 83–92. Springer, 2014.

[S8] Ville Heikkilä, Kristian Rautiainen, and Slinger Jansen. A revelatory case study on scaling agile release planning. In *Proceedings of the 36th EUROMICRO Conference on Software Engineering and Advanced Applications (EUROMICRO-SEAA)*, pages 289–296. Institute of Electrical and Electronics Engineers (IEEE), September 2010.

[S9] Mashal Alqudah and Rozilawati Razali. A review of scaling agile methods in large software development. *International Journal on Advanced Science, Engineering and Information Technology*, 6(6):828–837, 2016.

[S10] Mohammad Shameem, Bibhas Chandra, Rakesh Ranjan Kumar, and Chiranjeev Kumar. A systematic literature review to identify human related challenges in globally distributed agile software development: towards a hypothetical model for scaling agile methodologies. In *Proceedings of the 4th International Conference on Computing Communication and Automation (ICCCA)*, pages 1–7. Institute of Electrical and Electronics Engineers (IEEE), December 2018.

[S11] Robin Duijs, Pascal Ravesteyn, and Marlies van Steenbergen. Adaptation of enterprise architecture efforts to an agile environment. In *Proceedings of the 31st Bled eConference*, pages 389–400, June 2018.

[S12] Abdallah Salameh and Julian M. Bass. Spotify tailoring for promoting effectiveness in cross-functional autonomous squads. In *Proceedings*

*of the 2nd International Workshop on Autonomous Teams (XP)*, pages 20–28. Springer, May 2019.

[S13] Maria Paasivaara. Adopting safe to scale agile in a globally distributed organization. In *Proceedings of the 12th IEEE International Conference on Global Software Engineering (ICGSE)*, pages 36–40. Institute of Electrical and Electronics Engineers (IEEE), May 2017.

[S14] Julian M. Bass. Agile method tailoring in distributed enterprises: Product owner teams. In *Proceedings of the 8th IEEE International Conference on Global Software Engineering (ICGSE)*, pages 154–163. Institute of Electrical and Electronics Engineers (IEEE), August 2013.

[S15] Bettina Horlach, Ingrid Schirmer, and Paul Drews. Agile portfolio management: Design goal and principles. In *Proceedings of the 27th European Conference on Information Systems (ECIS)*. AIS Electronic Library (AISeL), June 2019.

[S16] Aivars Šāblis and Darja Šmite. Agile teams in large-scale distributed context: Isolated or connected? In *Proceedings of the Scientific Workshop of XP2016*, pages 1–5. Association for Computing Machinery (ACM), May 2016.

[S17] Sven Theobald, Anna Schmitt, and Philipp Diebold. Comparing scaling agile frameworks based on underlying practices. In *Proceedings of the 7th International Workshop on Large-Scale Agile (XP)*, pages 88–96. Springer, May 2019.

[S18] Julian M. Bass. Artefacts and agile method tailoring in large-scale offshore software development programmes. *Information and Software Technology*, 75:1–16, 2016.

[S19] Oktay Turetken, Igor Stojanov, and Jos JM Trienekens. Assessing the adoption level of scaled agile development: a maturity model for scaled agile framework. *Journal of Software: Evolution and Process*, 29(6): e1796, 2017.

[S20] Tomas Gustavsson. Assigned roles for inter-team coordination in large-scale agile development: A literature review. In *Proceedings of the 5th International Workshop on Large-Scale Agile Development (XP)*, pages 1–5. Association for Computing Machinery (ACM), May 2017.

[S21] Christoph Fuchs and Thomas Hess. Becoming agile in the digital transformation: The process of a large-scale agile transformation. In *Proceedings of the 39th International Conference on Information Systems (ICIS)*, December 2018.

[S22] Kim Dikert, Maria Paasivaara, and Casper Lassenius. Challenges and success factors for large-scale agile transformations: A systematic literature review. *Journal of Systems and Software*, 119:87–108, 2016.

[S23] Nils Brede Moe, Bjørn Dahl, Viktoria Stray, Lina Sund Karlsen, and Stine Schjødt-Osmo. Team autonomy in large-scale agile. In *Proceedings of the 52nd Hawaii International Conference on System Sciences (HICSS)*. Hawaii International Conference on System Sciences, January 2019.

[S24] Petri Kettunen and Maarit Laanti. Combining agile software projects and large-scale organizational agility. *Software Process: Improvement and Practice*, 13(2):183–193, 2008.

[S25] Antonio Martini, Lars Pareto, and Jan Bosch. Communication factors for speed and reuse in large-scale agile software development. In *Proceedings of the 17th international software product line conference (SPLC)*, pages 42–51. Association for Computing Machinery (ACM), August 2013.

[S26] Felix Evbota, Eric Knauss, and Anna Sandberg. Scaling up the planning game: Collaboration challenges in large-scale agile product development. In Helen Sharp and Tracy Hall, editors, *Proceedings of the 17th International Conference on Agile Software Development (XP)*, pages 28–38. Springer, May 2016.

[S27] Ville Heikkilä, Maria Paasivaara, Casper Lassenius, and Christian Engblom. Continuous release planning in a large-scale scrum development organization at ericsson. In *Proceedings of the 14th International Conference of Agile Software Development (XP)*, pages 195–209. Springer, June 2013.

[S28] Torgeir Dingsøyr, Nils Brede Moe, and Eva Amdahl Seim. Coordinating knowledge work in multi-team programs: Findings from a large-scale agile development program. *Project Management Journal*, 49(6): 64—77, 2018.

[S29] Saskia Bick, Kai Spohrer, Rashina Hoda, Alexander Scheerer, and Armin Heinzl. Coordination challenges in large-scale software development: A case study of planning misalignment in hybrid settings. *IEEE Trans. Software Eng.*, 44(10):932–950, 2018.

[S30] Alexander Scheerer, Tobias Hildenbrand, and Thomas Kude. Coordination in large-scale agile software development: A multiteam systems perspective. In *Proceedings of the 47th Hawaii International Conference on System Sciences (HICSS)*, pages 4780–4788. Institute of Electrical and Electronics Engineers (IEEE), January 2014.

[S31] Torgeir Dingsøyr, Knut Rolland, Nils Brede Moe, and Eva Amdahl Seim. Coordination in multi-team programmes: An investigation of the group mode in large-scale agile software development. *Procedia Computer Science*, 121:123–128, 2017.

[S32] Sina Katharina Weiss and Philipp Brune. Crossing the boundaries – agile methods in large-scale, plan-driven organizations: A case study from the financial services industry. In Eric Dubois and Klaus Pohl, editors, *Proceedings of the 29th International Conference Advanced Information Systems Engineering (CAiSE)*, pages 380–393. Springer, June 2017.

[S33] Helena Holmström Olsson, Jan Bosch, and Hiva Alahyari. Customer-specific teams for agile evolution of large-scale embedded systems. In *Proceedings of the 39th Euromicro Conference on Software Engineering and Advanced Applications (EUROMICRO-SEAA)*, pages 82–89, September 2013.

[S34] Maria Paasivaara, Sandra Durasiewicz, and Casper Lassenius. Distributed agile development: Using scrum in a large project. In *Proceedings of the 3rd IEEE International Conference on Global Software Engineering, (ICGSE)*, pages 87–95. Institute of Electrical and Electronics Engineers (IEEE), August 2008.

[S35] Ömer Uludağ, Nina-Mareike Harders, and Florian Matthes. Documenting recurring concerns and patterns in large-scale agile development. In *Proceedings of the 24th European Conference on Pattern Languages of Programs (EuroPLoP)*, July 2019.

[S36] Tomas Gustavsson. Dynamics of inter-team coordination routines in large-scale agile software development. In *Proceedings of the 27th European Conference on Information Systems (ECIS)*, June 2019.

[S37] Muhammad Usman, Ricardo Britto, Lars-Ola Damm, and Jürgen Börstler. Effort estimation in large-scale software development: An industrial case study. *Information and Software Technology*, 99:21–40, 2018.

[S38] Maria Paasivaara and Casper Lassenius. Empower your agile organization: Community-based decision making in large-scale agile development at ericsson. *IEEE Software*, 36(2):64–69, 2019.

[S39] Jan Henrik Gundelsby. Enabling autonomous teams in large-scale agile through architectural principles. In *Proceedings of the 1st International Workshop on Autonomous Agile Teams (XP Companion)*, page 17. Association for Computing Machinery (ACM), May 2018.

[S40] Ramesh Lal and Tony Clear. Enhancing product and service capability through scaling agility in a global software vendor environment. In *Proceedings of the 13th Conference on Global Software Engineering (ICGSE)*, pages 59–68. Association for Computing Machinery (ACM), May 2018.

[S41] Ömer Uludağ, Sascha Nägele, and Matheus Hauder. Establishing architecture guidelines in large-scale agile development through institutional pressures: A single-case study. In *Proceedings of the 25th Americas Conference on Information Systems (AMCIS)*, pages 551 – 560, August 2019.

[S42] Maria Paasivaara, Ville Heikkilä, and Casper Lassenius. Experiences in scaling the product owner role in large-scale globally distributed scrum. In *Proceedings of the 7th International Conference on Global Software Engineering (ICGSE)*, pages 174–178. Institute of Electrical and Electronics Engineers (IEEE), August 2012.

[S43] Torgeir Dingsøyr, Nils Brede Moe, Tor Erlend Fægri, and Eva Amdahl Seim. Exploring software development at the very large-scale: a revelatory case study and research agenda for agile method adaptation. *Empirical Software Engineering*, 23(1):490–520, 2018.

[S44] Viviane Santos, Alfredo Goldman, and Cleidson RB De Souza. Fostering effective inter-team knowledge sharing in agile software development. *Empirical Software Engineering*, 20(4):1006–1051, 2015.

[S45] Daniel Gerster, Christian Dremel, Walter Brenner, and Prashant Kelker. How enterprises adopt agile structures: A multiple-case study. In *Proceedings of the 52nd Hawaii International Conference on System Sciences (HICSS)*, pages 1–10, January 2019.

[S46] Ömer Uludağ, Martin Kleehaus, Christoph Caprano, and Florian Matthes. Identifying and structuring challenges in large-scale agile

development based on a structured literature review. In *Proceedings of the 22nd IEEE International Enterprise Distributed Object Computing Conference (EDOC)*, pages 191–197. Institute of Electrical and Electronics Engineers (IEEE), October 2018.

[S47] Ömer Uludağ and Florian Matthes. Identifying and documenting recurring concerns and best practices of agile coaches and scrum masters in large-scale agile development. In *Proceedings of the 26th International Conference on Pattern Languages of Programs (PLoP)*, volume 26, pages 191–197. Hillside Group, 2019.

[S48] Tomas Gustavsson. Impacts on team performance in large-scale agile software development. In *Proceedings of the 2018 Joint of the 17th Business Informatics Research Short Papers, Workshops and Doctoral Consortium (BIC)*, pages 421–431. CEUR-WS, September 2018.

[S49] Kieran Conboy and Noel Carroll. Implementing large-scale agile frameworks: Challenges and recommendations. *IEEE Software*, 36 (2):44–50, 2019.

[S50] Maria Paasivaara, Casper Lassenius, Ville Heikkilä, Kim-Karol Dikert, and Christian Engblom. Integrating global sites into the lean and agile transformation at ericsson. In *Proceedings of the 8th IEEE International Conference on Global Software Engineering, (ICGSE)*, pages 134–143. Institute of Electrical and Electronics Engineers (IEEE), August 2013.

[S51] Saskia Bick, Alexander Scheerer, and Kai Spohrer. Inter-team coordination in large agile software development settings: Five ways of practicing agile at scale. In *Proceedings of the Scientific Workshops of XP2016*, page 4. Association for Computing Machinery (ACM), May 2016.

[S52] Finn Olav Bjørnson, Julia Wijnmaalen, Christoph Johann Stettina, and Torgeir Dingsøyr. Inter-team coordination in large-scale agile development: A case study of three enabling mechanisms. In *Proceedings of the 19th International Conference on Agile Software Development (XP)*, pages 216–231. Springer, May 2018.

[S53] Maria Paasivaara, Casper Lassenius, and Ville T Heikkilä. Inter-team coordination in large-scale globally distributed scrum: Do scrum-of-scrums really work? In *Proceedings of the 6th ACM-IEEE international symposium on Empirical software engineering and measurement (ESEM)*, pages 235–238. Association for Computing Machinery (ACM), September 2012.

[S54] Helga Nyrud and Viktoria Stray. Inter-team coordination mechanisms in large-scale agile. In *Proceedings of the Scientific Workshops of XP2017*, pages 1–6. Association for Computing Machinery (ACM), May 2017.

[S55] Ömer Uludağ, Martin Kleehaus, Niklas Dreymann, Christian Kabelin, and Florian Matthes. Investigating the adoption and application of large-scale scrum at a german automobile manufacturer. In *Proceedings of the 14th International Conference on Global Software Engineering (ICGSE)*, pages 22–29. Institute of Electrical and Electronics Engineers (IEEE), May 2019.

[S56] Ömer Uludağ, Henderik A Proper, and Florian Matthes. Investigating the establishment of architecture principles for supporting large-scale agile transformations. In *Proceedings of the 23rd IEEE International Enterprise Distributed Object Computing Conference (EDOC)*, pages 41–50. Institute of Electrical and Electronics Engineers (IEEE), October 2019.

[S57] Ömer Uludağ, Martin Kleehaus, Xian Xu, and Florian Matthes. Investigating the role of architects in scaling agile frameworks. In *Proceedings of the 21st IEEE International Enterprise Distributed Object Computing Conference (EDOC)*, pages 123–132. Institute of Electrical and Electronics Engineers (IEEE), October 2017.

[S58] Yngve Lindsjørn, Gunnar R. Bergersen, Torgeir Dingsøyr, and Dag I. K. Sjøberg. Teamwork quality and team performance: Exploring differences between small and large agile projects. In *Proceedings of the 19th International Conference on Agile Software Development (XP)*, pages 267–274. Springer, 2018.

[S59] Finn Olav Bjørnson and Kathrine Vestues. Knowledge sharing and process improvement in large-scale agile development. In *Proceedings of the Scientific Workshops of XP2016*, page 7. Association for Computing Machinery (ACM), May 2016.

[S60] Maria Paasivaara, Benjamin Behm, Casper Lassenius, and Minna Hallikainen. Large-scale agile transformation at ericsson: a case study. *Empirical Software Engineering*, 23(5):2550–2596, 2018.

[S61] Julian M. Bass. Large-scale offshore agile tailoring: Exploring product and service organisations. In *Proceedings of the Scientific Workshops of XP2016*, page 8. Association for Computing Machinery (ACM), May 2016.

[S62] Torgeir Dingsøyr, Marius Mikalsen, Anniken Solem, and Kathrine Vestues. Learning in the large - an exploratory study of retrospectives in large-scale agile development. In *Proceedings of the 19th International Conference on Agile Software Development (XP)*, pages 191–198. Springer, May 2018.

[S63] Ville T. Heikkilä, Maria Paasivaara, Casper Lassenius, Daniela E. Damian, and Christian Engblom. Managing the requirements flow from strategy to release in large-scale agile development: a case study at ericsson. *Empirical Software Engineering*, 22(6):2892–2936, 2017.

[S64] Antonio Martini, Viktoria Stray, and Nils Brede Moe. Technical-, social- and process debt in large-scale agile: An exploratory case-study. In Rashina Hoda, editor, *Proceedings of the 7th International Workshop on Large-Scale Agile (XP)*, pages 112–119, Cham, May 2019. Springer.

[S65] Muhammad Faisal Abrar, Muhammad Sohail Khan, Sikandar Ali, Umar Ali, Muhammad Faran Majeed, Amjad Ali, Bahrul Amin, and Nasir Rasheed. Motivators for large-scale agile adoption from management perspective: A systematic literature review. *IEEE Access*, 7: 22660–22674, 2019.

[S66] Georgios Papadopoulos. Moving from traditional to agile software development methodologies also on large, distributed projects. *Procedia-Social and Behavioral Sciences*, 175:455–463, 2015.

[S67] Nils Brede Moe, Darja Šmite, Aivars Šāblis, Anne-Lie Börjesson, and Pia Andréasson. Networking in a large-scale distributed agile project. In *Proceedings of the 8th ACM/IEEE International Symposium on Empirical Software Engineering and Measurement (ESEM)*, pages 1–8. Association for Computing Machinery (ACM), September 2014.

[S68] Ville Heikkilä, Maria Paasivaara, Kristian Rautiainen, Casper Lassenius, Towo Toivola, and Janne Järvinen. Operational release planning in large-scale scrum with multiple stakeholders - A longitudinal case study at f-secure corporation. *Information and Software Technology*, 57:116–140, 2015.

[S69] Deepika Badampudi, Samuel A Fricker, and Ana M Moreno. Perspectives on productivity and delays in large-scale agile projects. In *Proceedings of the 14th International Conference on Agile Software Development (XP)*, pages 180–194. Springer, June 2013.

[S70] Viktoria Stray. Planned and unplanned meetings in large-scale projects. In *Proceedings of the 6th International Workshop on Large-scale Agile Development (XP Companion)*, pages 1–5. Association for Computing Machinery (ACM), May 2018.

[S71] Tomas Gustavsson. Practices for vertical and horizontal coordination in the scaled agile framework. In *Proceedings of the 27th International Conference on Information Systems Development (ISD)*, August 2018.

[S72] Mohammad Shameem, Rakesh Ranjan Kumar, Chiranjeev Kumar, Bibhas Chandra, and Arif Ali Khan. Prioritizing challenges of agile process in distributed software development environment using analytic hierarchy process. *Journal of Software: Evolution and Process*, 30(11), 2018.

[S73] Knut H. Rolland, Brian Fitzgerald, Torgeir Dingsøyr, and Klaas-Jan Stol. Problematizing agile in the large: Alternative assumptions for large-scale agile development. In *Proceedings of the 37th International Conference on Information Systems (ICIS)*, December 2016.

[S74] Tor Erlend Fægri and Nils Brede Moe. Re-conceptualizing requirements engineering: findings from a large-scale, agile project. In *Proceedings of the 1st International Workshop on Requirements Engineering in Agile Development (READ)*, page 4. Association for Computing Machinery (ACM), May 2015.

[S75] Magne Jørgensen. Relationships between project size, agile practices, and successful software development: Results and analysis. *IEEE Software*, 36(2):39–43, 2019.

[S76] Knut H. Rolland. Scaling across knowledge boundaries: A case study of A large-scale agile software development project. In *Proceedings of the Scientific Workshops of XP2016*, page 5. Association for Computing Machinery (ACM), May 2016.

[S77] Philipp Diebold, Anna Schmitt, and Sven Theobald. Scaling agile: how to select the most appropriate framework. In *Proceedings of the 6th International Workshop on Large-scale Agile Development (XP Compan-*

*ion)*, pages 1–4. Association for Computing Machinery (ACM), May 2018.

[S78] Martin Kalenda, Petr Hyna, and Bruno Rossi. Scaling agile in large organizations: Practices, challenges, and success factors. *Journal of Software: Evolution and Process*, 30(10):e1954, 2018.

[S79] Maria Paasivaara and Casper Lassenius. Scaling scrum in a large globally distributed organization: A case study. In *Proceedings of the 11th IEEE International Conference on Global Software Engineering (ICGSE)*, pages 74–83. Institute of Electrical and Electronics Engineers (IEEE), August 2016.

[S80] Julian M. Bass. Scrum master activities: Process tailoring in large enterprise projects. In *Proceedings of the 9th IEEE International Conference on Global Software Engineering (ICGSE)*, pages 6–15. Institute of Electrical and Electronics Engineers (IEEE), August 2014.

[S81] Ricardo Britto, Darja Šmite, and Lars-Ola Damm. Software architects in large-scale distributed projects: An ericsson case study. *IEEE Software*, 33(6):48–55, 2016.

[S82] Helena Tendedez, Maria Angela MAF Ferrario, and Jon Whittle. Software development and cscw: Standardization and flexibility in large-scale agile development. pages 1–23. Association for Computing Machinery (ACM), November 2018.

[S83] Darja Šmite, Nils Brede Moe, Aivars Šāblis, and Claes Wohlin. Software teams and their knowledge networks in large-scale software development. *Information and Software Technology*, 86:71–86, 2017.

[S84] Darja Šmite, Nils Brede Moe, Georgiana Levinta, and Marcin Floryan. Spotify guilds: How to succeed with knowledge sharing in large-scale agile organizations. *IEEE Software*, 36(2):51–57, 2019.

[S85] Marthe Berntzen, Nils Brede Moe, and Viktoria Stray. The product owner in large-scale agile: An empirical study through the lens of relational coordination theory. In *Proceedings of the 20th International Conference on Agile Software Development (XP)*, pages 121–136. Springer, May 2019.

[S86] Maria Paasivaara, Outi Väättänen, Minna Hallikainen, and Casper Lassenius. Supporting a large-scale lean and agile transformation by defining common values. In *Proceedings of the XP 2014 International Workshops (XP)*, pages 73–82. Springer, May 2014.

[S87] Ömer Uludağ, Matheus Hauder, Martin Kleehaus, Christina Schimpfle, and Florian Matthes. Supporting large-scale agile development with domain-driven design. In *Proceedings of the 19th International Conference on Agile Software Development (XP)*, pages 232–247. Springer, May 2018.

[S88] Kristian Rautiainen, Joachim von Schantz, and Jarno Vähäniitty. Supporting scaling agile with portfolio management: Case paf.com. In *Proceedings of the 44th Hawaii International Conference on System Sciences (HICSS)*, pages 1–10. Institute of Electrical and Electronics Engineers (IEEE), January 2011.

[S89] Mohammad Shameem, Chiranjeev Kumar, Bibhas Chandra, and Arif Ali Khan. Systematic review of success factors for scaling agile methods in global software development environment: A client-vendor perspective. In *Proceedings of the 24th Asia-Pacific Software Engineering Conference Workshops (APSEC)*, pages 17–24. Institute of Electrical and Electronics Engineers (IEEE), December 2017.

[S90] Julian M. Bass and Andy Haxby. Tailoring product ownership in large-scale agile projects: Managing scale, distance, and governance. *IEEE Software*, 36(2):58–63, 2019.

[S91] Nelson Sekitoleko, Felix Evbota, Eric Knauss, Anna Sandberg, Michel Chaudron, and Helena Holmström Olsson. Technical dependency challenges in large-scale agile software development. In *Proceedings of the 15th International Conference on Agile Software Development (XP)*, pages 46–61. Springer, May 2014.

[S92] Kai Petersen and Claes Wohlin. The effect of moving from a plan-driven to an incremental software development approach with agile practices: An industrial case study. *Empirical Software Engineering*, 15(6):654–693, 12 2010.

[S93] Alexander Scheerer, Saskia Bick, Tobias Hildenbrand, and Armin Heinzl. The effects of team backlog dependencies on agile multiteam systems: A graph theoretical approach. In *Proceedings of the 48th Hawaii International Conference on System Sciences (HICSS)*, pages 5124–5132. Institute of Electrical and Electronics Engineers (IEEE), January 2015.

[S94] Lina Lagerberg, Tor Skude, Pär Emanuelsson, Kristian Sandahl, and Daniel Stahl. The impact of agile principles and practices on large-scale software development projects: A multiple-case study of two projects at ericsson. In *Proceedings of the 7th ACM / IEEE International Symposium on Empirical Software Engineering and Measurement (ESEM)*, pages 348–356. Institute of Electrical and Electronics Engineers (IEEE), October 2013.

[S95] Jan Pries Heje and Malene M Krohn. The safe way to the agile organization. In *Proceedings of the 5th International Workshop on Large-Scale Agile Development (XP)*, page 18, May 2017.

[S96] Hilary Berger and Paul Beynon-Davies. The utility of rapid application development in large-scale, complex projects. *Information Systems Journal*, 19(6):549–570, 2009.

[S97] Nils Brede Moe, Torgeir Dingsøyr, and Knut Rolland. To schedule or not to schedule? an investigation of meetings as an inter-team coordination mechanism in large-scale agile software development. *IJISPM - International Journal of Information Systems and Project Management*, 6(3):45–59, 2018.

[S98] Helena Holmström Olsson and Jan Bosch. Towards continuous validation of customer value. In *Proceedings of the 3rd International Workshop on Large-Scale Agile Development (XP)*, page 3. Association for Computing Machinery (ACM), May 2015.

[S99] Maria Paasivaara, Benjamin Behm, Casper Lassenius, and Minna Hallikainen. Towards rapid releases in large-scale xaas development at ericsson: A case study. In *Proceedings of the 9th IEEE International Conference on Global Software Engineering (ICGSE)*, pages 16–25. Institute of Electrical and Electronics Engineers (IEEE), August 2014.

[S100] Miloš Jovanović, Antònia Mas, Antoni-Lluís Mesquida, and Bojan Lalić. Transition of organizational roles in agile transformation process: A grounded theory approach. *Journal of Systems and Software*, 133:174–194, 2017.

[S101] Wasim Alsaqaf, Maya Daneva, and Roel Wieringa. Understanding challenging situations in agile quality requirements engineering and their solution strategies: Insights from a case study. In *Proceedings of the 26th IEEE International Requirements Engineering Conference (RE)*, pages 274–285. Institute of Electrical and Electronics Engineers (IEEE), August 2018.

[S102] B Veeresh Thummadi, Vishal D. Khapre, and Rosalie J. Ocker. Unpacking agile enterprise architecture innovation work practices: A qualitative case study of a railroad company. In *Proceedings of the 23rd Americas Conference on Information Systems (AMCIS)*, pages 3782 – 3791, August 2017.

[S103] Maria Paasivaara, Sandra Durasiewicz, and Casper Lassenius. Using scrum in a globally distributed project: a case study. *Software Process: Improvement and Practice*, 13(6):527–544, 2008.

[S104] Ömer Uludağ, Martin Kleehaus, Soner Erçelik, and Florian Matthes. Using social network analysis to investigate the collaboration between architects and agile teams: A case study of a large-scale agile development program in a german consumer electronics company. In *Proceedings of the 20th International Conference on Agile Software Development (XP)*, pages 137–153, May 2019.

[S105] Torgeir Dingsøyr, Tor Erlend Fægri, and Juha Itkonen. What is large in large-scale? a taxonomy of scale for agile software development. In *Proceedings of the 15th International Conference on Product-Focused Software Process Improvement (PROFES)*, pages 273–276. Springer, December 2014.

[S106] Ömer Uludağ, Martin Kleehaus, Niklas Reiter, and Florian Matthes. What to expect from enterprise architects in large-scale agile development? A multiple-case study. In *Proceedings of the 25th Americas Conference on Information Systems (AMCIS)*, pages 2683 – 2692, August 2019.

[S107] Antonio Martini, Lars Pareto, and Jan Bosch. Towards introducing agile architecting in large companies: The caffea framework. In *Proceedings of the 16th International Conference on Agile Software Development (XP)*, pages 218–223. Springer, May 2015.

[S108] Necmettin Ozkan and Ayca Tarhan. A review of scaling approaches to agile software development models. *Software Quality Professional*, 21 (4):11–20, 2019.

[S109] Marius Mikalsen, Magne Næsje, Erik André Reime, and Anniken Solem. Agile autonomous teams in complex organizations. In *Proceedings of the 2nd International Workshop on Autonomous Teams (XP)*, pages 55–63. Springer, May 2019.

[S110] Robert L. Nord, Ipek Ozkaya, and Philippe Kruchten. Agile in distress: Architecture to the rescue. In *Proceedings of the XP 2014 International Workshops*, pages 43–57. Springer, May 2014.

[S111] Brian Hobbs and Yvan Petit. Agile methods on large projects in large organizations. *Project Management Journal*, 48(3):3–19, 2017.

[S112] Christoph Johann Stettina and Jeannette Hörz. Agile portfolio management: An empirical perspective on the practice in use. *International Journal of Project Management*, 33(1):140 – 152, 2015.

[S113] Noel Carroll and Kieran Conboy. Applying normalization process theory to explain large-scale agile transformations. In *Proceedings of the 14th International Research Workshop on IT Project Management (IRWITPM)*, January 2019.

[S114] Yngve Lindsjørn and Roza Moustafa. Challenges with lack of trust in agile projects with autonomous teams and fixed-priced contracts. In *Proceedings of the 1st International Workshop on Autonomous Agile Teams (XP Companion)*, pages 1–5, May 2018.

[S115] Tomas Gustavsson. Changes over time in a planned inter-team coordination routine. In *Proceedings of the 7th International Workshop on Large-Scale Agile (XP)*, pages 105–111. Springer, May 2019.

[S116] Tomas Gustavsson. Voices from the teams - impacts on autonomy in large-scale agile software development settings.

[S117] Jaana Nyfjord, Sameer Bathallath, and Harald Kjellin. Conventions for coordinating large agile projects. In *Proceedings of the XP 2014 International Workshops (XP)*, pages 58–72. Springer, May 2014.

[S118] Darja Šmite, Nils Brede Moe, Jonas Wigander, and Hendrik Esser. Corporate-level communities at ericsson: Parallel organizational structure for fostering alignment for autonomy. In *Proceedings of the 20th International Conference on Agile Software Development (XP)*, pages 173–188. Springer, May 2019.

[S119] Magne Jørgensen. Do agile methods work for large software projects? In Juan Garbajosa, Xiaofeng Wang, and Ademar Aguiar, editors, *Proceedings of the 19th International Conference on Agile Software Development (XP)*, pages 179–190. Springer, May 2018.

[S120] Yvan Petit and Carl Marnewick. Earn your wings: A novel approach to deployment governance. In Rashina Hoda, editor, *Proceedings of the 2nd International Workshop on Autonomous Teams (XP)*, pages 64–71. Springer, May 2019.

[S121] Leonor Barroca, Helen Sharp, Torgeir Dingsøyr, Peggy Gregory, Katie Taylor, and Raid AlQaisi. Enterprise agility: A balancing act - a local government case study. In *Proceedings of the 20th International Conference on Agile Softeware Development (XP)*, pages 207–223. Springer, May 2019.

[S122] Teemu Karvonen, Helen Sharp, and Leonor Barroca. Enterprise agility: Why is transformation so hard? In *Proceedings of the 19th International Conference on Agile Softeware Development (XP)*, pages 131–145. Springer, May 2018.

[S123] Petri Kettunen, Maarit Laanti, Fabian Fagerholm, Tommi Mikkonen, and Tomi Männistö. Finnish enterprise agile transformations: A survey study. In *Proceedings of the 7th International Workshop on Large-Scale Agile (XP)*, pages 97–104. Springer, May 2019.

[S124] Abheeshta Putta, Maria Paasivaara, and Casper Lassenius. How are agile release trains formed in practice? a case study in a large financial corporation. In *Proceedings of the 20th International Conference on Agile Software Development (XP)*, pages 154–170. Springer, May 2019.

[S125] Sven Theobald and Philipp Diebold. Interface problems of agile in a non-agile environment. In *Proceedings of the 19th International Conference on Agile Software Development (XP)*, pages 123–130. Springer, May 2018.

[S126] Bjørnar Tessem and Frank Maurer. Job satisfaction and motivation in a large agile team. In *Proceedings of the 8th International Conference on Agile Software Development (XP)*, pages 54–61. Springer, June 2007.

[S127] Roger Sweetman and Kieran Conboy. Portfolios of agile projects: A complex adaptive systems' agent perspective. *Project Management Journal*, 49(6):18–38, 2018.

[S128] Christoph Johann Stettina and Lennard Schoemaker. Reporting in agile portfolio management: Routines, metrics and artefacts to maintain an effective oversight. In *Proceedings of the 19th International Conference on Agile Software Development (XP)*, pages 199–215. Springer, May 2018.

[S129] Maarit Laanti and Petri Kettunen. Safe adoptions in finland: A survey research. In *Proceedings of the 7th International Workshop on Large-Scale Agile (XP)*, pages 81–87. Springer, May 2019.

[S130] Iris Figalist, Christoph Elsner, Jan Bosch, and Helena Holmström Olsson. Scaling agile beyond organizational boundaries: Coordination challenges in software ecosystems. In *Proceedings of the 20th International Conference on Agile Software Development (XP)*, pages 189–206. Springer, May 2019.

[S131] Nirnaya Tripathi, Pilar Rodríguez, Muhammad Ovais Ahmad, and Markku Oivo. Scaling kanban for software development in a multisite organization: Challenges and potential solutions. In *Proceedings of the 16th International Conference on Agile Software Development (XP)*, pages 178–190. Springer, May 2015.

[S132] Abheeshta Putta, Maria Paasivaara, and Casper Lassenius. Benefits and challenges of adopting the scaled agile framework (safe): Preliminary results from a multivocal literature review. In *Product-Focused Software Process Improvement*, pages 334–351, Cham, November 2018. Springer.

[S133] Günther Schuh, Eric Rebentisch, Christian Dölle, Christian Mattern, Georgiy Volevach, and Alexander Menges. Defining scaling strategies for the improvement of agility performance in product development projects. *Procedia CIRP*, 70:29–34, May 2018. 28th CIRP Design Conference 2018, 23-25 May 2018, Nantes, France.

[S134] Wasim Alsaqaf, Maya Daneva, and Roel Wieringa. Quality requirements in large-scale distributed agile projects – a systematic literature review. In *Requirements Engineering: Foundation for Software Quality*, pages 219–234, Cham, February 2017. Springer.

[S135] Hina Saeeda, Hannan Khalid, M. Ahmed, Abu Sameer, and Fahim Arif. Systematic literature review of agile scalability for large scale projects. *International Journal of Advanced Computer Science and Applications*, 6(9):63–75, 2015.

[S136] Abbas Moshref Ravazi and Rodina Ahmad. Agile development in large and distributed environments: A systematic literature review on organizational, managerial and cultural aspects. In *Proceedings of the 8th Malaysian Software Engineering Conference (MySEC)*, pages 216–221. Institute of Electrical and Electronics Engineers (IEEE), September 2014.

## Selected workshop summaries

[W1] Torgeir Dingsøyr and Nils Brede Moe. Research challenges in large-scale agile software development. *ACM SIGSOFT Software Engineering Notes*, 38(5):38–39, 2013.

[W2] Torgeir Dingsøyr and Nils Brede Moe. Towards principles of large-scale agile development - A summary of the workshop at XP2014 and a revised research agenda. In *Proceedings of the XP 2014 International Workshops (XP)*, pages 1–8. Springer, May 2014.

[W3] Nils Brede Moe, Helena Holmström Olsson, and Torgeir Dingsøyr. Trends in large-scale agile development: A summary of the 4th workshop at XP2016. In *Proceedings of the Scientific Workshop of XP2016*, page 1. Association for Computing Machinery (ACM), May 2016.

[W4] Nils Brede oe and Torgeir Dingsøyr. Emerging research themes and updated research agenda for large-scale agile development: a summary of the 5th international workshop at XP2017. In *Proceedings of the 5th International Workshop on Large-Scale Agile Development (XP)*, pages 1–4. Association for Computing Machinery (ACM), May 2017.

[W5] Torgeir Dingsøyr, Nils Brede Moe, and Helena Holmström Olsson. Towards an understanding of scaling frameworks and business agility: a summary of the 6th international workshop at XP2018. In *Proceedings of the 6th International Workshop on Large-scale Agile Development (XP Companion)*, pages 1–4. Association for Computing Machinery (ACM), May 2018.

[W6] Julian M. Bass. Future trends in agile at scale: A summary of the 7th international workshop on large-scale agile development. In *Proceedings of the 7th International Workshop on Large-Scale Agile (XP)*, pages 75–80. Springer, May 2019.

[W7] Leonor Barroca, Torgeir Dingsøyr, and Marius Mikalsen. Agile transformation: A summary and research agenda from the first international workshop. In *Proceedings of the 1st International Workshop on Agile Transformation (XP)*, pages 3–9. Springer, May 2019.

[W8] Viktoria Stray, Nils Brede Moe, and Rashina Hoda. Autonomous agile teams: challenges and future directions for research. In *Proceedings of the 1st International Workshop on Autonomous Agile Teams (XP Companion)*, pages 1–5. Association for Computing Machinery (ACM), May 2018.

[W9] Nils Brede Moe, Viktoria Stray, and Rashina Hoda. Trends and updated research agenda for autonomous agile teams: A summary of the second international workshop at XP2019. In *Proceedings of the 2nd International Workshop on Autonomous Teams (XP)*, pages 13–19. Springer, May 2019.